\documentclass[prx,twocolumn,showpacs,amsmath,amssymb,superscriptaddress,floatfix,reprint, nofootinbib]{revtex4-2}
\usepackage{amsmath}
\usepackage{amssymb}
\usepackage{amsthm}
\usepackage{amsfonts}
\usepackage{dsfont}
\usepackage{listings}
\usepackage{macros}
\usepackage{enumerate}
\usepackage{latexsym}
\usepackage{hyperref}

\usepackage{bbold}

\usepackage{psfrag}
\usepackage{xcolor}
\usepackage{float}

\usepackage{comment}

\usepackage{graphicx}
\usepackage{subfigure}

\makeatletter
\def\l@subsection#1#2{}
\def\l@subsubsection#1#2{}
\makeatother

\begin{document}

\tolerance 10000
\title{Exhaustive Characterization of Quantum Many-Body Scars using Commutant Algebras}
\author{Sanjay Moudgalya}
\email{sanjay.moudgalya@gmail.com}
\affiliation{Department of Physics and Institute for Quantum Information and Matter,
California Institute of Technology, Pasadena, California 91125, USA}
\affiliation{Walter Burke Institute for Theoretical Physics, California Institute of Technology, Pasadena, California 91125, USA}
\affiliation{School of Natural Sciences, Technische Universit\"{a}t M\"{u}nchen (TUM), James-Franck-Str. 1, 85748 Garching, Germany}
\affiliation{Munich Center for Quantum Science and Technology (MCQST), Schellingstr. 4, 80799 M\"{u}nchen, Germany}
\author{Olexei I. Motrunich}
\affiliation{Department of Physics and Institute for Quantum Information and Matter,
California Institute of Technology, Pasadena, California 91125, USA}
\begin{abstract}
We study Quantum Many-Body Scars (QMBS) in the language of commutant algebras, which are defined as symmetry algebras of \textit{families} of local Hamiltonians.
This framework explains the origin of dynamically disconnected subspaces seen in models with exact QMBS, i.e., the large ``thermal" subspace and the small ``non-thermal" subspace, which are attributed to the existence of unconventional non-local conserved quantities in the commutant; hence this unifies the study of conventional symmetries and weak ergodicity breaking phenomena into a single framework.
Furthermore, this language enables us to use the von Neumann Double Commutant Theorem (DCT) to formally write down the exhaustive algebra of \textit{all} Hamiltonians with a desired set of QMBS, which demonstrates that QMBS survive under large classes of local perturbations.
We illustrate this using several standard examples of QMBS, including the spin-1/2 ferromagnetic, AKLT, spin-1 XY $\pi$-bimagnon, and the electronic $\eta$-pairing towers of states; and in each of these cases we explicitly write down a set of generators for the full algebra of Hamiltonians with these QMBS. 
Understanding this hidden structure in QMBS Hamiltonians also allows us to recover results of previous ``brute-force" numerical searches for such Hamiltonians.
In addition, this language clearly demonstrates the equivalence of several unified formalisms for QMBS proposed in the literature, and also illustrates the connection between two apparently distinct classes of QMBS Hamiltonians -- those that are captured by the so-called Shiraishi-Mori construction, and those that lie beyond.
Finally, we show that this framework motivates a precise definition for QMBS that automatically implies that they violate the conventional Eigenstate Thermalization Hypothesis (ETH), and we discuss its implications to dynamics.
\end{abstract}
\date{\today}
\maketitle
%


\tableofcontents

\section{Introduction}
\label{sec:intro}
The dynamics of isolated quantum systems has been a subject of much recent interest.
Such systems evolve unitarily, hence all the information of the dynamics of the system can be deduced from the eigenstates of the time-evolution operator, e.g., the Hamiltonian. 
In generic non-integrable systems, where any initial state at finite energy-density is expected to thermalize under time-evolution, the eigenstates are themselves expected to be thermal, which leads to the Eigenstate Thermalization Hypothesis (ETH)~\cite{deutsch1991quantum, srednicki1994chaos, rigol2008thermalization, d2016quantum, mori2018thermalization}. 
The conventional form of this hypothesis is violated for all the eigenstates in systems that do not thermalize, e.g., in integrable or many-body localized systems~\cite{alba2015eth, nandkishore2015many, abanin2019mbl}.
An additional possibility was recently discovered in non-integrable systems,  where some ``anomalous" eigenstates that do not satisfy the conventional form of ETH exist amidst most eigenstates that do satisfy ETH.
Such systems are said to exhibit ``weak ergodicity breaking," which is further categorized into the phenomena of quantum many-body scarring or Hilbert space fragmentation~\cite{serbyn2020review, papic2021review, moudgalya2021review, chandran2022review}, depending on the scaling of the number of anomalous eigenstates with system size.
These weak ergodicity breaking phenomena have gathered much attention due to their natural occurrence in several experimentally relevant contexts, e.g., quantum many-body scarring is responsible for long-lived revivals in several Rydberg and cold atom experiments~\cite{bernien2017probing, turner2017quantum, turner2018quantum,  su2022observation,  desaules2022weak, halimeh2022robust}, and Hilbert space fragmentation plays a role in slow dynamics in the presence of a strong electric field~\cite{sala2020fragmentation, khemani2020localization, moudgalya2019thermalization, guardado2020subdiffusion, scherg2020observing, sengupta2021phases, kohlert2021experimental}, as realized in cold atoms in tilted lattices. 
In this work, we will be focusing on quantum many-body scarring, where the anomalous eigenstates, referred to as Quantum Many-Body Scars (QMBS)~\cite{turner2017quantum}, constitute a vanishing fraction of the full Hilbert space.
There has been a lot of recent theoretical progress in understanding systems that exhibit QMBS, with the discovery of several non-integrable models with exactly solvable QMBS eigenstates.
These include exact equally-spaced towers of QMBS in several well-known systems, starting from the one-dimensional AKLT model~\cite{moudgalya2018a, moudgalya2018b}, to several spin models~\cite{schecter2019weak, iadecola2020quantum, chattopadhyay2019quantum, shibata2020onsager}, also including the Hubbard models and their deformations in any number of dimensions~\cite{vafek2017entanglement, moudgalya2020eta, mark2020eta}; as well as several examples of isolated QMBS~\cite{shiraishi2017eth, ok2019topological, lee2020exact, surace2020weakergodicity, moudgalya2021review}, including in the well-studied PXP models in one and higher dimensions~\cite{lin2019exact, lin2020quantum}.
For a comprehensive review of the literature on this subject, we refer readers to the recent reviews~\cite{serbyn2020review, papic2021review, moudgalya2021review, chandran2022review}.
Due to the abundance of examples of QMBS, it is highly desirable to try to capture them in a single framework or obtain a systematic procedure for their construction.
Progress in this direction has been made for certain classes of isolated QMBS, where large classes of QMBS eigenstates can be systematically embedded into spectra of non-integrable models using so-called Shiraishi-Mori construction~\cite{shiraishi2017eth}.
In addition, there have been several attempts to unify models exhibiting towers of QMBS into a single framework, with varying degrees of success~\cite{shiraishi2017eth, mark2020unified, moudgalya2020eta, mark2020unified, pakrouski2020many, ren2020quasisymmetry, odea2020from, pakrouski2021group, ren2021deformed, rozon2022constructing, rozon2023broken} (see \cite{moudgalya2021review} for a broad overview of some of these approaches). 
Moreover, it has been noted that even for a given set of QMBS states, multiple ``parent Hamiltonians" with those states as eigenstates can be constructed. 
While some such Hamiltonians can be systematically understood using the unified formalisms, several of them, including the AKLT model~\cite{moudgalya2018a, moudgalya2018b}, are yet to be satisfactorily understood within any of these systematic constructions.
A striking feature of all these exact examples is that the QMBS eigenstates appear to be uncorrelated from the rest of the spectrum to a large extent.
For example, in many cases, it is known that the QMBS eigenstates can be made to move in and out of the bulk energy spectrum by tuning parameters in the Hamiltonian, and in some cases they can also be the ground states of their respective Hamiltonians~\cite{mark2020unified, moudgalya2020large, moudgalya2020eta, mark2020eta, odea2020from}.
This is reminiscent of eigenstates within quantum number sectors of conventional symmetries, where level crossings can occur between eigenstates in different sectors by tuning a parameter in a Hamiltonian, for example as a function of a magnetic field in $SU(2)$-symmetric systems.
Indeed, in models exhibiting exact QMBS, the Hilbert space $\mH$ is said to ``fracture" into dynamically disconnected blocks as~\cite{bull2020quantum, serbyn2020review, papic2021review, moudgalya2021review}
\begin{equation}
    \mH = \mHtherm \oplus \mHscar,
\label{eq:scarfactorize}
\end{equation}
where $\mHtherm$ and $\mHscar$ are ``large" and ``small" subspaces\footnote{In particular, $\Dtherm \defn \dim(\mHtherm)$, $\Dscar \defn \dim(\mHscar)$, and $D \defn \dim(\mH)$ satisfy $\Dtherm/D \rightarrow 1$ and $\Dscar/D \rightarrow 0$ as system size $L \rightarrow \infty$. This can happen either when $\Dscar \sim \mathcal{O}(1)$ (e.g., isolated QMBS), $\Dscar \sim \poly(L)$ (e.g., towers of QMBS), or when $\Dscar \sim \exp(c L)$ (e.g., weak fragmentation)~\cite{moudgalya2021review}.} that are invariant under the action of $H$, and the subspaces are such that eigenstates of $H$ in $\mHtherm$ typically satisfy the conventional form of ETH, whereas eigenstates in $\mHscar$ have anomalous properties and are the QMBS.
While a decomposition of the form of Eq.~(\ref{eq:scarfactorize}) is expected if eigenstates in $\mHscar$ and $\mHtherm$ differ by some symmetry quantum numbers, in the typical model realizations the QMBS do not differ from the rest of the spectrum under any obvious symmetries. 
On the other hand, for any finite-dimensional Hilbert space and a given Hamiltonian $H$, Eq.~(\ref{eq:scarfactorize}) is trivially true, since one can always use eigenbasis of $H$ to split the Hilbert space in multiple ways. 
This necessitates a more precise definition for the blocks $\mHtherm$ and $\mHscar$ in Eq.~(\ref{eq:scarfactorize}).
Important progress in this direction has been made in the literature in two works~\cite{shiraishi2017eth, pakrouski2020many}.
First, Shiraishi and Mori introduced an embedding formalism in \cite{shiraishi2017eth, shiraishimorireply2018}, where the QMBS were part of a ``target space" that was annihilated by a set of strictly local operators, a condition that is typically satisfied by tensor network states~\cite{cirac2021matrix}.
This property was then used to construct families of Hamiltonians with those QMBS as eigenstates, hence for these Hamiltonians $\mHscar$ in Eq.~(\ref{eq:scarfactorize}) refers to the target space. 
More recently, following realization how particularly simple and familiar states -- so-called $\eta$-pairing states~\cite{yang1989eta, vafek2017entanglement} -- can appear as scars in deformed Hubbard models~\cite{mark2020eta, moudgalya2020eta}, Pakrouski {\it et.~al.} noted in \cite{pakrouski2020many, pakrouski2021group} that QMBS in certain systems can be understood as singlets (i.e., one-dimensional representations) of certain Lie algebras, and this perspective brought to the fore the spatial structure (in fact, lack thereof in any dimension) in these QMBS.
In particular, they constructed sets of local operators that are $D$-dimensional representations of the generators of a semisimple Lie algebra, where $D = \dim(\mH)$, and their unique decomposition into smaller-dimensional irreducible representations (irreps) splits the Hilbert space into blocks that transform under various irreps.
Reference \cite{pakrouski2020many} used this structure to systematically construct families of local Hamiltonians that preserve the states in the Hilbert space that transform under one-dimensional irreps (i.e., the singlets) as QMBS, while mixing all other states.
This resulted in models where Eq.~(\ref{eq:scarfactorize}) holds, where $\mHscar$ is the subspace spanned by the Lie group singlets.
Apart from these classes of systems, a non-trivial definition of the scar subspace $\mHscar$, i.e., one that does not directly refer to the individual eigenstates themselves, does not exist for other examples of QMBS, and it is still not clear if all examples of QMBS can be understood within the frameworks proposed in \cite{shiraishi2017eth,pakrouski2020many}.
This calls for a more general understanding of the fracture of Hilbert space into dynamically disconnected blocks such as Eq.~(\ref{eq:scarfactorize}).
A similar question arises in systems exhibiting Hilbert space fragmentation~\cite{sala2020fragmentation, khemani2020localization, moudgalya2019thermalization, yang2019hilbertspace}, where the Hilbert space ``fragments" into exponentially many dynamically disconnected blocks, as opposed to two in Eq.~(\ref{eq:scarfactorize}).
Recently, in \cite{moudgalya2021hilbert}, we showed that the blocks in fragmented systems can be understood by studying the local and non-local conserved quantities that commute with \textit{each term} of the Hamiltonian.
The algebra of all such conserved quantities was referred to as the ``commutant algebra," which is the centralizer of the algebra generated by the terms of the Hamiltonian, or more generally the individual parts of a family of Hamiltonians; and the latter algebra was referred to as the ``bond algebra"~\cite{nussinov2009bond, cobanera2010unified, cobenera2011bond} when the individual parts are strictly local operators, or more generally as a ``local algebra," when the individual parts can include extensive local (i.e., sums of strictly local) operators.  
In a parallel work~\cite{moudgalya2022from}, we applied this formalism to understand conserved quantities of several standard Hamiltonians, including the spin-1/2 Heisenberg model, several free-fermion models, and the Hubbard model.
There we showed that this captures all of the conventional on-site unitary symmetries of those models, and in some cases it also revealed examples of unconventional ``non-local" symmetries that manifest themselves in degeneracies of eigenstates that are not captured by on-site unitary symmetries. 
In this work, we extend the formalism of commutant algebras developed in \cite{moudgalya2021hilbert, moudgalya2022from} to understand Eq.~(\ref{eq:scarfactorize}) in models that exhibit QMBS.
As we will show, the unified formalisms for QMBS in \cite{shiraishi2017eth, pakrouski2020many} can be recast in terms of bond and local algebras, their commutants, and their singlets, which also elucidates their connections to other proposed unified formalisms in \cite{odea2020from, ren2020quasisymmetry}.
Further, stating them in this language has multiple benefits, which we briefly summarize below.
First, it allows the application of the Double Commutant Theorem (DCT) that enables the construction of (the exhaustive algebra of) all local Hamiltonians that possess a given set of QMBS as eigenstates, which is the analog of constructing the algebra of all symmetric operators corresponding to conventional symmetries, discussed in \cite{moudgalya2022from}.
This circumvents the ``guess-work" or ``brute-force" approaches used for such purposes in earlier works~\cite{mark2020unified, moudgalya2020large, mark2020eta, moudgalya2020eta, odea2020from}, and enables the construction of numerous local perturbations that exactly preserve a given set of QMBS, demonstrating that it a much less fine-tuned property.
Further, the local/commutant algebra formalism also allows us to conjecture certain constraints on the spectra of local Hamiltonians that contain QMBS, e.g., that certain sets of QMBS necessarily appear as equally spaced towers in the spectrum of any Hamiltonian containing them as eigenstates.
Second, this formalism reveals the distinction between two types of Hamiltonians with QMBS that appear in the literature, the ``Shiraishi-Mori" type that can be captured by the Shiraishi-Mori construction~\cite{shiraishi2017eth} and the ``as-a-sum" type that lie beyond the Shiraishi-Mori construction; in the algebra language these correspond to two distinct types of ``symmetric" Hamiltonians that can be constructed starting from a set of strictly local generators of a bond algebra, which we refer to as Type I and Type II symmetric Hamiltonians respectively.
Hamiltonians believed to be of the latter type include the Dzyaloshinskii-Moriya Hamiltonian~\cite{mark2020eta} and the AKLT Hamiltonian~\cite{moudgalya2018a, moudgalya2018b, mark2020unified, moudgalya2020large, odea2020from}, and we demonstrate their distinctions and connections to Hamiltonians obtained from the Shiraishi-Mori formalism via the exhaustive algebra of QMBS Hamiltonians obtained from the commutant language, hence resolving a previously open question~\cite{mark2020unified, odea2020from, tang2020multi}.
Third, this language also motivates a concrete definition of QMBS eigenstates, and we propose that they are always simultaneous eigenstates of multiple non-commuting local operators. 
As we will discuss, this definition automatically implies that these states violate ETH, due to the non-uniqueness of local Hamiltonian reconstruction from the state~\cite{garrison2018does, qi2019determininglocal}. 
Finally, perhaps most importantly, this formalism hence provides a very general framework for understanding systems with QMBS and elucidates the precise connections of systems exhibiting QMBS to those with other conventional symmetries and/or Hilbert space fragmentation, allowing to incorporate several phenomena involving dynamically disconnected subspaces~\cite{moudgalya2021review} into a single framework.
This paper is organized as follows.
In Sec.~\ref{sec:commutants}, we briefly review the concepts of bond, local, and commutant algebras and the DCT, and in Sec.~\ref{subsec:QMBSfamilies} we use these concepts to formulate the main ideas presented in this work. 
In Sec.~\ref{sec:QMBSunified}, we revisit previously proposed symmetry-based unified frameworks for understanding QMBS, and describe them in the language of local and commutant algebras. 
In Sec.~\ref{sec:scarexamples}, we study several standard examples of QMBS, and we construct the full algebras of Hamiltonians that possess these QMBS as eigenstates, and discuss implications of the DCT. 
Then in Sec.~\ref{sec:eth} we propose a definition for QMBS motivated from this framework, and we discuss implications for thermalization and dynamics.
We conclude with open questions in Sec.~\ref{sec:conclusion}.
\section{Recap of Local and Commutant Algebras}\label{sec:commutants}
We now briefly review some key concepts of bond, local, and commutant algebras relevant for this work, and we refer to \cite{moudgalya2021hilbert, moudgalya2022from} for more detailed discussions on the general properties of these algebras.
\subsection{Definition}
Focusing on systems with a $D$-dimensional tensor product Hilbert space $\mH$ of local degrees of freedom on some lattice, we are interested in Hamiltonians of the form 
\begin{equation}
    H = \sumal{\alpha}{}{J_{\alpha} \hH_{\alpha}}, 
\label{eq:genhamil}
\end{equation}
where $\{\hH_\alpha\}$ is some set of local operators, either \textit{strictly local} with support on a few nearby sites on the lattice or \textit{extensive local}, i.e., a sum of such terms, and $\{J_\alpha\}$ is an arbitrary set of coefficients.
Corresponding to this family, we can define the local and commutant algebras $\mA$ and $\mC$ as
\begin{equation}
    \mA = \lgen \{\hH_\alpha\} \rgen,\;\;\;\mC = \{\hO :\;\;[\hH_\alpha, \hO] = 0\;\;\;\forall \alpha\}.
\label{eq:localcommalgebradefn}
\end{equation}
Here $\mC$ and $\mA$ can be viewed as the ``symmetry algebra" and the algebra of all ``symmetric operators" respectively, and we use $\lgen \cdots \rgen$ to denote the associative algebra generated by (linear combinations with complex coefficients of arbitrary products of) the enclosed elements and the identity operator $\mathds{1}$, also assumed to be closed under Hermitian conjugation of operators (``$\dagger$-algebra"), which is natural in our setting.
As a concrete example, we consider $SU(2)$ symmetry for one-dimensional spin-$1/2$ systems with $L$ sites.
In the commutant language, this symmetry can be expressed in terms of the pair of algebras~\cite{moudgalya2022from}
\begin{equation}
    \mA_{SU(2)} = \lgen \{\vec{S}_j \cdot \vec{S}_{j+1}\}\rgen,\;\;\;\mC_{SU(2)} = \lgen S^x_{\tot}, S^y_{\tot}, S^z_{\tot}\rgen,
\label{eq:SU2AC}
\end{equation}
where $S^\alpha_{\tot} \defn \sum_j{S^\alpha_j}$. 
In other words, starting with a family of Heisenberg models of the form of Eq.~(\ref{eq:genhamil}) with the generators $\{\hH_\alpha\} = \{\vec{S}_j\cdot\vec{S}_{j+1}\}$, the commutant $\mC_{SU(2)}$ can be derived using Eq.~(\ref{eq:localcommalgebradefn}); hence $SU(2)$ is the \textit{complete} symmetry of the family of Heisenberg models~\cite{moudgalya2022from}.
Since the Heisenberg term can also be related to the two-site permutation term that acts as $P_{j,j+1}\ket{\sigma \tau}_{j,j+1} = \ket{\tau \sigma}_{j,j+1}$, $\mA_{SU(2)}$ is also the group algebra of the permutation group $S_L$ on $L$ sites. 

\subsection{Hilbert Space Decomposition}\label{subsec:Hilbertdecomp}
Given such $\dagger-$algebras $\mA$ and $\mC$ that are centralizers of each other in the algebra of all operators on the full Hilbert space $\mH$, their irreps can be used to construct a bipartition~\cite{zanardi2001virtual, bartlett2007reference, readsaleur2007, moudgalya2021hilbert, moudgalya2022from} of the Hilbert space, i.e., a basis where the operators $\hh_{\mA}$ and $\hh_{\mC}$ in $\mA$ and $\mC$ respectively act as
\begin{equation}
    \hh_\mA = \bigoplus_{\lambda} (M^\lambda_{D_\lambda}(\hh_\mA) \otimes \mathds{1}_{d_\lambda}),\;\;\;\hh_{\mC} = \bigoplus_{\lambda} (\mathds{1}_{D_\lambda} \otimes N^\lambda_{d_\lambda}(\hh_\mC)),
\label{eq:Hilbertdecomp}
\end{equation}
where $D_\lambda$ and $d_\lambda$ are the dimensions of the irreps of $\mA$ and $\mC$, $M^\lambda_{D_\lambda}(\hh_\mA)$ and $N^\lambda_{d_\lambda}(\hh_\mC)$ are $D_\lambda$-dimensional and $d_\lambda$-dimensional matrices respectively, and arbitrary such matrices are realized in the corresponding algebras.
Equation~(\ref{eq:Hilbertdecomp}) can be simply viewed as the matrix forms of the operators in the basis in which all the operators in $\mA$ or $\mC$ are simultaneously (maximally) block-diagonal.
Since the Hamiltonians we are interested in are part of $\mA$, this decomposition can be used to precisely define dynamically disconnected ``Krylov"  subspaces for all Hamiltonians in the family~\cite{moudgalya2021hilbert}.
Hence Eq.~(\ref{eq:Hilbertdecomp}) implies the existence of $d_\lambda$ number of identical $D_\lambda$-dimensional Krylov subspaces for each $\lambda$.
In systems with only conventional symmetries such as $U(1)$ or $SU(2)$, these correspond to regular quantum number sectors~\cite{moudgalya2022from}, whereas in fragmented systems they are the exponentially many Krylov subspaces~\cite{moudgalya2021hilbert}.
For example, in the case of the $SU(2)$ algebras of Eq.~(\ref{eq:SU2AC}), we have $0 \leq \lambda \leq L/2$ where $\lambda(\lambda + 1)$ is the eigenvalue of $S^2 \defn \sum_\alpha{(S^\alpha_{\tot})^2}$, and $(D_\lambda, d_\lambda) = (\binom{L}{L/2 + \lambda} - \binom{L}{L/2+\lambda+1}, 2\lambda + 1)$ for even $L$, which denote the sizes of the quantum number sectors and their degeneracies respectively.
Note that $\{D_\lambda\}$ and $\{d_\lambda\}$ are the dimensions of the irreducible representations of the permutation group $S_L$ (and hence that of $\mA_{SU(2)}$) and the group $SU(2)$ (and hence that of $\mC_{SU(2)}$) respectively.
\subsection{Singlets}\label{subsec:singlets}
In the decomposition of Eq.~(\ref{eq:Hilbertdecomp}), it is sometimes possible to have $D_\lambda = 1$ for some $\lambda$, which means the existence of simultaneous eigenstates of all the operators in the algebra $\mA$. 
We refer to these eigenstates as ``singlets" of the algebra $\mA$, and in \cite{moudgalya2022from} we discussed examples of singlets that appear in standard Hamiltonians. 
For the bond algebra $\mA_{SU(2)}$, the singlets are the ferromagnetic multiplet of states~\cite{moudgalya2022from} given by
\begin{equation}
    \ket{\Psi_n} \defn (S^-_{\tot})^n\ket{F},\;\;\;\; \ket{F} \defn \ket{\uparrow \cdots \uparrow},
\label{eq:FMtower}
\end{equation}
where $S^-_{\tot} \defn \sum_j{S^-_j}$.
Since $\mA_{SU(2)}$ is the group algebra of the permutation group $S_L$, its singlets are simply the states invariant under arbitrary permutations of sites, which are spanned by $\{\ket{\Psi_n}\}$.
Since these are states with $S^2$ eigenvalue $\frac{L}{2}(\frac{L}{2} + 1)$, following the discussion in Sec.~\ref{subsec:Hilbertdecomp} they appear in the decomposition of Eq.~(\ref{eq:Hilbertdecomp}) with $(D_\lambda, d_\lambda) = (1, L+1)$.
In general, $\mA$ could have many sets of singlets that are degenerate within each set and non-degenerate between the sets, e.g., when it has irreps such that $D_\lambda = D_{\lambda'} = 1$ for some $\lambda \neq \lambda'$, and the different sets of singlets differ by their eigenvalues under some operators in $\mA$, while all singlets within a set have the same eigenvalue under all operators in $\mA$.
The projectors onto the singlet states are all in the commutant algebra $\mC$, and are thus examples of eigenstate projectors that can be viewed as conserved quantities of the family of Hamiltonians we are interested in.
For the case of degenerate singlets $\ket{\psi}$ and $\ket{\psi'}$, ``ket-bra" operators $\ketbra{\psi}{\psi'}$ are also in the commutant $\mC$.
As we will discuss in Secs.~\ref{sec:QMBSunified} and \ref{sec:scarexamples}, the singlets of local algebras will define the subspace $\mHscar$ of Eq.~(\ref{eq:scarfactorize}).
\subsection{Double Commutant Theorem (DCT)}
An important property satisfied by $\mA$ and $\mC$ is the Double Commutant Theorem (DCT)~\cite{landsman1998lecture, harlow2017, kabernik2021reductions, moudgalya2022from}, which for our purposes is the following statement.
\begin{theorem}[DCT]
\label{thm:DCT}
Given a finite-dimensional Hilbert space $\mH$ and an algebra $\mA = \lgen \{\hH_\alpha\} \rgen$ where $\{\hH_\alpha\}$ is a set of Hermitian operators, and its centralizer $\mC$, then $\mA$ is the centralizer of $\mC$.
\end{theorem}
\noindent In other words, the DCT implies that $\mA$ and $\mC$ are centralizers of each other in the space of all operators in the Hilbert space $\mH$.
The DCT has some deep implications when applied to local and commutant algebras, and in principle allows us to exhaustively construct all operators that commute with some set of conserved quantities~\cite{moudgalya2022from}.
In particular, given a set of conserved quantities $\{Q_k\}$ that generate a commutant algebra $\mC = \lgen \{Q_k\} \rgen$, if we are able to determine a set of local generators $\{\hH_\alpha\}$ for its centralizer $\mA$, i.e., if $\mA = \lgen \{\hH_\alpha\} \rgen$, then we can in principle exhaustively construct all local operators that commute with the conserved quantities $\{Q_k\}$ starting from $\{\hH_\alpha\}$. 
In the case of $SU(2)$ symmetry, where the algebras are shown in Eq.~(\ref{eq:SU2AC}), this is the statement that \textit{all} $SU(2)$-symmetric can be expressed in terms of the Heisenberg terms $\{\vec{S}_j \cdot \vec{S}_{j+1}\}$, which are the generators of $\mA_{SU(2)}$.
Locality considerations bring new aspects to the application of the DCT.
First, given a commutant $\mC$, there is the obvious question whether $\mA$ can be generated by local operators; however, we are not concerned about this in this work, since we start from a local algebra $\mA$ and then determine its commutant $\mC$.
More importantly for us, given an extensive local operator in $\mA$, i.e., a Hamiltonian symmetric under operators in $\mC$, we wish to express it in terms of the local generators of $\mA$.
The DCT guarantees that an expression exists, and in several situations we can use it to constrain the allowed forms of extensive local operators in $\mA$.
For example, in \cite{moudgalya2022from} we showed that when $\mC$ is completely generated by on-site unitary symmetries, such as in the $SU(2)$ case of Eq.~(\ref{eq:SU2AC}), all extensive local operators in the corresponding bond algebra $\mA$ (i.e., all symmetric Hamiltonians) can be expressed as sums of symmetric strictly local operators.
Further, in systems with dynamical symmetries, we proved that \textit{all} extensive local operators contain equally spaced towers of states in their spectra. 
\section{Families of Models with QMBS in the Commutant Algebra Language}\label{subsec:QMBSfamilies}
\begin{figure*}
\includegraphics[scale=0.7]{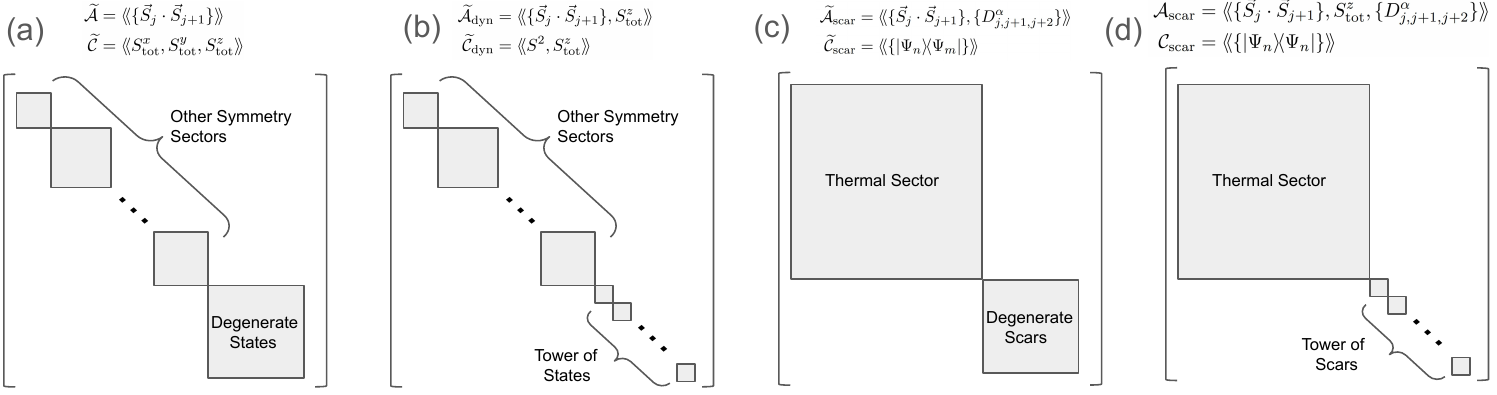}
\caption{
Summary of the local and commutant algebras and associated block decompositions that appear enroute the exhaustive description of towers of QMBS.
We show explicit algebras for the ferromagnetic states $\{\ket{\Psi_n}\}$ as QMBS, but similar block decompositions hold for other examples we study.
Particular Hamiltonians that realize these block decompositions have been studied previously, but the algebra language provides an exhaustive characterization of such Hamiltonians.
This framework is also much more general, treats symmetries and scars on similar footing, and captures examples such as the AKLT tower of QMBS that do not fit into previous frameworks.
(a) Pre-bond algebra generated by a set of strictly local terms that have a degenerate set of common eigenstates, which are the ``target states."
Such an algebra usually has a larger commutant which leads to other symmetry sectors in the block decomposition; this could be a conventional non-Abelian symmetry such as $SU(2)$, although not always.
(b) An extensive local ``lifting operator" added to the generators of the pre-bond algebra lifts degeneracies between the target states, while still preserving some symmetry sectors.
In the ferromagnetic example, this corresponds to a dynamical $SU(2)$ symmetry, and the states are split into an equally spaced tower in any local Hamiltonian from this algebra.
(c) Additional symmetries of the pre-bond algebra can be broken by the inclusion of terms that preserve the degenerate target states while mixing the remaining symmetry sectors into a single large thermal block; the target states are now examples of degenerate scars.
(d) The lifting operator and the terms that break other symmetries of the pre-bond algebra can be added to obtain the typical decomposition in the case of QMBS systems, into a thermal block and a scar block composed of non-degenerate scar states.
In the ferromagnetic example, we conjecture that the scar states appear as an equally spaced tower in any local Hamiltonian in this algebra.
This is the algebra that ultimately exhaustively characterizes the QMBS, while the algebras in (a)-(c) are simply motivating steps.
}
\label{fig:blocks}
\end{figure*}
Having reviewed the formalism of local algebras and their commutants, we can now provide an overview of the approach to defining QMBS and determining exhaustive families of models with exact scars.
As a concrete example for illustration throughout this section and the next, we consider models where the QMBS are the ferromagnetic tower of states $\{\ket{\Psi_n}\}$ of Eq.~(\ref{eq:FMtower}). 
Several Hamiltonians with these states as QMBS have been studied in the literature~\cite{mark2020eta, moudgalya2021review, dooley2021robust}, e.g.,
\begin{equation}
    H_{\scar} = \sumal{j = 1}{L}{J_j \vec{S}_j\cdot\vec{S}_{j+1}} + h \sumal{j = 1}{L}{S^z_j} + D \sumal{j = 1}{L}{(\vec{S}_j \times \vec{S}_{j+1})\cdot \hat{z}}.
\label{eq:FMscar}
\end{equation}
$\{\ket{\Psi_n}\}$ are exact eigenstates of $H_{\scar}$, and for generic values of $\{J_j\}$ and $D$, they appear as a tower of QMBS with splitting of $2h$; hence when $h = 0$ they are examples of \textit{degenerate} QMBS.\footnote{Throughout this work, whenever we refer to a given set of states as QMBS of a family of Hamiltonians, we simply mean that the states in that set are \textit{eigenstates} for all Hamiltonians in that family. In particular, we are not concerned about when such states are in the middle of the full spectrum of a given Hamiltonian from the family, since there is likely a generic choice of couplings for which that is the case.}
We refer to the first sum in Eq.~(\ref{eq:FMscar}) as the Heisenberg Hamiltonian, and the last sum as the Dzyaloshinskii-Moriya Interaction (DMI) Hamiltonian~\cite{mark2020eta, kunimi2023proposal}.
For the sake of brevity, we directly state the key results here, while detailed justifications and proofs can be found in Sec.~\ref{sec:scarexamples} and the appendices.
\subsection{General Structure of the Commutants}
Our primary aim in this work is to show that for many examples of QMBS, algebras $\mA_{\scar}$, generated by a set of local operators, with commutants $\mC_{\scar}$, spanned by projectors or ket-bra operators of QMBS eigenstates, exist and can be explicitly constructed.
Corresponding to a set of QMBS eigenstates and their degeneracies, say $\{\ket{\psi_{n,\alpha}}\}$, where all the $\ket{\psi_{n,\alpha}}$'s for fixed $n$ are degenerate, we wish to construct a local algebra $\mA_{\scar}$ such that its commutant is given by
\begin{equation}
\mC_{\scar} = \lgen \{\ketbra{\psi_{n,\alpha}}{\psi_{n,\beta}}\}\rgen ~.
\label{eq:scargencom}
\end{equation}
The operators $\{\ketbra{\psi_{n,\alpha}}{\psi_{n,\beta}}\}$ are then examples of ``non-local" conserved quantities of Hamiltonians in $\mA_{\scar}$, and the exhaustive set of Hamiltonians with these QMBS eigenstates can be constructed using the local generators of $\mA_{\scar}$.
In some situations, the construction of physically relevant Hamiltonians might sometimes call for commutants consisting of both the QMBS and some other natural conserved quantities $\{Q_\mu\}$ such as $U(1)$ spin conservation, in which case we would be interested in constructing local algebras $\mA_{\text{sym}-\scar}$ corresponding to commutants such as $\mC_{\text{sym}-\scar} = \lgen \{\ketbra{\psi_{n, \alpha}}{\psi_{n, \beta}}\}, \{Q_\mu\} \rgen$.
\subsection{Degeneracies and Lifting Operators}\label{subsec:degnondegQMBS}
Note that in cases with multiple QMBS, there can sometimes be some arbitrariness in the algebras we are interested in, depending on which set of degeneracies among the scar states we choose to preserve in the Hamiltonians we are interested in.
For example, for the same set of QMBS $\{\ket{\Psi_n}\}$ of Eq.~(\ref{eq:FMtower}), we could construct two distinct algebras $\tmA_{\scar}$ and $\mA_{\scar}$ which correspond to the commutants $\tmC_{\scar} = \lgen \{\ketbra{\Psi_m}{\Psi_n}\}\rgen$ and $\mC_{\scar} = \lgen \{\ketbra{\Psi_n}\}\rgen$ respectively, and hence $\{\ket{\Psi_n}\}$ are degenerate or non-degenerate eigenstates respectively.
The expressions for these algebras are given by
\begin{align}
    &\tmA^{\FM}_{\scar} = \lgen \{\vec{S}_j\cdot\vec{S}_{j+1}\}, \{D^\alpha_{j,j+1, j+2}\}\rgen,\nn \\ 
    &\mA^{\FM}_{\scar} = \lgen \{\vec{S}_j\cdot\vec{S}_{j+1}\},\{D^\alpha_{j-1, j,j+1}\},  S^z_{\tot}\rgen, 
\label{eq:FMalgebras}
\end{align}
where $D^\alpha_{j_1, j_2, j_3} \defn \sum_{k = 1}^3{(\vec{S}_{j_k} \times \vec{S}_{j_{k+1}})\cdot\widehat{\alpha}}$ is the three-site DMI term, where the sum over $k$ is modulo $3$ (i.e., the three sites are considered as forming a loop hosting the DMI term).
Note that $\tmA^{\FM}_{\scar}$ is an example of a bond algebra since it is generated by a set of strictly local terms, whereas $\mA^{\FM}_{\scar}$ is not, since its generators include an extensive local operator $S^z_{\tot}$, which we refer to as a \textit{lifting operator}. 
\begin{definition}
Given a set of QMBS, we refer to any extensive local operator that lifts the degeneracies of the QMBS as a \textit{lifting operator}. 
\end{definition}
\noindent We find that this is a general feature of the local algebras for which the QMBS states are non-degenerate, i.e., they cannot be generated by strictly local operators.
Examples of lifting operators in various other examples of QMBS are shown in Tab.~\ref{tab:scarexamples}, and we discuss them in more detail in Sec.~\ref{sec:scarexamples} and the appendices.
\subsection{Hilbert Space Decomposition}
Denoting the span of the QMBS states $\{ \ket{\psi_{n,\alpha}} \}$ as $\mH_{\scar}$ and its dimension as $D_{\scar}$, the structure of the commutant $\mC_{\scar}$ such as in Eq.~(\ref{eq:scargencom}) implies that $\mA_{\scar}$ acts irreducibly in the orthogonal complement to $\mH_{\scar}$, which is then naturally viewed as $\mH_\text{therm}$ from Eq.~(\ref{eq:scarfactorize}).
Indeed, in this case $\mA_{\scar}$ can realize any operator acting in $\mH_\text{therm}$, and it is natural to expect a generic Hamiltonian from $\mA_{\scar}$ to be ``thermal'' (i.e., with random-matrix-like level spacings) in this subspace.
In the decomposition in Eq.~(\ref{eq:Hilbertdecomp}), the subspace $\mH_\therm$ corresponds to the block $\lambda_\therm$ which is the only block other than those of the algebra singlets; the corresponding $D_{\therm} = D - D_{\scar}$ and $d_{\therm} = 1$, thus connecting the decompositions of Eqs.~(\ref{eq:scarfactorize}) and (\ref{eq:Hilbertdecomp}).
For the ferromagnetic tower of QMBS of Eq.~(\ref{eq:FMtower}), we hence have $\mH_{\scar} = \text{span}\{\ket{\Psi_n}\}$, $D_{\scar} = L + 1$, and $D_{\therm} = 2^L - L - 1$, as shown in Fig.~\ref{fig:blocks}(d).
Note that the commutants in general also contain information about degeneracies among the scar states for a given family of Hamiltonians, which is a finer characterization than just the statement of the fracture in Eq.~(\ref{eq:scarfactorize}). 
Pictorially, the distinction between the block decompositions with degenerate and non-degenerate scars is shown in Fig.~\ref{fig:blocks}(c) and \ref{fig:blocks}(d).
Further, in cases where the commutants consist of QMBS ket-bra operators as well as other conventional conserved quantities, operators in the local algebra do not act irreducibly in the complement of $\mH_{\scar}$, and instead have smaller blocks within $\mH_{\therm}$ corresponding to also the non-scar conserved quantities.
Nevertheless, we expect generic Hamiltonians within each of those blocks to be ``thermal" in the conventional sense.
\subsection{Type I and Type II Symmetric Hamiltonians}\label{subsec:symhamtypes}
Locality considerations for the construction of symmetric extensive local operators or Hamiltonians are more challenging in the QMBS and other weak ergodicity breaking problems where the corresponding ``symmetries" are highly unconventional, generically non-local and non-on-site.
In particular, in any bond algebra corresponding to commutants with QMBS, we find qualitatively new types of symmetric Hamiltonians which are forbidden for conventional commutants corresponding to on-site unitary symmetries.
In general, given a bond algebra $\mA \defn \lgen \{\hH_\alpha\} \rgen$, i.e., where $\{\hH_\alpha\}$ are strictly local, we can make a clear distinction between two types of symmetric extensive local operators or symmetric Hamiltonians in $\mA$.
\begin{definition}
An extensive local operator in a bond algebra $\mA$ is a Type I (Type II) operator if it can (cannot) be expressed as a sum of strictly local operators also in the same bond algebra $\mA$.
\end{definition}
While the DCT guarantees that the Type II Hamiltonians can in principle be produced from the strictly local generators in the algebra sense, such procedure necessarily involves highly non-local expressions in terms of those generators.
Lemma II.2 in \cite{moudgalya2022from} shows that for commutants generated by on-site unitary operators, all symmetric Hamiltonians are of Type I, hence Type II Hamiltonians can only exist for unconventional commutants as examplified by the kinds we consider in this work. 
As an example, for the bond algebra $\tmA_{\scar}$ corresponding to the commutant $\tmC_{\scar} = \lgen \{\ketbra{\Psi_n}{\Psi_m} \}\rgen$ that contains the states of Eq.~(\ref{eq:FMtower}) as degenerate QMBS, we will show that the Heisenberg Hamiltonian in Eq.~(\ref{eq:FMscar}) is a Type I operator, whereas the DMI Hamiltonian is a Type II operator.
Examples of Type II operators for other instances of QMBS are shown in Tab.~\ref{tab:scarexamples}, and discussed in detail in Sec.~\ref{sec:scarexamples} and the appendices.
With these definitions, we can make a few simple observations on the nature of these operators that we use in this work. 
First, note that Type I and Type II properties of an operator are invariant under the addition of Type I operators. 
This allows us to define \textit{equivalence classes} of Type II operators, where two Type II operators are equivalent if they differ by the addition of a Type I operator.
For extensive local operators that are sums of strictly local operators of a maximum range $r_{\max}$, Type I symmetric operators form a vector space that is a subspace of the space of all symmetric operators of that range, hence the set of equivalence classes of Type II operators has a natural quotient space structure.
An advantage of studying the equivalence classes instead of the operators directly is that the number of linearly independent equivalence classes of Type II operators of range at most $r_{\max}$ can be extracted numerically in a rather straightforward manner~\cite{moudgalya2022numerical}.
Second, the Type II property depends on the local algebra in question and in general, a Type II operator w.r.t.\ one algebra might be a Type I operator w.r.t.\ another.  
Nevertheless, given two algebras $\mA_1 \subseteq \mA_2$, we can always say that any Type I operator in $\mA_1$ is also a Type I operator in $\mA_2$.
Equivalently, any operator in $\mA_1$ that is a Type II operator w.r.t.\  $\mA_2$ is also a Type II operator w.r.t.\ $\mA_1$, i.e., any extensive local operator in $\mA_1$ that cannot be written as a linear combination of strictly local operators in $\mA_2$ cannot be written as a linear combination of strictly local operators in $\mA_1$.
Note that the distinction between Type I and Type II symmetric Hamiltonians is not so clear in local algebras that are not bond algebras, i.e., if one of the generators is necessarily extensive local, due to the arbitrariness in the choice of generators of a local algebra.
For example, without the restriction of strict locality that is natural in bond algebras, the extensive local Hamiltonian itself can be made a generator in a local algebra, wiping out the distinction between the types of symmetric Hamiltonians.
Hence whenever we refer to an operator as Type I or Type II Hamiltonians, we implicitly assume that there is a bond algebra involved.
Following the discussion in Sec.~\ref{subsec:degnondegQMBS}, this usually means that the distinction can only be made in cases of isolated QMBS or degenerate QMBS, and not in the cases with non-degenerate towers of QMBS.
\subsection{Structure of Local Hamiltonians with QMBS}\label{subsec:structure}
While locality considerations for bond algebras corresponding to degenerate QMBS lead to distinctions between Type I and Type II Hamiltonians, locality considerations for local algebras corresponding to non-degenerate QMBS also lead to certain constraints on the structure of Hamiltonians.
As discussed in Sec.~\ref{subsec:degnondegQMBS}, the algebras $\mA_{\scar}$ of operators with non-degenerate QMBS usually involve an extensive local ``lifting operator" in its generators.
Heuristically, in the expression of any local operator in $\mA_{\scar}$ whose generation necessarily involves the lifting operator, it should appear either in a linear combination or in a commutator with another local operator; all other combinations are generically non-local. 
However, in any operator where it appears as a commutator, the QMBS remain degenerate, since they are eigenstates of all operators in $\mA_{\scar}$.
Hence in many examples, we conjecture that any Hamiltonian in $\mA_{\scar}$ is a \textit{linear combination} of the extensive local operator and an operator from $\tmA_{\scar}$ for which the QMBS are degenerate.
Of course the validity of this conjecture depends on the details of the models and the operators involved, nevertheless we can conjecture a more precise statement for the algebras $\tmA^{\FM}_{\scar}$ and $\mA^{\FM}_{\scar}$ of Eq.~(\ref{eq:FMalgebras}), which contain the ferromagnetic states $\{\ket{\Psi_n}\}$ of Eq.~(\ref{eq:FMtower}) as degenerate and non-degenerate QMBS respectively.
\begin{conjecture}\label{conj:exhaustivechars}
\textit{Any} extensive local Hamiltonian with the ferromagnetic states $\{\ket{\Psi_n}\}$ as eigenstates, i.e., any Hamiltonian in the algebra $\mA^{\FM}_{\scar}$, is a linear combination of the lifting operator $S^z_{\tot}$ and the Type I or II Hamiltonian from the bond algebra $\tmA^{\FM}_{\scar}$.
\end{conjecture}
\noindent The local algebras in several other examples of towers of QMBS we study in Sec.~\ref{sec:scarexamples} have similar structures, as shown in Tab.~\ref{tab:scarexamples}, and we also make similar conjectures for them.
Since the lifting operator is simply $S^z_{\tot}$, this conjecture has an immediate corollary on the equal spacing of the spectra of Hamiltonians with these states as QMBS.
\begin{conjecture}\label{conj:equalspacing}
Any local Hamiltonian with the ferromagnetic states $\{\ket{\Psi_n}\}$ as QMBS necessarily has them as an equally spaced tower of states in the spectrum.
\end{conjecture}
\noindent Note that this is analogous to the claim we proved in \cite{moudgalya2022from} on the spectra of Hamiltonians with the dynamical $SU(2)$ symmetry, but here we have been unable to prove it.
\section{From Unified Formalisms To Exhaustive Algebras}\label{sec:QMBSunified}
We now discuss some unified formalisms of QMBS that potentially capture several examples of QMBS in a single framework; an overview can be found in the reviews on this subject~\cite{moudgalya2021review, papic2021review, chandran2022review}.
Particularly, the Shiraishi-Mori (SM) formalism introduced in \cite{shiraishi2017eth} and the closely related Group-Invariant (GI) formalism introduced in \cite{pakrouski2020many} motivate a concrete route to constructing the exhaustive algebra $\mA_{\scar}$ of Hamiltonians with a given set of QMBS.
Identifying these exhaustive algebras then allows us to directly connect all the ``symmetry-based" unified formalisms of QMBS to the commutant language, which in turn allows for generalizations that apply to many more examples of QMBS.
\subsection{Shiraishi-Mori (SM) Formalism}\label{subsec:SMformalism}
\subsubsection{Original formulation}
Reference \cite{shiraishi2017eth} introduced a formalism for embedding exact eigenstates into the spectra of non-integrable Hamiltonians, which provides a way of explicitly constructing Hamiltonians with ETH-violating eigenstates. 
In particular, they considered a target space $\mT$, spanned by a set of states that are all annihilated by a set of (generically non-commuting) local projectors $\{P_{[j]}\}$, where $P_{[j]}$ denotes a projector with support in the vicinity of a site $j$, i.e., 
\begin{equation}
    \mT = \{\ket{\psi} :\;P_{[j]}\ket{\psi} = 0\;\;\forall j\}.
\label{eq:targetdefn}
\end{equation}
Given a target space $\mT$, SM considered Hamiltonians of the form
\begin{equation}
    H_{\textrm{SM}} = \sumal{j}{}{P_{[j]} h_{[j]} P_{[j]}} + H_0,\;\;\;[H_0, P_{[j]}] = 0\;\;\forall j,
\label{eq:SMHamil}
\end{equation}
where $h_{[j]}$ is a sufficiently general (e.g., ``randomly chosen'') local operator with support in the vicinity of site $j$ that might have a support distinct from $P_{[j]}$, and $H_0$ is a local operator.
The $\{P_{[j]} h_{[j]} P_{[j]}\}$ terms of $H_{\SM}$ in Eq.~(\ref{eq:SMHamil}) vanish on the states in $\mT$, and $H_0$ leaves the target space $\mT$ invariant as a consequence of the imposed commutation conditions. 
Hence eigenstates of $H_{\SM}$ can be constructed from within the target space $\mT$, and they are generically in the middle of the spectrum.  
Since in many examples $\mT$ can be completely spanned by low-entanglement states,\footnote{In fact, the results of \cite{yao2022bounding} can be used to show that the Entanglement Entropy (EE) of any state in $\mT$ over a subregion $R$ is bounded by the logarithm of the ground state degeneracy of the Hamiltonian $\sum_{j \in R}{}{P_{[j]}}$ (where the sum only includes $P_{[j]}$'s completely within the region $R$), which is simply $\log \dim(\mT_R)$, where $\mT_R$ is the common kernel of $P_{[j]}$'s completely within the region $R$.
As we discuss in Apps.~\ref{app:SMexistence} and \ref{subsec:singleQMBS}, for a system with a tensor product Hilbert space and local Hilbert space dimension $d$, and an extensive contiguous subregion of size $L_R$, $\dim(\mT_R) \leq (d p)^{L_R} $ for some $p < 1$ [see Eq.~(\ref{eq:fracTbound})], hence the EE of any state in $\mT$ over this subregion is bounded by $L_R \log (p d)$, which is always less than the Page value~\cite{page1993} $L_R \log d - \mathcal{O}(1)$ expected for a generic infinite-temperature eigenstate of a local Hamiltonian. See also \cite{movassagh2010unfrustrated} for some results on the entanglement of states in the common kernel of local projectors.} e.g., states with an MPS form, and $H_{\textrm{SM}}$ is generically non-integrable, these eigenstates are said to be QMBS of $H_{\textrm{SM}}$~\cite{shiraishi2017eth, serbyn2020review, papic2021review, moudgalya2021review}. 
For example, this formalism can be used to construct Hamiltonians with the ferromagnetic states $\{\ket{\Psi_n}\}$ of Eq.~(\ref{eq:FMtower}) as QMBS, using the fact that they are the singlets of $\mA_{SU(2)}$ as discussed in Sec.~\ref{subsec:singlets}, and hence are in the common kernel of the set of strictly local projectors 
\begin{equation}
    P_{j,j+1} \defn \frac{1}{4} - \vec{S}_j\cdot\vec{S}_{j+1}.
\label{eq:FMtowerproj}
\end{equation}
Then any Hamiltonian of the form of Eq.~(\ref{eq:SMHamil}) with $P_{[j]} = P_{j,j+1}$ and $H_0 = \sum_j{S^\alpha_j}$ (that satisfies the requirements) has $\{\ket{\Psi_n}\}$ as QMBS.
To cast the SM formalism in terms of local and commutant algebras, we start with the bond algebra $\tmA = \lgen \{P_{[j]}\} \rgen$ generated by the aforementioned projectors.
$\mT$ is then a subspace spanned by one set of the degenerate singlets of $\tmA$, namely by the ones on which all the projectors $P_{[j]}$ vanish.
The block decomposition typical for such bond algebras is depicted in Fig.~\ref{fig:blocks}(a).
According to Eq.~(\ref{eq:SMHamil}), $H_0$ is then a local operator that belongs to $\tmC$, the commutant of $\tmA$. 
Thus we see that the singlets of any local algebra can be made into QMBS of some Hamiltonian of the form of Eq.~(\ref{eq:SMHamil}), provided the $\mT$ and $H_0$ that satisfy the required conditions exist.
We will sometimes refer to $\tmA$ and $\tmC$ as the ``pre-bond" and ``pre-commutant" algebras respectively, and QMBS can be constructed from several pre-bond algebras, e.g., all of those discussed in \cite{moudgalya2022from}.
In the example of the ferromagnetic QMBS discussed above, we simply have $\tmA = \lgen \{P_{j,j+1}\} \rgen = \mA_{SU(2)}$ and $\tmC = \mC_{SU(2)}$ of Eq.~(\ref{eq:SU2AC}), which explains the choice of $H_0$ there.
\subsubsection{Immediate Generalizations}
While in the SM framework all states in the target subspace are singlets that are annihilated under the local projectors $\{P_{[j]}\}$, we can embed other sets of singlets with Hamiltonians of the form
\begin{equation}
    H_{\SM\text{-gen}} = \sumal{j}{}{\tP_{[j]} h_{[j]} \tP_{[j]}} + H_0, 
\label{eq:SMgeneral}
\end{equation}
where $\tP_{[j]}$'s are some operators in the vicinity of site $j$ that need not commute with each other and need not be projectors,\footnote{In principle, we can always choose a different set of local \textit{projectors} like in the original SM formulation instead of the $\tP_{[j]}$'s, since for any local operator $\hat{O}$ with eigenvalues $o_1, o_2, \dots, o_M$, we can construct a local projector annihilating the (assumed target) subspace of states with eigenvalue $o_1$ as $P \defn \mathds{1} - \prod_{k=2}^M \frac{\hat{O} - o_k}{o_1 - o_k}$. However, this would complicate their expressions and also obfuscate the fact that only the singlet structure of the local pre-algebra is important.
In either case, the families of Hamiltonians we construct in the end would be unchanged since we intend to use sufficiently general $h_{[j]}$'s.} $h_{[j]}$'s are arbitrary Hermitian operators, and $H_0$ is any operator that leaves the target space $\mT$ (now defined as the common kernel of all the $\{\tP_{[j]}\}$) invariant, which can also be any operator in the commutant of the pre-bond algebra $\lgen \{\tP_{[j]}\} \rgen$ or any other local operator that commutes with the projector onto $\mT$.  
In this generalized setting, an arbitrary degenerate set of singlets of a bond algebra $\tmA \defn \lgen \{A_{[j]}\} \rgen$, e.g., those that satisfy $A_{[j]}\ket{\psi} = a_{[j]} \ket{\psi}$ with some fixed set $\{a_{[j]}\}$, can be made into QMBS of $H_{\SM\text{-gen}}$ by choosing $\tP_{[j]} = A_{[j]} - a_{[j]}\mathds{1}$.
Another obvious generalization is to target two sets of singlets of the pre-bond algebra $\tmA = \lgen \{A_{[j]}\} \rgen$, one described by the generator eigenvalue set $\{a_{[j]} \}$ and the other by $\{a_{[j]}' \}$, by using $\tP_{[j]} = (A_{[j]} - a_{[j]}) (A_{[j]} - a_{[j]}')$.
In this case, a possible choice for $H_0$ is any operator from $\lgen \{A_{[j]}\} \rgen$ that splits the degeneracy between the two sets; this need not belong to the commutant of $\lgen \{\tP_{[j]}\}\rgen$ and hence is an example where local $H_0$ preserves the target space but does not commute with the $\tP_{[j]}$'s that was required in the original SM formalism.
\subsubsection{Exhaustive Algebra of QMBS Hamiltonians}\label{subsubsec:SMexhaustive}
While these approaches provide a way to construct one family of Hamiltonians with QMBS, we are primarily interested in exhaustively characterizing \textit{all} Hamiltonians with a given set of QMBS, and we now show that the SM formalism appropriately extended and interpreted provides a way to do so.  
We start by analyzing the Hamiltonians $H_{\SM}$ by focusing on the first term in Eq.~(\ref{eq:SMHamil}), and we consider the bond algebra $\tmA_{\SM} = \lgen \{P_{[j]} h_{[j]} P_{[j]}\} \rgen$ where we choose sufficiently general operators $h_{[j]}$ with an appropriate support (implicitly allowing several generators associated with each $[j]$ if needed).
It is natural to expect that most operators in the pre-commutant $\tmC$ no longer commute with general Hamiltonians built out of $\tmA_{\SM}$.
Nevertheless the states in the target space $\mT$ are still annihilated by the generators of $\tmA_{\SM}$, hence ket-bra operators formed by those states are in $\tmC_{\SM}$, the centralizer of $\tmA_{\SM}$.
For sufficiently general $h_{[j]}$ we expect these to be the \textit{only} operators in $\tmC_{\SM}$, hence we obtain the bond and commutant algebra pair
\begin{equation}
\tmA_{\SM} = \lgen \{P_{[j]} h_{[j]} P_{[j]}\} \rgen,\;\;\tmC_{\SM} = \lgen \{\ketbra{\psi_m}{\psi_n}\}\rgen,
\label{eq:SMdegbondalg}
\end{equation}
where $\ket{\psi_m},\ket{\psi_n} \in \mT$; we sometimes refer to bond algebras of this form as ``Shiraishi-Mori" bond algebras. 
A proof of this statement depends on the specific details of the operators and the target spaces, but it can be verified for several examples we discuss in Sec.~\ref{sec:scarexamples} using numerical methods we present in \cite{moudgalya2022numerical}. 
The existence of this pair of algebras is equivalent to the statement that $\tmA_{\SM}$ is irreducible in $\mT^{\perp}$, the orthogonal complement of the target space $\mT$ of Eq.~(\ref{eq:targetdefn}).
While it is not a priori clear that $\tmA_{\SM}$ can be generated by strictly local terms, in App.~\ref{app:SMexistence} we are able to prove the following Lemma that guarantees the existence of such an algebra as long as a target space $\mT$ of the form of Eq.~(\ref{eq:targetdefn}) exists.
\begin{restatable}{lem}{smsuf}\label{lem:smsuf}
Consider the target space $\mT = \{\ket{\psi},\;\;P_{[j]}\ket{\psi} = 0\}$, where $P_{[j]}$'s  are strictly local projectors of range at most an $L$-independent number $r_{\max}$. 
Then, we can always construct a bond algebra $\tmA_{\SM} = \lgen \{\wh_{[j]}\} \rgen$ where $\wh_{[j]}$'s are strictly local terms of range bounded by some $L$-independent number $r'_{\max} \geq r_{\max}$, such that it is irreducible in $\mT^\perp$, the orthogonal complement of the target space. 
\end{restatable}
\noindent Hence such $\tmA_{\SM}$ is an example of $\tmA_{\scar}$ discussed in Sec.~\ref{subsec:QMBSfamilies}, in particular it contains all Hamiltonians that have the QMBS $\{\ket{\psi_n}\}$ as degenerate eigenstates.
The block decomposition corresponding to such algebras is depicted in Fig.~\ref{fig:blocks}(c).
Note that while Lem.~\ref{lem:smsuf} provides an upper bound on the range of the generators of $\tmA_{\SM}$, in many examples discussed in Sec.~\ref{sec:scarexamples} we are able to use the structure of the states in $\mT$ to reduce this range.
To understand all Hamiltonians that simply have the states $\{\ket{\psi_n}\}$ as eigenstates, including the ones that lift the degeneracies among them, we can simply add \textit{a single} $H_0$ to the generators of $\tmA_{\SM}$.  
That is, assuming the existence of an $H_0$ that lifts all the degeneracies among the states in $\mT$,\footnote{Note that this depends on the precise model in hand.
The required $H_0$ might lie outside $\tmC$, since the only required condition is that it leaves the target space $\mT$ invariant, hence $H_0$ can be any local operator that commutes with the projector onto $\mT$.
There might also not exist any local $H_0$ that lifts all the degeneracies, in which case the commutant would also contain ket-bra operators of states that remain degenerate under $H_0$.}  it is clear that we can write down a local and commutant algebra pair of the form
\begin{equation}
    \mA_{\SM} = \lgen \{P_{[j]} h_{[j]} P_{[j]}\}, H_0 \rgen,\;\;\;\mC_{\SM} = \lgen \{\ketbra{\psi_n}\} \rgen,
\label{eq:SMnondegalg}
\end{equation}
where $\ket{\psi_n}$ now refer to the eigenstates of $H_0$ in the scar space.
This completes the construction of a local algebra with the projectors onto the QMBS eigenstates completely determining its commutant, hence all Hamiltonians with these QMBS can be constructed from the generators of $\mA_{\SM}$, which then gives the algebra $\mA_{\scar}$ that we envisioned in Sec.~\ref{subsec:QMBSfamilies}.
The block decomposition corresponding to this algebra is depicted in Fig.~\ref{fig:blocks}(d).
{\it In summary, for any set of states $\{\ket{\psi_n}\}$ that span the complete kernel of a set of strictly local projectors $\{P_{[j]}\}$ (or equivalently, that can be expressed as the ground state subspace of a frustration-free Hamiltonian), a locally generated algebra of Hamiltonians for which these states are QMBS is guaranteed to exist.} 
Since the ferromagnetic QMBS $\{\ket{\Psi_n}\}$ of Eq.~(\ref{eq:FMtower}) can be understood as the common kernel of the set of strictly local projectors $\{P_{j,j+1}\}$, the exhaustive algebras with these states as degenerate or non-degenerate QMBS, $\tmA^{\FM}_{\scar}$ and $\mA^{\FM}_{\scar}$ can respectively be written as
\begin{align}
    \tmA^{\FM}_{\scar} &= \lgen \{P_{j,j+1} h_{[j]} P_{j,j+1}\}\rgen,\nn \\
    \mA^{\FM}_{\scar} &=  \lgen \{P_{j,j+1} h_{[j]} P_{j,j+1}, S^z_{\tot}\}\rgen,
\label{eq:FMSMalgebra}
\end{align}
where $h_{[j]}$ is a term of range at most $4$ (proved analytically), although we numerically find that terms of range $3$ are sufficient.
As we will discuss with examples in Sec.~\ref{sec:scarexamples}, this is also true for several if not all examples of QMBS studied in the literature, and this allows us to identify the appropriate algebras that contain the exhaustive set of local Hamiltonians that have these states as QMBS.
Note that although the generators of $\tmA_{\SM}$ and $\mA_{\SM}$ as motivated from the SM construction include the ``randomly chosen" operators $\{h_{[j]}\}$, the algebras as a whole are $h_{[j]}$-independent since they are the centralizers of $h_{[j]}$-independent algebras $\tmC_{\SM}$ and $\mC_{\SM}$. 
Indeed, it is possible to choose a set of ``nice" $h_{[j]}$-independent generators for $\tmA_{\SM}$, which are more useful in systematically constructing local operators in this algebra.
For example, for the ferromagnetic tower of QMBS $\{\ket{\Psi_n}\}$ this exhaustive algebra of Eq.~(\ref{eq:FMSMalgebra}) can equivalently be expressed as shown in Eq.~(\ref{eq:FMalgebras}). 
\subsubsection{Nature of Shiraishi-Mori Hamiltonians}
\label{subsubsec:SMnature}
We now emphasize a few aspects of the class of Hamiltonians of the form of Eq.~(\ref{eq:SMHamil}).
The main change of interpretation compared to considering individual instances of $H_{\SM}$ done in prior works is that here we are exhaustively characterizing the family of Hamiltonians with the given QMBS, and then expect that reasonably generic instances from this family will have the exact QMBS inside otherwise thermal spectrum.
As a consequence, the algebra $\mA_{\SM}$ includes Hamiltonians that are not of the form of $H_{\SM}$ of Eq.~(\ref{eq:SMHamil}) but nevertheless contain the same QMBS as $H_{\SM}$.
As discussed in Sec.~\ref{subsec:symhamtypes}, there are two types of symmetric Hamiltonians that can in principle occur for any bond algebra, the symmetry here being the QMBS commutant $\tmC_{\SM}$ and the bond algebra being $\tmA_{\SM}$. 
It is easy to see that all Hamiltonians of the form of Eq.~(\ref{eq:SMHamil}), or even the immediate generalizations in Eq.~(\ref{eq:SMgeneral}), are a linear combination of a \textit{Type I operator} in the algebra $\tmA_{\SM}$ that leaves the QMBS degenerate and a lifting operator that lifts the degeneracy between the QMBS. 
However, the most general Hamiltonian with the same set of QMBS could be a \textit{Type II operator} in the algebra $\tmA_{\SM}$, along with a linear combination of the lifting operator and Type I operator. 
This allows us to explain QMBS in Hamiltonians that are considered to be ``beyond" the Shiraishi-Mori formalism.
For example, in the case of the ferromagnetic tower of QMBS, the DMI term is a Type II operator, and in Sec.~\ref{sec:scarexamples} we show that the Hamiltonian of Eq.~(\ref{eq:FMscar}) with $D \neq 0$ cannot be expressed as Eq.~(\ref{eq:SMgeneral}), while with $D = 0$ it can.
\subsection{Group-Invariant (GI) Formalism}\label{subsec:GIformalism}
\subsubsection{Original formulation}
A closely related formalism, which we refer to as the Group Invariant (GI) formalism, was introduced and developed by Pakrouski, Pallegar, Popov, and Klebanov in \cite{pakrouski2020many, pakrouski2021group}, where they proposed that QMBS are singlets of certain Lie groups. 
Given a set of operators $\{T_a\}$ that are generators of a Lie group $G$, the singlets of the group are states that are invariant under the action of any element in $G$.
Since the elements of the group are unitaries of the form $\text{exp}(i\sum_a{\alpha_a T_a})$, the singlets are annihilated by all the generators.
Defining the space of singlets as $\mT = \text{span}\{\ket{\psi}:\;\;T_a\ket{\psi} = 0\;\forall a\}$, Ref.~\cite{pakrouski2020many} showed that the states in $\mT$ are QMBS of Hamiltonians of the form
\begin{equation}
    H_{\GI} = \sumal{a}{}{O_{a} T_{a} + H_0}, \;\;\;[H_0, C^2_G] = W C^2_G,
\label{eq:GIhamil}
\end{equation}
where $O_{a}$ are arbitrary operators chosen such that $H_{\GI}$ is Hermitian, $C^2_G$ is the quadratic Casimir of the Lie group $G$, and $W$ can be any operator. 
Reference \cite{pakrouski2020many} found examples where the generators $\{T_a\}$ of the Lie group $G$ can be chosen to be strictly local operators, allowing $H_{\GI}$ to be a local Hamiltonian. 
In particular, when $T_a$'s are quadratic fermion operators in an $N$-site spinful electron system---e.g., hopping terms, on-site chemical potentials, or magnetic fields---they generate some Lie algebras (depending on the chosen set of generators) and the corresponding Lie groups $G$ are subgroups of $U(2N)$; see \cite{pakrouski2020many, pakrouski2021group, moudgalya2022from} for detailed discussions with several examples.
Further, the condition on $H_0$ in Eq.~(\ref{eq:GIhamil}) is equivalent to stating that it leaves the subspace $\mT$ invariant.\footnote{
We wish to show that $[H_0, C^2_G] = W C^2_G \iff H_0 \mT \subseteq \mT$.
To show the $\Longrightarrow$ direction, note that $\mT = \textrm{span}\{ \ket{\psi}, C^2_G\ket{\psi} = 0 \}$. 
Hence the condition on $H_0$ implies $C_G^2 H_0\ket{\psi} = (H_0 - W) C_G^2\ket{\psi} = 0$, i.e., $H_0 \ket{\psi} \in \mT\;\;\forall \ket{\psi} \in \mT$ and hence $H_0 \mT \subseteq \mT$. 
To show the $\Longleftarrow$ direction, we separate the Hilbert space into the space of singlets of $G$ and its orthogonal complement $\mTperp$, $\mH = \mT \oplus \mTperp$, and work in the corresponding basis diagonalizing $C_G^2$. 
In this basis, we can express $C^2_G = 0_{\mT} \oplus C_{\mTperp}$, where $0_{\mT}$ denotes a zero matrix, and $C_{\mTperp}$ is a diagonal matrix with non-zero entries. 
Since $H_0$ leaves $\mT$ invariant, $H_0 = H_\mT \oplus H_{\mTperp}$, and we can then express the commutator as $[H_0, C_G^2] = 0_{\mT} \oplus B_{\mTperp} = (0_{\mT} \oplus B_{\mTperp} C_{\mTperp}^{\text{-}1})(0_{\mT} \oplus C_{\mTperp}) \defn W C^2_G$ for some matrices $B_{\mTperp}$ and $W$.
}
Hence, with a choice of strictly local $\{T_a\}$ and the substitutions $T_a \rightarrow \tP_{[j]}$ and $O_a \rightarrow \tP_{[j]} h_{[j]}$, Eq.~(\ref{eq:GIhamil}) is equivalent to the generalized SM Hamiltonian of Eq.~(\ref{eq:SMgeneral}).\footnote{Indeed, consider the (assumed) strictly local Hermitian operator $O_a T_a$ on its support and diagonalize it on this few-site Hilbert space.  The kernel of $T_a$ on this Hilbert space belongs to the zero eigenvalue subspace of $O_a T_a$.  Since the Hermitian $T_a$ is invertible on the orthogonal complement to its kernel, we can write $O_a T_a = T_a h_a T_a$ with some $h_a$ on the same few-sites Hilbert space.}
Note that similar to the SM formalism, this can be generalized further to include singlets of $\tmA$ that satisfy $T_a\ket{\psi} = t_a \ket{\psi}$ by substituting $T_a \rightarrow T_a - t_a$ in the GI construction. 
While the states embedded this way are not ``group-invariant" in the original sense, they are still invariant under the action of elements of $G$ up to an overall phase. 
\subsubsection{Extensions to local and commutant algebras}
With this mapping to the (generalized) SM formalism, all of the exhaustive algebra results of Secs.~\ref{subsubsec:SMexhaustive} and \ref{subsubsec:SMnature} go through here.
The target space $\mT$ here is spanned by the singlets of the group $G$, which are also the singlets of the bond algebra $\tmA = \lgen \{T_a\} \rgen$.
The families of Hamiltonians with these singlets as eigenstates can be constructed in direct analogy with the SM formalism, e.g., the algebra $\tmA_{\GI} = \lgen \{O_a T_a \}\rgen$ is the analog of $\tmA_{\SM}$ and leaves the singlets degenerate while breaking symmetries of $\tmA$, and $\mA_{\GI} = \lgen \{O_a T_a\}, H_0 \rgen$ is the analog of $\mA_{\SM}$ and lifts (some of) the degeneracies of the singlets.
Further, we can also apply Lemma~\ref{lem:smsuf} to show that as long as $T_a$'s are strictly local operators, we can construct algebras that provide an exhaustive description of all Hamiltonians with singlets of $G$ as eigenstates.
Similar to the SM formalism, Hamiltonians of $H_{\GI}$ are linear combinations of a Type I operator in $\tmA_{\GI}$ and a lifting operator, whereas the most general Hamiltonian with these group singlets as eigenstates could be a Type II operator in $\tmA_{\GI}$ and a lifting operator.
Note that the interpretation of these states as being group-invariant or singlets of $\tmA$ is not necessary for characterizing the final algebra $\tmA_{\GI}$. 
In fact, multiple choices of the group $G$ or the pre-bond algebra $\tmA$ can have the same set of singlets (e.g., see \#2 and \#3a, \#3b in Tab.~II in \cite{moudgalya2022from}), and all such pre-algebras would give rise to the same $\mA_{\GI}$ under the construction discussed above.
Nevertheless, starting from ``well-known" pre-bond algebras, e.g., any of the algebras discussed in \cite{moudgalya2022from}, provides a convenient route to construct the final bond algebra of interest. 
\subsubsection{Features of the QMBS revealed by this framework}
The group-invariant interpretation illustrates several non-trivial features of QMBS, as emphasized in \cite{pakrouski2020many}.
Since the pre-bond algebra $\tmA = \lgen \{T_a\} \rgen$ is generated by the generators of a Lie group $G$, and since the QMBS are singlets of $\tmA$, their projectors are a part of its commutant $\tmC$.
This also means that the QMBS projectors, and hence the states themselves, are ``symmetric" under the group $G$.
For example, as discussed in \cite{pakrouski2020many}, several QMBS that have group-invariant interpretations (e.g., the tower of $\eta$-pairing states in the Hubbard model~\cite{mark2020eta, moudgalya2020eta}) are invariant (i.e., symmetric [understood more generally to include cases with very specific sign factors under the action of the symmetry operations]) under the permutation of sites, since the permutation group is a subgroup of $G$ in those cases.
However, the presence of the permutation group within a bond algebra for QMBS does not require parent Lie group structure and occurs much more generally.
For example, the ferromagnetic tower of QMBS $\{\ket{\Psi_n}\}$ are singlets of $\mA_{SU(2)}$, which is the group algebra of the permutation group $S_L$ that is not a Lie group.
From this perspective the states $\{\ket{\Psi_n}\}$ are invariant under permutation of sites of the lattice, which can be readily verified from their expressions in Eq.~(\ref{eq:FMtower}).
Hence the commutant language is also useful in generalizing key ideas from the GI approach.
\subsection{Particular Breaking of Symmetries or Tunnels to Towers Formalism}\label{subsec:tunnelstowers}
With the understanding of the exhaustive algebra motivated by the SM and GI formalisms, we now discuss other unified frameworks in the algebra language, and demonstrate how they lead to constructions of the QMBS algebra $\mA_{\scar}$.
References \cite{mark2020eta} and \cite{odea2020from} introduced a mechanism which can be viewed as a particular removal of symmetries that preserves an original symmetry-dictated multiplet and dubbed the ``Tunnels-Towers" mechanism in \cite{odea2020from}.
This is a three-step process to construct Hamiltonians with QMBS, which we now summarize and describe in the commutant language.
First, they start with a model with a non-Abelian symmetry under which the ``target" QMBS eigenstates are degenerate.
This can be a model from the pre-bond algebra $\tmA$ which has a non-Abelian commutant $\tmC$, where the potential QMBS states are the singlets of $\tmA$, as shown in Fig.~\ref{fig:blocks}(a).  
For the ferromagnetic states $\{\ket{\Psi_n}\}$ of Eq.~(\ref{eq:FMtower}) we have $\tmA = \mA_{SU(2)}$ of Eq.~(\ref{eq:SU2AC}), which has the $SU(2)$ symmetry of $\tmC = \mC_{SU(2)}$.
Second, terms are added to this Hamiltonian that lift the degeneracy between the potential QMBS states, while the Hamiltonian preserves (a part of) the original symmetry.
Such a term can be like $H_0$ from the SM or GI constructions that preserves the target space, e.g., a local operator from the pre-commutant $\tmC$. 
Addition of this term to the generators results in an algebra of the form $\tmA_{\text{dyn}}$, which is the algebra of Hamiltonians for which the singlets of $\tmA$ are eigenstates, albeit not necessarily degenerate, as shown in Fig.~\ref{fig:blocks}(b).
If $H_0$ is chosen from the pre-commutant $\tmC$ and added to the generators of $\tmA$ to construct $\tmA_{\dynam}$, the commutant of $\tmA_{\dynam}$ would be at least as large as the center $\tmZ$ of $\tmA$ and $\tmC$, i.e., $\tmC_{\dynam} \supseteq \tmZ$.
In the ferromagnet example, $H_0$ can be chosen to be any operator from $\mC_{SU(2)}$, e.g., $S^z_{\tot}$, which results in the algebra $\tmA_{\dynam} = \mA_{\dynam-SU(2)} = \lgen \{\vec{S}_j\cdot \vec{S}_{j+1}\}, S^z_{\tot}\rgen$.
Hamiltonians from $\mA_{\dynam-SU(2)}$ exhibit a dynamical $SU(2)$ symmetry~\cite{moudgalya2022from}, i.e., the commutant is $\mC_{\dynam-SU(2)} = \lgen \vec{S}^2, S^z_{\tot} \rgen$,\footnote{We refer readers to \cite{moudgalya2022from} for a more detailed discussion of dynamical symmetries in the commutant language.} and the degeneracies among the states in the ferromagnetic tower are lifted. 
Third, Hamiltonians with QMBS are constructed by breaking even this restricted (dynamical) symmetry $\tmC_{\dynam}$ while preserving the target manifold of states. 
In the commutant language, this step corresponds to enlarging the algebra $\mA_{\dynam}$ to $\mA_{\scar}$, which then coincide with the exhaustive algebra $\mA_{\SM}$ or $\mA_{\GI}$ constructed from the Shiraishi-Mori or Group-Invariant formalisms respectively.
In the ferromagnet example, this corresponds to adding terms that preserve $\{\ket{\Psi_n}\}$ as eigenstates but break the dynamical $SU(2)$ symmetry $\mC_{\dynam-SU(2)}$, e.g., strictly local such terms such as $\{P_{j,j+1} h_{[j]} P_{j,j+1}\}$ that appear in the Shiraishi-Mori formalism; this ultimately leads to the algebra $\mA^{\FM}_{\scar}$ of the form of Eq.~(\ref{eq:FMSMalgebra}) or (\ref{eq:FMalgebras}).
In all, this formalism constructs QMBS Hamiltonians by sequentially constructing Hamiltonians that realize the block decompositions shown in Fig.~\ref{fig:blocks}(a,b,d).
The description of this formalism in the local and commutant algebra language provides additional insights.
First, the original formulation relied on starting from QMBS states that transform under a particular representation of a conventional non-Abelian symmetry such as $SU(2)$. 
However, in the algebra language these can be the degenerate singlets of \text{any} locally generated pre-bond algebra $\tmA$.
Second, in the original formulation in each of these steps, the terms with the right properties are determined either by guesswork or brute-force numerical searches.
However, a systematic way to derive these terms is only evident in the local and commutant algebra language.  
Third, in the final step of this construction, \cite{mark2020eta, odea2020from} noted that two distinct types of terms can be added that break the dynamical symmetry while preserving QMBS, one which annihilated the QMBS locally and one which annihilated the QMBS ``as-a-sum".
Once these steps are described in the algebra language, the origin of these two types of terms can be traced back to the existence of Type I and Type II extensive local operators in the corresponding algebras, as discussed in Sec.~\ref{subsec:symhamtypes}.
\subsection{Quasisymmetry Formalism}\label{subsec:QSformalism}
Similarly, Ref.~\cite{ren2020quasisymmetry} illustrated a mechanism for constructing QMBS models, introducing the idea of a quasisymmetry, which can also be understood clearly in the algebra language.
To summarize, quasisymmetries are symmetries only on a part of the Hilbert space, and they lead to degeneracies in the spectrum of the Hamiltonian that cannot be understood as a consequence of conventional on-site symmetries.
For example, when the pre-commutant $\tmC$ consists of a regular non-Abelian symmetry [e.g., when $(\tmA, \tmC) = (\mA_{SU(2)}, \mC_{SU(2)})$], the operators in $\tmA_{\SM}$ or $\tmA_{\GI}$ are considered to exhibit a quasisymmetry, since the singlets of $\tmA$ [e.g., the ferromagnetic manifold $\{\ket{\Psi_n}\}$] are their degenerate eigenstates, and this degeneracy can be understood as a consequence of the original non-Abelian symmetry restricted to the space of singlets.  
Hamiltonians with non-degenerate QMBS are then constructed by adding appropriate terms to lift these degeneracies, e.g., terms such as $H_0$ in the SM or GI constructions.
In the commutant language, $\tmA_{\SM}$ or $\tmA_{\GI}$ is the bond algebra of quasisymmetric operators for which the QMBS are degenerate, as depicted in Fig.~\ref{fig:blocks}(c), and the addition of $H_0$ to the generators of this algebra leads to $\mA_{\scar}$, which coincides with $\mA_{\SM}$ or $\mA_{\GI}$.
In the example of the ferromagnetic states, $\tmA^{\FM}_{\scar}$ is the algebra with a quasisymmetry, and adding $H_0 = S^z_{\tot}$ results in the algebra $\mA^{\FM}_{\scar}$ that exhaustively characterizes Hamiltonians with the ferromagnetic states as QMBS.
Hence the quasisymmetry framework sequentially constructs particular Hamiltonians that realize the block decompositions of Fig.~\ref{fig:blocks}(a,c,d).
However, similar to the previous unified formalisms, in the original quasisymmetry formulation, the states that have the quasisymmetry transform under a particular representation of a conventional non-Abelian symmetry such as $SU(2)$, and terms with the required properties are determined by brute force numerics or guess work.
The algebra language generalizes these conditions and provides a systematic way to understand such terms.
In addition, \cite{ren2021deformed} found two distinct types of operators that can be added to a symmetric operator to make it ``quasisymmetric," which in the algebra language correspond to Type I and Type II operators in $\tmA_{\scar}$.
Moreover, the ``quasisymmetry" that preserves the degeneracy between states need not originate from any conventional symmetry -- in the algebra language the degeneracy simply arises from the fact that these are the degenerate singlets of some pre-bond algebra $\tmA$. 
\section{Examples}\label{sec:scarexamples}
\begin{table*}[]
    \centering
    \renewcommand{\arraystretch}{1.5}
    \begin{tabular}{|c|c|c|c|c|c|c|c|c|}
        \hline
        \# & {\bf QMBS} & \multicolumn{2}{c|}{$(\boldsymbol{\tmA_{\text{{\bf scar}}}}, \boldsymbol{\tmC_{\text{{\bf scar}}}})$}  & \multicolumn{2}{c|}{\bf SM Projectors $\boldsymbol{\{P_{[j]}\}}$} & \multicolumn{2}{c|}{\bf Type II Op.} &  {\bf Lift Op. $\boldsymbol{H_0}$} 
        \\
        \hline
        \#1 & AKLT Ground State(s) & $(\tmA^{\AKLT}_{\scar}, \tmC^{\AKLT}_{\scar})$ & Eq.~(\ref{eq:AKLTSMpair}) & $\{P^{\AKLT}_{j,j+1}\}$ & Eq.~(\ref{eq:AKLTprojdefn}) & $S^z_{\tot}$ & App.~\ref{subsec:Sztotimpossible} & - \\
        \hline
        \#2 & Spin-1/2 ferromagnet & $(\tmA^{\FM}_{\scar}, \tmC^{(\FM)}_{\scar})$ & Eq.~(\ref{eq:FMscarsimplegens})  & $\{P_{j,j+1}\}$ & Eq.~(\ref{eq:FMtowerproj}) & $H_{\alpha-\DMI}$ & Eq.~(\ref{eq:DMIHamil}) & $S^\alpha_{\tot}$\\
        \hline
        \#3 & PBC AKLT QMBS & $(\tmA^{(p)}_{\scar}, \tmC^{(p)}_{\scar})$ & Eq.~(\ref{eq:PBCAKLTtowerpair})   & $\{\Pi_{[j,j+2]}\}$ & App.~\ref{subsec:PBCAKLT} & $\tH^{(p)}_{\AKLT}$ & Eq.~(\ref{eq:AKLTdeghamil}) &  $S^z_{\tot}$ \\
        \hline
        \#4 & OBC AKLT QMBS & $(\tmA^{(o)}_{\scar}, \tmC^{(o)}_{\scar})$ &  Eq.~(\ref{eq:ACpairobc})  & $\{\Pi^{(l)}_{1,2}, \{\Pi_{[j,j+2]}\}, \Pi^{(r)}_{L-1,L}\}$ & App.~\ref{subsec:OBCAKLT} & $\tH^{(o)}_{\AKLT}$ & Eq.~(\ref{eq:AKLTdeghamil}) &  $S^z_{\tot}$\\
        \hline
        \#5 & Spin-1 XY $\pi$-bimagnon & $(\tmA^{(\XY)}_{\scar}, \tmC^{(\XY)}_{\scar})$ &  Eq.~(\ref{eq:spin1XYpair})  & $\{P^{(\XY, \pi)}_{j,j+1}\}$ & Eq.~(\ref{eq:XYSMproj}) & \#12 & Eq.~(\ref{eq:S1asasum}) &  $S^z_{\tot}$\\
        \hline
        \#6 & Hubbard $\eta$-pairing & $(\tmA^{(\hub)}_{\scar}, \tmC^{(\hub)}_{\scar})$ & Eq.~(\ref{eq:etapair})  & \multicolumn{2}{c|}{Tab.~III in \cite{mark2020eta}} & \multicolumn{2}{c|}{\#12 in Tab.~III~\cite{mark2020eta}} &  $N_{\tot}$\\
        \hline
    \end{tabular}
    \caption{
    A summary of QMBS examples studied in this work. 
    \#1 is an isolated QMBS, whereas \#2--\#6 are towers of QMBS.
    In \#2-\#6, $\tmA_{\scar}$ is the exhaustive (bond) algebra of Hamiltonians that contain the QMBS as \textit{degenerate} eigenstates, and $\tmC_{\scar}$ is the symmetry (commutant) algebra for those Hamiltonians.
    Note that the choices of generators of the bond algebras are not unique, and we point to the simplest choice we are able to derive.
    The existence of the bond algebras is guaranteed by Lem.~\ref{lem:smsuf} if strictly local ``Shiraishi-Mori" (SM) projectors $\{P_{[j]}\}$ can be found with their common kernel spanned by the QMBS states; these can be constructed in all cases we study.
    These projectors can be used to construct Type I symmetric Hamiltonians with degenerate QMBS using the Shiraishi-Mori construction, but the bond algebras also contain Type II symmetric Hamiltonians that lie beyond the Shiraishi-Mori construction, and we show one example in each case. 
    For the towers of QMBS in \#2-\#6, exhaustive algebras $\mA_{\scar}$ of Hamiltonians with the QMBS as potentially non-degenerate eigenstates are obtained by adding the ``lift operator" $H_0$ to the corresponding bond algebras $\tmA_{\scar}$. 
    In such cases, we conjecture that \textit{any} local Hamiltonian in $\mA_{\scar}$ is a \textit{linear combination} of the lift operator and the Type I or II Hamiltonian from $\tmA_{\scar}$ (Conj.~\ref{conj:exhaustivechars}).
    This implies that the QMBS necessarily appear as \textit{equally spaced} towers in the spectra of Hamiltonians containing them (Conj.~\ref{conj:equalspacing}).
    }
    \label{tab:scarexamples}
\end{table*}
We now illustrate examples of systems where the commutant algebra picture is useful in understanding the QMBS. 
The discussion broadly follows the template presented in Sec.~\ref{subsec:QMBSfamilies}.
In particular, for some of the well-known examples of QMBS, (i) we show that there is a locally generated algebra $\mA_{\scar}$ corresponding to the commutants $\mC_{\scar}$ with ket-bra operators of QMBS; (ii) we illustrate Type I and Type II operators with QMBS, which are related to Hamiltonians beyond the Shiraishi-Mori formalisms; (iii) we derive constraints on extensive local Hamiltonians with QMBS using the DCT and locality considerations.
We particularly associate the following examples with the Shiraishi-Mori formalism, since, as we discussed in Sec.~\ref{subsubsec:SMexhaustive}, identifying the strictly local projectors that annihilate the QMBS guarantees the existence of local algebras with the desired commutants.
However, we will also use inspiration from the other formalisms to construct nicer expressions for the local algebras.
We summarize the examples and results in Tab.~\ref{tab:scarexamples}.

Note that in the following, whenever we are working with examples with multiple QMBS, we will use the notation $\tmA_{\scar}$ and $\tmC_{\scar}$ with appropriate superscripts to denote the local and commutant algebras for which the QMBS eigenstates are degenerate.
Similarly, we use $\mA_{\scar}$ and $\mC_{\scar}$ with appropriate superscripts to denote the local and commutant algebras for which the QMBS are non-degenerate to the extent possible with local operators.
\subsection{Embedding Matrix Product States}\label{subsec:MPSscar}
We start with the embedding of Matrix Product States in the middle of the spectrum, as envisioned by Shiraishi and Mori in \cite{shiraishi2017eth}.
Although it was clear in the earlier literature that Hamiltonians with MPS as QMBS exist, the exhaustive algebras of Hamiltonians with a given MPS as QMBS, including ones that are not of the Shiraishi-Mori form of Eq.~(\ref{eq:SMHamil}), was not discussed.
\subsubsection{AKLT Ground State}
For the purpose of illustration we start with the unique AKLT ground state $\ket{G}$ with PBC; analogous results can be derived for the four OBC AKLT ground states, and we refer readers to App.~\ref{subsec:OBCAKLTGS} for detailed discussions.
We also refer to earlier literature~\cite{aklt1987rigorous,schollwock2011density, moudgalya2018a} for detailed discussions on the AKLT state and its properties.
The AKLT ground state can be expressed as the unique state in the kernel of nearest-neighbor projectors $\{P^{\AKLT}_{j,j+1}\}$
, where the projectors are defined as~\cite{aklt1987rigorous}
\begin{equation}
P^{\AKLT}_{j,j+1} \defn \frac{1}{3} + \frac{1}{2}(\vec{S}_j \cdot \vec{S}_{j+1}) + \frac{1}{6}(\vec{S}_j\cdot \vec{S}_{j+1})^2,
\label{eq:AKLTprojdefn}
\end{equation}
where $\vec{S}_j$ is the spin-1 operator on site $j$ [see Eq.~(\ref{eq:AKLTHamil}) for an equivalent definition in terms of total angular momentum states on the two sites].
Hence the AKLT state can be viewed as 
unique singlet of the pre-bond algebra $\tmA^{\AKLT} \defn \lgen\{P^{\AKLT}_{j,j+1}\}\rgen$.
To construct a bond algebra with this singlet projector 
as completely generating the commutant, we can use ideas from the Shiraishi-Mori construction and consider the algebra generated by $\{P^{\AKLT}_{j,j+1} h_{[j]} P^{\AKLT}_{j,j+1}\}$ for a generic strictly local term $h_{[j]}$ with support in the vicinity of $j$.
As guaranteed by Lem.~\ref{lem:smsuf}, for sufficiently large but finite range of $h_{[j]}$, there exists the bond and commutant pair 
\begin{equation}
    \tmA^{\AKLT}_{\scar} = \lgen \{P^{\AKLT}_{j,j+1} h_{[j]} P^{\AKLT}_{j,j+1}\} \rgen,\;\;
    \tmC^{\AKLT}_{\scar} = \lgen \ketbra{G} \rgen.
\label{eq:AKLTSMpair}
\end{equation}
We numerically observe that for system size $L \geq 3$, $h_{[j]}$ can be chosen to be a sufficiently generic nearest-neighbor term for Eq.~(\ref{eq:AKLTSMpair}) to be true.
In App.~\ref{app:SMalgebra}, we use this observation to prove an equivalent statement, namely that the algebra $\tmA^{\AKLT}_{\scar}$ generated with generic nearest-neighbor $h_{[j]}$ is irreducible in the space orthogonal to $\ket{G}$. 
This is different from the general proof of existence of the Shiraishi-Mori bond algebra presented in App.~\ref{app:SMexistence} since here we use the structure of $\ket{G}$ to show that the required Shiraishi-Mori bond algebra can be generated by \textit{nearest-neighbor} terms.
\subsubsection{DCT and Type II Operators}\label{subsubsec:DCTAKLT}
Using the DCT, we can then infer that all operators that commute with $\tmC^{\AKLT}_{\scar}$, i.e., all operators with $\ket{G}$ as an eigenstate, are in the algebra $\tmA^{\AKLT}_{\scar}$ (remembering that the identity operator is always included in our bond algebras).
Hence, $\tmA^{\AKLT}_{\scar}$ is the algebra of \textit{all} parent Hamiltonians of the AKLT states (not requiring the states to be ground states).
This includes Hamiltonian terms comprised of longer range projectors that annihilate the AKLT states as well as extensive local operators such as $S^z_{\tot}$,  which vanishes on $\ket{G}$.  
While it is highly non-obvious to see that $S^z_{\tot}$ can be expressed in terms of the generators $\{P^{\AKLT}_{j,j+1} h_{[j]} P^{\AKLT}_{j,j+1}\}$, the existence of such an expression can be argued for using the irreducibility of $\tmA^{\AKLT}_{\scar}$ in the non-singlet space, as we discuss in App.~\ref{app:SMalgebra}. 
However, we have not been able to obtain a compact expression for $S^z_{\tot}$ in terms of the generators of $\tmA^{\AKLT}_{\scar}$, and we suspect any such expression is tedious and non-local.
Indeed, in App.~\ref{subsec:Sztotimpossible}, we use the MPS structure of $\ket{G}$ to prove that $S^z_{\tot}$ in $\tmA^{\AKLT}_{\scar}$ for PBC is an example of a Type II symmetric operator defined in Sec.~\ref{subsec:symhamtypes}, i.e., it cannot be expressed as a sum of strictly local bounded-range operators in $\tmA^{\AKLT}_{\scar}$.
These arguments also directly extend to $S^\alpha_{\tot}$ for $\alpha \in \{x, y, z\}$,  and indeed we numerically observe that the number of linearly independent equivalence classes of Type II operators of range $r_{\max} = 1$ in $\tmA^{\AKLT}_{\scar}$ for PBC is 3,  which are the classes containing $S^x_{\tot}$,  $S^y_{\tot}$, and $S^z_{\tot}$ respectively.\footnote{This example of Type II symmetric operator applies to so-called PXP model (Rydberg-blockaded atom chain) with respect to its known exact QMBS MPS eigenstates found in Ref.~\cite{lin2019exact}.
Those states have a precise relation to the AKLT ground state~\cite{lin2019exact,shiraishi2019connection}, and the PXP model can be cast as a sum of two-site terms that annihilate these scars and a term proportional to $S^\alpha_{\text{tot}}$ under the appropriate relation.}
Moreover, the number of independent equivalence classes for range $r_{\max} = 2$ grows to $8$, which suggests that the number of independent equivalence classes grows with range $r_{\max}$, but we defer a more detailed study to future work.
The existence of non-trivial classes of Type II operators points to an important difference between the commutants here and those generated by on-site unitary symmetries, discussed in detail in \cite{moudgalya2022from}, where Type II symmetric operators are forbidden. 
We believe this difference is due to the ``non-locality" or ``non-onsite" property of the conserved quantities in $\tmC_{\scar}^{\AKLT}$, but we defer a systematic exploration of this issue for future work. 
\subsubsection{General MPS}
The AKLT ground state scar construction can be directly extended to arbitrary Matrix Product States (MPS), since the projectors $\{P^{\AKLT}_{j,j+1}\}$ can also be constructed starting from the MPS representation of the AKLT state and the parent Hamiltonian construction~\cite{perezgarcia2007matrix, moudgalya2020large}.
For a general MPS $\ket{\psi}$ for PBC, if it is injective as in the AKLT case, it can be expressed as the unique state in a kernel of a set of local projectors~\cite{perezgarcia2007matrix, schuch2010peps} of a range $r$ that depends on the bond dimension of the MPS, say $\{\Pi_{[j, j+r-1]}\}$.
Then due to Lem.~\ref{lem:smsuf}, we are guaranteed the bond algebra and commutant pair
\begin{align}
    \tmA^{\MPS}_{\scar} &= \lgen \{P_{[j, j+ r- 1]} h_{[j]} P_{[j, j+ r -1]}\} \rgen, \nn \\
    \tmC^{\MPS}_{\scar} &= \lgen \ketbra{\psi} \rgen,
\label{eq:injectivealgpair}
\end{align}
for some generic choice of strictly local $h_{[j]}$.
$\tmA^{\MPS}$ is then also the algebra of all Hamiltonians that have the MPS $\ket{\psi}$ as an eigenstate, which includes both Type I operators such as the parent Hamiltonians used regularly in the literature, as well as potential Type II operators that could exist. 
Similar results also hold if the MPS $\ket{\psi}$ is not injective but is so-called $G$-injective~\cite{schuch2010peps}.
It can be expressed as a part of a larger manifold of states $\{\ket{\psi_\alpha}\}$ that span the common kernel of a set of strictly local projectors, and by Lem.~\ref{lem:smsuf}, we are guaranteed to have a bond algebra with the commutant $\tmC_{\SM} = \lgen \{\ketbra{\psi_\alpha}{\psi_\beta}\} \rgen$.
We have checked numerically that this is the case for the Majumdar-Ghosh states~\cite{majumdar1969next} with $r=3$, and Hamiltonians with these states as QMBS were constructed in \cite{shiraishi2017eth}.
In some cases the degeneracy between these states $\{\ket{\psi_\alpha}\}$ can be lifted using some extensive local lifting operator (as demonstrated for the MG states in \cite{shiraishi2017eth}), although its existence is not guaranteed in general.
\subsection{Spin-1/2 Ferromagnetic Scar Tower}\label{subsec:FMtowerscar}
We now methodically discuss Hamiltonians for which the multiplet of spin-1/2 ferromagnetic states $\{\ket{\Psi_n}\}$ of Eq.~(\ref{eq:FMtower}) is the QMBS subspace; we stated the key results for this case as immediate illustrations of various concepts in Secs.~\ref{subsec:QMBSfamilies} and \ref{sec:QMBSunified}.
Several examples of such Hamiltonians have been constructed, e.g., see \cite{choi2018emergent, mark2020unified, mark2020eta, moudgalya2021review}, and many of them can be understood within the Shiraishi-Mori formalism, i.e., they are of the form of Eq.~(\ref{eq:SMHamil}).
This interpretation is possible because as discussed in Sec.~\ref{subsec:SMformalism}, the ferromagnetic multiplet can be expressed as the common kernel of a set of spin-1/2 projectors $\{P_{j,j+1}\}$ defined in Eq.~(\ref{eq:FMtowerproj}) or equivalently, as the unique degenerate singlets of the pre-bond algebra $\tmA = \mA_{SU(2)} = \lgen \{\vec{S}_j \cdot \vec{S}_{j+1}\} \rgen$.
Note that while we focus on the one-dimensional case, this discussion directly generalizes to higher dimensions.
\subsubsection{Local Algebras}
As discussed in Sec.~\ref{subsec:SMformalism}, the bond algebra $\tmA^{\FM}_{\scar}$ with the commutant $\tmC^{\FM}_{\scar} = \lgen \ketbra{\Psi_m}{\Psi_n}\rgen$, which contains all Hamiltonians with the ferromagnetic multiplet as degenerate eigenstates, can be directly constructed following the Shiraishi-Mori prescription as in Sec.~\ref{subsubsec:SMexhaustive}.
As mentioned in Eq.~(\ref{eq:FMSMalgebra}) and following Lem.~\ref{lem:smsuf}, the generators of the bond algebra corresponding to $\{\ket{\Psi_n}\}$ can be chosen to be of the form $\{P_{j,j+1} h_{[j]} P_{j,j+1}\}$.
Note that $h_{[j]}$ cannot be a nearest-neighbor term with support only on sites $\{j, j+1\}$ since $P_{j,j+1}$ is a projector of rank 1, hence $P_{j,j+1} h_{j,j+1} P_{j,j+1} \propto P_{j,j+1}$, which would lead to $\tmA^{\FM}_{\scar} = \mA_{SU(2)}$.
We numerically observe that generic choices of $h_{[j]}$ with a support of at least 3 sites in the vicinity of $j$ are sufficient to yield the necessary commutant; in particular for system sizes $L \geq 5$ we find (irrespective of the boundary conditions)
\begin{align}
    \tmA^{\FM}_{\scar} &= \lgen \{h_{j-1} P_{j,j+1}\}, \{P_{j,j+1} h_{j+2}\}\rgen, \nn
    \\
    \tmC^{\FM}_{\scar} &= \lgen \{\ketbra{\Psi_n}{\Psi_m}\}\rgen, 
\label{eq:FMdegalgebra}
\end{align}
where $h_{j-1}$ and $h_{j+2}$ are some generic terms. 
In App.~\ref{subsec:FMirreducible} we show that $\tmA^{\FM}_{\scar}$ generated by such three-site terms acts irreducibly in the orthogonal complement of the ferromagnetic multiplet states, which proves Eq.~(\ref{eq:FMdegalgebra}), which is tighter than the general result of Lem.~\ref{lem:smsuf} since we use the specific structure of $\{\ket{\Psi_n}\}$.
In addition, as discussed in App.~\ref{subsec:FMsimple}, we are able to obtain a simpler set of generators for this algebra, which reads
\begin{equation}
    \tmA^{\FM}_{\scar} = \lgen \{\vec{S}_j\cdot\vec{S}_{j+1}\}, \{D^\alpha_{j,j+1, j+2}\}\rgen, \;\;\;\tmC^{\FM}_{\scar} = \lgen \{\ketbra{\Psi_n}{\Psi_m}\}\rgen,
\label{eq:FMscarsimplegens}
\end{equation}
where $D^\alpha_{j_1, j_2, j_3} \defn \sum_{k = 1}^3{(\vec{S}_{j_k} \times \vec{S}_{j_{k+1}})\cdot\widehat{\alpha}}$ is the three-site DMI term, where the sum over $k$ is modulo $3$ (i.e., the three sites are considered as forming a loop hosting the DMI term).
As discussed in Sec.~\ref{subsubsec:SMexhaustive}, a local operator from the pre-commutant, i.e., commutant of the pre-bond algebra $\tmA$, e.g.,  $S^x_{\tot}$ or $S^z_{\tot}$, can be added to the algebra $\tmA^{\FM}_{\scar}$ to break the degeneracy among the ferromagnetic states.
For example, if we add $S^z_{\tot}$, we have the local algebra and commutant pair
\begin{align}
  \mA^{\FM}_{\scar} &= \lgen \{\vec{S}_j\cdot\vec{S}_{j+1}\},\{D^\alpha_{j-1, j,j+1}\},  S^z_{\tot}\rgen, \nn \\
  \mC^{\FM}_{\scar} &= \lgen \{\ketbra{\Psi_m}{\Psi_m}\} \rgen,
\label{eq:FMlocalcommpair}
\end{align}
and $S^z_{\tot}$ is a lifting operator as defined in Sec.~\ref{subsec:degnondegQMBS}.
Note that there are several different choices for the generators of $\mA^{\FM}_{\scar}$, and also several choices of lifting operators that lift the degeneracies between the QMBS, and we have chosen a simple natural set.
\subsubsection{DCT and Type II Operators for Degenerate Scars}\label{subsubsec:FMQMBSTypeII}
We now discuss a few aspects of constructing local operators in the local algebras, starting with those in $\tmA^{\FM}_{\scar}$.
Strictly local operators with support in a contiguous region $R$, when required to commute with the ket-bra operators or projectors in $\tmC^{\FM}_{\scar}$, necessarily commute with these algebras restricted to the region $R$, i.e., $\tmC^{\FM}_{\scar, R} \defn \lgen \{\ket{\Psi_n}_R\bra{\Psi_m}_R\} \rgen$, where $\ket{\Psi_n}_R \defn (S^{-}_{\tot, R})^n\ket{F}_R$ and $S^{-}_{\tot, R}$ and $\ket{F}_R$ are the restrictions of $S^{-}_{\tot}$ and $\ket{F}$ to the region $R$, which are well-defined in the obvious way.
Since $\tmC^{\FM}_{\scar, R}$ has the same structure as $\tmC^{\FM}_{\scar}$, the corresponding local algebra is generated by restricting the generators of $\tmA^{\FM}_{\scar}$ to the region $R$; hence all strictly local operators in $\tmA^{\FM}_{\scar}$ within the region $R$ can be expressed in terms of these generators restricted to the same region.\footnote{Strictly speaking, the validity of this statement in dimensions greater than one depends on the precise shape of the region $R$ and its coverage by specific three-site generators used. Nevertheless, symmetric strictly local terms in any region $R$ can be generated from local generators from within or the close vicinity of $R$.}
Moving on to extensive local operators, there are indeed lots of Type I operators that can be constructed by simple linear combinations of strictly local terms in $\tmA^{\FM}_{\scar}$.
However, there are also Type II operators that are not of this form and yet have the ferromagnetic states as degenerate eigenstates, e.g., the Dzyaloshinskii-Moriya Hamiltonian with PBC, which reads
\begin{equation}
    H_{\alpha-\DMI} \defn \sumal{j = 1}{L}{(\vec{S}_j \times \vec{S}_{j+1})\cdot \widehat{\alpha}},  
\label{eq:DMIHamil}
\end{equation}
where the site labels are modulo $L$.
Hamiltonians of this type, first derived in \cite{mark2020eta}, where the QMBS are not eigenstates of individual terms, were referred to as ``as-a-sum" Hamiltonians~\cite{odea2020from}, and are considered to be ``beyond" the SM formalism~\cite{mark2020eta,odea2020from, tang2020multi}.
In agreement with the DCT, in App.~\ref{subsec:fullDMI}, we explicitly show that the $H_{\alpha-\DMI}$ can be expressed in terms of the generators of $\tmA^{\FM}_{\scar}$, although the expression that we find for the PBC $H_{\alpha-\DMI}$ in terms of these local generators involves manifestly non-local constructions. 
In fact, in App.~\ref{subsec:DMIimpossible}, we prove that there \textit{does not} exist a rewriting of $H_{\alpha-\DMI}$ as a sum of strictly local symmetric terms of range bounded by some fixed number independent of system size, which is proof that it is a Type II symmetric Hamiltonian discussed in Sec.~\ref{subsec:symhamtypes}.
Given the Type II operators, we also numerically observe that there are $3$ linearly independent equivalence classes, defined in Sec.~\ref{subsec:symhamtypes}, for operators of range at most $r_{\max} = 2$, which correspond to classes containing $H_{\alpha-\DMI}$ for $\alpha \in \{x, y, z\}$. 
Similar to the AKLT case in Sec.~\ref{subsec:MPSscar}, the number of independent equivalence classes grows with the range, and we observe that there are $8$ such classes for $r_{\max} = 3$; we defer a detailed study of these to future work.
Such extensive local operators cannot exist in the case of  commutants generated by on-site unitary operators~\cite{moudgalya2022from}, and this appears to be a feature of the non-onsite nature of the commutant $\tmC^{\FM}_{\scar}$.
\subsubsection{Non-Degenerate Scars and the Equal Spacing Conjecture}
Locality considerations can also be applied to local operators in $\mA^{\FM}_{\scar}$ that includes $S^z_{\tot}$.
In particular, any strictly local operators in a contiguous region $R$ necessarily commute with the ket-bra operators formed using the Schmidt states of $\{\ket{\Psi_n}\}$ over the region $R$.\footnote{One can see this by noting that these Schmidt states are labelled by $S^z_{\tot,R}$ and are equal-weight superpositions of all configurations in $R$ with fixed $S^z_{\tot,R}$, hence are the same as the ferromagnetic multiplet on the region $R$, and the same is true for the Schmidt states over the complement region.}
The algebra generated by these operators is precisely $\tmC^{\FM}_{\scar, R}$ defined previously, hence strictly local operators within $\mA^{\FM}_{\scar}$ can actually be expressed in terms of generators of $\tmA^{\FM}_{\scar}$ that are within the region $R$; note that they have ``more" symmetry than desired.
We can also comment on the structure of the extensive local operators constructed using the generators of $\mA^{\FM}_{\scar}$.
In \cite{moudgalya2022from}, we showed that any extensive local operators in the local algebra corresponding to a dynamical $SU(2)$ symmetry, i.e., $\mA_{\dyn SU(2)} \defn \lgen \{\vec{S}_j \cdot \vec{S}_{j+1}\}, S^z_{\tot} \rgen$, is always a \textit{linear combination} of $S^z_{\tot}$ and an operator from the $SU(2)$ bond algebra $\mA_{SU(2)} \defn \lgen \{\vec{S}_j\cdot\vec{S}_{j+1}\}\rgen$.
This structure implied that any Hamiltonian with a dynamical $SU(2)$ symmetry necessarily contains equally spaced towers of states in its spectrum.
Since $\mA^{\FM}_{\scar}$ is an extension of $\mA_{\dyn SU(2)}$, we make the \textit{conjecture} of Conj.~\ref{conj:exhaustivechars} that any such operator is a \textit{linear combination} of $S^z_{\tot}$ and an operator from $\tmA^{\FM}_{\scar}$. 
Since the latter leaves the ferromagnetic states degenerate and the former splits their degeneracy into an equally spaced tower, we obtain conjecture Conj.~\ref{conj:equalspacing}.
Finally, note that the Hamiltonians of the (generalized) Shiraishi-Mori form of Eq.~(\ref{eq:SMgeneral}) that contain $\{\ket{\Psi_n}\}$ as QMBS are necessarily a linear combination of a Type I Hamiltonian from $\mA^{\FM}_{\scar}$ and a lifting operator $S^z_{\tot}$, see discussion in Sec.~\ref{subsubsec:SMnature}. 
The exhaustive algebra analysis and associated conjectures imply that the only additional class of Hamiltonians with the same set of QMBS $\{\ket{\Psi_n}\}$ are linear combinations of a Type II operator from $\tmA^{\FM}_{\scar}$, e.g., the DMI term of Eq.~(\ref{eq:DMIHamil}),  and a lifting operator such as $S^z_{\tot}$.
\subsection{AKLT Scar Tower}\label{subsec:AKLTscar}
Next, we discuss the tower of QMBS in the one-dimensional AKLT models. 
For simplicity, we restrict to the PBC Hamiltonian $H^{(p)}_{\AKLT} = \sum_{j = 1}^L{P^{\AKLT}_{j,j+1}}$, where $P^{\AKLT}_{j,j+1}$ is defined in Eq.~(\ref{eq:AKLTprojdefn}).
The QMBS eigenstates were first derived in \cite{moudgalya2018a, moudgalya2018b}, and the same states were subsequently shown to be eigenstates of a large family of models in \cite{mark2020eta, moudgalya2020large, odea2020from}.
We review key results below and refer readers to App.~\ref{subsec:AKLTtowerreview} for a more detailed discussion on the Hamiltonians.
\subsubsection{QMBS Eigenstates}
We briefly review the AKLT tower of QMBS in the AKLT and related Hamiltonians.
For PBC in one dimension, we start with the unique AKLT ground state $\ket{G}$ discussed in Sec.~\ref{subsec:MPSscar}. 
For even system sizes, a tower of exact eigenstates $\{\ket{\psi_n}\}$ of the AKLT and related models can be constructed from $\ket{G}$, defined as
\begin{equation}
    \ket{\psi_n} \defn (Q^\dagger)^n \ket{G},\;\;Q^\dagger \defn \sum_j{(-1)^j (S^+_j)^2},
\label{eq:psiPBCQdagdefn}
\end{equation}
where $(S^+_j)$ is the spin-1 raising operator on site $j$. 
Given the QMBS eigenstates, we wish to construct the local algebra $\tmA^{(p)}_{\scar}$ and $\mA^{(p)}_{\scar}$ (superscript ``$p$'' standing for PBC) of the form discussed in Sec.~\ref{subsec:QMBSfamilies} such that their commutants are completely spanned by ket-bra operators or projectors onto the desired QMBS eigenstates respectively, i.e.,  $\{\ket{\psi_n}\}$.
We construct the algebras by first identifying a set of strictly local projectors such that their common kernel is completely spanned \textit{only} by the QMBS eigenstates $\ket{\psi_n}$, and then proceeding via the route discussed in Sec.~\ref{subsubsec:SMexhaustive}.
As we discuss below, we do not always find such a choice of projectors, and we sometimes find that any choice of such strictly local projectors necessarily contains more states in the common kernel.
Nevertheless, we can consider these extra states as valid examples of QMBS as long as they are eigenstates of the AKLT and related Hamiltonians, which we find is the case; hence we can use this information to construct a local algebra containing those Hamiltonians.
We outline the construction below, and refer readers to App.~\ref{subsec:PBCAKLT}.
We also discuss analogous constructions for the OBC case in App.~\ref{subsec:OBCAKLT}.
\subsubsection{Shiraishi-Mori Projectors}
To construct local algebras with the commutants $\tmC^{(p)}_{\scar} \defn \lgen \{\ketbra{\psi_n}{\psi_m}\} \rgen$ and $\mC^{(p)}_{\scar} \defn \lgen \{\ketbra{\psi_n}\}\rgen$, which contain Hamiltonians with $\{\ket{\psi_n}\}$ as degenerate or non-degenerate QMBS respectively, we wish to construct a set of strictly local ``Shiraishi-Mori" projectors whose common kernel is completely spanned by $\{\ket{\psi_n}\}$. 
As a naive guess, we start with two-site projectors $\{\Pi_{j,j+1}\}$ that vanish on the AKLT towers of states, which can be inferred from results in \cite{mark2020unified,moudgalya2020large}; the exact expressions are shown in Eq.~(\ref{eq:2siteproj}).
Using these projectors, we numerically observe that the dimension of their common kernel grows exponentially with system size [see Eq.~(\ref{eq:2sitekernelPBC})], hence this kernel contains many more states than the tower of states $\{\ket{\psi_{n}}\}$.
We also numerically check that the extra states are not eigenstates of the AKLT model, hence these projectors cannot be used for the construction of the desired $\tmA^{(p)}_{\scar}$.
We then systematically construct three-site projectors $\{\Pi_{[j,j+2]}\}$ that vanish on the tower of states.
As we discuss in detail in App.~\ref{subsec:PBCAKLT}, their expressions can be derived directly from the MPS structure of $\{\ket{\psi_n}\}$, by first computing the total linear span of all the Schmidt states that appear on sites $\{j,j+1,j+2\}$ from all $\{\ket{\psi_n}\}$;  $\Pi_{[j,j+2]}$ is the projector onto the orthogonal complement of that subspace.
This linear span of Schmidt states turns out to be an 8-dimensional subspace spanned by states listed in Eq.~(\ref{eq:3sitesubspace}), hence $\Pi_{[j,j+2]}$ is a projector onto its orthogonal 19-dimensional subspace of the Hilbert space of three spin-1's, spanned by states listed in Eq.~(\ref{eq:19dimspan}).
The same projectors were also found numerically recently in \cite{yao2022bounding} in a different context.
We numerically find that the common kernel of these projectors is spanned by the tower of states $\{\ket{\psi_n}\}$ and one or two additional states [depending on the system size, see Eq.~(\ref{eq:3sitekernelPBC}) and App.~\ref{subsec:AKLTanalyticPBC} for a partial analytical proof], which we denote by $\{\ket{\phi_n}\}$.
\subsubsection{Exhaustive Algebras}
Lem.~\ref{lem:smsuf} then implies that there exists a bond algebra generated by finite-range terms that is irreducible in the orthogonal complement of this common kernel, i.e., we obtain bond and commutant algebra pairs of the form
\begin{align}
    \tmA^{(p)}_{\scar} &= \lgen \{\Pi_{[j,j+2]} h_{[j]} \Pi_{[j,j+2]}\} \rgen, \nn \\
    \tmC^{(p)}_{\scar} &= \lgen \{\ketbra{\psi_n}{\psi_m}\}, \{\ketbra{\psi_n}{\phi_m}\}, \{\ketbra{\phi_n}{\phi_m}\}\rgen,
\label{eq:PBCAKLTtowerpair}
\end{align}
where $h_{[j]}$ is a generic  (e.g., randomly chosen) term in the vicinity of site $j$.
Indeed, we can verify this numerically for small system sizes using methods we discuss in \cite{moudgalya2022numerical}, and we find that a generic three-site term $h_{[j]}$ is sufficient.
While this commutant $\tmC^{(p)}_{\scar}$ is larger than naively desired (which would be $\lgen \ketbra{\psi_n}{\psi_m} \rgen$), we can analytically determine that the extra states $\{\ket{\phi_m}\}$
are either the ferromagnetic state $\ket{F}$ or spin-wave states $\ket{1_{k=\pm \pi/2}}$ that are eigenstates of the AKLT model obtained in \cite{moudgalya2018a} [see Eqs.~(\ref{eq:spin1ferromagnetic}) and (\ref{eq:1kstates})].
Hence all the states $\{\ket{\psi_n}\}$ and $\{\ket{\phi_m}\}$ are degenerate exact eigenstates of Hamiltonians such as $H^{(p)}_{\AKLT} - S^z_{\tot}$, in particular of the entire family of Hamiltonians shown in Eq.~(\ref{eq:degHamils}).
According to the DCT, $\tmA^{(p)}_{\scar}$ is the algebra of all Hamiltonians with these as degenerate eigenstates, hence $H^{(p)}_{\AKLT} - S^z_{\tot}$ and related family of Hamiltonians all belong to $\tmA^{(p)}_{\scar}$.
However, we have neither been able to obtain an analytical expression for these Hamiltonians in terms of the generators of $\tmA^{(p)}_{\scar}$ nor prove analytically the numerically observed irreducibility of this algebra, although we anticipate that a proof  similar to the ones for the AKLT ground state [App.~\ref{app:SMalgebra}] or the ferromagnetic tower states [App.~\ref{subsec:FMirreducible}] could work. 
The algebra of Hamiltonians with $\{\ket{\psi_n}\}$ and $\{\ket{\phi_m}\}$ as potentially non-degenerate eigenstates is obtained by adding an $S^z_{\tot}$ to the algebra, and we have the local and commutant algebra pair
\begin{align}
    \mA^{(p)}_{\scar} &= \lgen \{\Pi_{[j,j+2]} h_{[j]} \Pi_{[j,j+2]}\}, S^z_{\tot} \rgen\nn~, \\
    \mC^{(p)}_{\scar} &= \lgen \{\ketbra{\psi_n}\}, \{\ketbra{\phi_n}{\phi_m}\}\rgen.
    \label{eq:ACpairpbcnondeg}
\end{align}
Note that degeneracy between the two $\{\ket{\phi_m}\}$ states is not split by $S^z_{\tot}$ if $L = 4n$ [see App.~\ref{subsec:PBCAKLT} for details].
We verified this numerically for small system sizes using methods in \cite{moudgalya2022numerical} using a randomly chosen $h_{[j]}$ with support on sites $\{j,j+1,j+2\}$.
The PBC AKLT Hamiltonian $H^{(p)}_{\AKLT}$ and the related families should be a part of this algebra, although we have not been able to obtain an analytical expression for $H^{(p)}_{\AKLT}$ in terms of the generators of $\mA^{(p)}_{\scar}$.
\subsubsection{Locality Considerations and Type II operators}
We now discuss some aspects of locality considerations for the AKLT algebras, and several results are similar to the ferromagnetic tower case discussed in Sec.~\ref{subsec:FMtowerscar}. 
For example, strictly local operators in the bond algebra $\tmA^{(p)}_{\scar}$ within a contiguous region $R$ in the bulk of the system should commute with the algebra of ket-bra operators formed by the Schmidt states of $\{\ket{\psi_n}\}$ over the region $R$. 
This algebra can actually be shown to be $\tmC_{\scar, R} = \lgen \{\ket{\psi_{n, \sigma\sigma'}}_R\bra{\psi_{m, \tau\tau'}}_R\}\rgen$, where $\ket{\psi_{n, \sigma\sigma'}}_R$ is the state of the OBC tower within the contiguous region $R$ that starts from $\ket{G_{\sigma\sigma'}}_R$, the AKLT state restricted to $R$ and boundary spin configurations $\sigma$ and $\sigma'$ [see App.~\ref{subsec:OBCAKLT} for a discussion of the OBC tower].
Using the bond and commutant algebra pair for the OBC tower [see Eq.~(\ref{eq:alltowerpair})], we can conclude that all strictly local operators in $\tmA^{(p)}_{\scar}$ that are in the bulk contiguous region $R$ should be expressible in terms of the generators of $\tmA^{(p)}_{\scar}$ in Eq.~(\ref{eq:PBCAKLTtowerpair}) within that region.
Following similar arguments as for the ferromagnetic tower, it is easy to show that this also applies to strictly local operators within a bulk contiguous region $R$ in the algebra $\mA^{(p)}_{\scar}$ that includes $S^z_{\tot}$, hence all such strictly local operators necessarily have ``more" symmetry than desired. 
Moving on to extensive local operators, it is straightforward to construct Type I operators using the generators of the bond algebra $\tmA^{(p)}_{\scar}$.
In addition, the algebra also contains operators that are not easily expressible in terms of symmetric strictly local ones~\cite{odea2020from, tang2020multi}, a prime example being
\begin{equation}
    \tH^{(p)}_{\AKLT} = H^{(p)}_{\AKLT} - S^z_{\tot},
\label{eq:AKLTdeghamil}
\end{equation}
where $S^z_{\tot}$ has been subtracted from $H^{(p)}_{\AKLT}$ to ensure that the states $\{\ket{\psi_n}\}$ and $\{\ket{\phi_m}\}$ are \textit{degenerate} eigenstates, which by DCT guarantees that $\tH^{(p)}_{\AKLT} \in \tmA^{(p)}_{\scar}$.
We now show that $\tH^{(p)}_{\AKLT}$ is a Type II operator in the algebra $\tmA^{(p)}_{\AKLT}$.
Note that $\tmA^{(p)}_{\scar} \subset \tmA^{(\AKLT)}_{\scar}$, where $\tmA^{(\AKLT)}_{\scar}$ is defined in Eq.~(\ref{eq:AKLTSMpair}), hence any operator in $\tmA^{(p)}_{\scar}$ that is Type II w.r.t.\ $\tmA^{(\AKLT)}_{\scar}$ is also Type II operator w.r.t.\ $\tmA^{(p)}_{\scar}$. 
$H^{(p)}_{\AKLT} - S^z_{\tot}$ is a Type II operator w.r.t. $\tmA^{(\AKLT)}_{\scar}$, since $H^{(p)}_{\AKLT}$ is Type I and $S^z_{\tot}$ is Type II as shown in App.~\ref{subsec:Sztotimpossible}.
Hence it follows that this operator is Type II w.r.t. $\tmA^{(\AKLT)}_{\scar}$ as well, i.e., it is impossible to express it as a linear combination of operators in $\tmA^{(\AKLT)}_{\scar}$. 
Note that similar arguments can also be used to show that the entire family of Hamiltonians $\tH^{(\Upsilon)}_{\AKLT\text{-fam}}$ of Eq.~(\ref{eq:degHamils}) for both OBC and PBC (i.e., $\Upsilon \in \{p, o\}$) are Type II operators.
However, many of them are in the same equivalence class since they differ by Type I operators.
In addition, there could be many distinct equivalence classes of Type II operators in $\tmA^{(p)}_{\scar}$, and perhaps some of them capture Hamiltonians discussed in \cite{sanada2023quantum}, but we defer a systematic exploration of this to future work.
\subsubsection{Difference from Shiraishi-Mori Hamiltonians and Equal Spacing Conjecture}
The PBC AKLT Hamiltonian itself, which belongs to the local algebra of $\mA^{(p)}_{\scar}$ of Eq.~(\ref{eq:ACpairpbcnondeg}), is then an example of a Type II Hamiltonian plus a lifting operator that is the uniform magnetic field $S^z_{\tot}$.
This makes it different from QMBS Hamiltonians of the generalized Shiraishi-Mori form, which, as discussed in Sec.~\ref{subsubsec:SMnature}, is a linear combination of a Type I operator and a lifting operator.
This also makes it analogous to the DMI Hamiltonian for the ferromagnetic tower of QMBS, which, as discussed in Sec.~\ref{subsubsec:FMQMBSTypeII}, has the same form.
Finally, due the similar structures of the algebra $\mA^{(p)}_{\scar}$ to that of $\mA^{\FM}_{\scar}$ of Eq.~(\ref{eq:FMlocalcommpair}), similar to Conj.~\ref{conj:equalspacing}, we conjecture that any local Hamiltonian that preserves the AKLT QMBS $\{\ket{\psi_n}\}$ necessarily breaks the degeneracy among the states into an equally spaced tower.
\subsection{Spin-1 XY \texorpdfstring{$\boldsymbol{\pi}$}{}-bimagnon and electronic Hubbard \texorpdfstring{\boldsymbol{$\eta$}}{}-pairing Scar Towers}\label{subsec:spin1XYHubbardscar}
We now move on to the towers of QMBS in the spin-1 XY model~\cite{schecter2019weak, mark2020eta} and the deformations of the electronic Hubbard models~\cite{moudgalya2020eta, mark2020eta}.
Both these Hamiltonians host very similar towers of QMBS (the precise correspondence between them was established in \cite{mark2020eta}), and they also
closely resemble the ferromagnetic towers of QMBS discussed in Sec.~\ref{subsec:FMtowerscar}, e.g., they too exist in lattices in all dimensions, irrespective of boundary conditions.
We refer readers to App.~\ref{app:S1XY} for a quick recap of some of the results on the spin-1 XY QMBS.
\subsubsection{Towers of Spin-1 XY QMBS}
We start with the ``$\pi$-bimagnon" QMBS in the spin-1 XY Hamiltonian on $L$ sites in one dimension, given by
\begin{equation}
    \ket{\Phi_n} \defn (Q^\dagger)^n \ket{\bar{F}},\;\;\;Q^\dagger \defn \sumal{j}{}{(-1)^j (S^+_j)^2},
\label{eq:spin1XYQMBS}
\end{equation}
where $\ket{\bar{F}} \defn \ket{- - \cdots - -}$ is a spin-1 ``ferromagnetic" state polarized in the $(-\hat{z})$ direction.
Note that $\ket{\Phi_n}$ can also be expressed by acting $Q \defn (Q^\dagger)^\dagger$ multiple times on the ferromagnetic state $\ket{F} \defn \ket{+ + \cdots + +}$.
$\ket{F}$ and $\ket{\bar{F}}$ are the highest and lowest ladder states of the ``pseudospin" $SU(2)$ symmetry generated by the operators $Q^\dagger$ and $Q$, hence this tower of QMBS spans a complete multiplet of the corresponding $SU(2)$, similar to the ferromagnetic tower of QMBS discussed in Sec.~\ref{subsec:FMtowerscar}.
Indeed, the momentum $k = \pi$ ``$\pi$-bimagnon" tower of Eq.~(\ref{eq:spin1XYQMBS}) can be unitarily mapped onto momentum $k = 0$ ``$0$-bimagnon" tower obtained using $Q^\dagger_{k=0} \defn \sum_j (S_j^+)^2$, which is similar to the spin-1/2 ferromagnetic tower, and we summarize this mapping in App.~\ref{subsec:unitaryspin1xy}.
Another similarity to the ferromagnetic tower is that the operators that permute the sites of the lattice are in the bond algebra of the $0$-bimagnon QMBS (see App.~\ref{subsec:permspin1xy}), hence the $0$-bimagnon states lack a spatial structure, i.e., they are invariant under arbitrary permutations of the sites.
\subsubsection{Local algebras}
Similar to the previous examples, we wish to construct a local algebra whose commutant is spanned by ket-bra operators of the QMBS eigenstates, i.e.,  $\lgen \ketbra{\Phi_n}{\Phi_m} \rgen$ or $\lgen \ketbra{\Phi_n} \rgen$.
Since the spin-1 XY Hamiltonian can be completely understood within the Shiraishi-Mori formalism of Eq.~(\ref{eq:SMgeneral}) (see \cite{schecter2019weak,mark2020unified} for the details of this construction, and also App.~\ref{app:S1XY} for the expressions of the Shiraishi-Mori projectors), it is straightforward to repeat the procedure discussed earlier in Sec.~\ref{subsubsec:SMexhaustive} to construct the corresponding local algebras.
However, we illustrate a different approach here, which might be more useful for searching for physically relevant Hamiltonians with the same QMBS. 
In particular, we restrict our search to local algebras that have additional natural symmetries, e.g., with $U(1)$ symmetry, hence we are interested in constructing the local algebras $\tmA^{(\XY)}_{\scar}$ and $\mA^{(\XY)}_{\scar}$ such that the commutants are given by $\tmC^{(\XY)}_{\scar} \defn \lgen \{\ketbra{\Phi_m}{\Phi_n}\}, S^z_{\tot} \rgen$ and $\mC^{(\XY)}_{\scar} \defn \lgen \{\ketbra{\Phi_n}\}, S^z_{\tot} \rgen$. 
We guess a set of nice nearest neighbor terms that generate the desired bond algebra, and we conjecture that the bond and commutant algebra pair for all $L$ in OBC and even $L$ in PBC is given by
\begin{align}
    &\tmA^{(\XY)}_{\scar} = \lgen \{S^x_j S^x_{j+1} + S^y_j S^y_{j+1}\}, \{(S^z_j)^2\},\nn \\
    &\qquad\qquad\{(S^z_j + S^z_{j+1}) (1 - S^z_j S^z_{j+1})\} \rgen, \nn \\
    &\tmC^{(\XY)}_{\scar} = \lgen \{\ketbra{\Phi_m}{\Phi_n}\}, S^z_{\tot} \rgen,
\label{eq:spin1XYpair}
\end{align}
which we numerically verify for small system sizes using methods we discuss in \cite{moudgalya2022numerical}. 
Given this bond algebra, the degeneracy of the eigenstates can be lifted as in the ferromagnetic and AKLT tower examples, by the addition of $S^z_{\tot}$ to the generators of $\tmA_{\scar}$. 
Hence we then expect the local and commutant algebra pair
\begin{align}
    &\mA^{(\XY)}_{\scar} = \lgen \{S^x_j S^x_{j+1} + S^y_j S^y_{j+1}\}, \{(S^z_j)^2\},\nn \\
    &\qquad\qquad\{(S^z_j + S^z_{j+1}) (1 - S^z_j S^z_{j+1})\}, S^z_{\tot} \rgen, \nn \\
    &\mC^{(\XY)}_{\scar} = \lgen \{\ketbra{\Phi_n}\}, S^z_{\tot} \rgen.
\label{eq:spin1XYnondegpair}
\end{align}
It is straightforward to see that this algebra contains the standard spin-1 XY Hamiltonian discussed in \cite{schecter2019weak, moudgalya2021review, chandran2022review}.
Note that these algebras can be directly generalized to the case of $0$-bimagnon towers, as we discuss in App.~\ref{app:S1XY}.
\subsubsection{Locality Considerations and Type II Operators}\label{subsubsec:locspin1XY}
According to the DCT, all operators that annihilate the states $\{\ket{\Phi_n}\}$ should be a part of the algebra $\tmA^{(\XY)}_{\scar}$.
This can be used to understand the structure of $U(1)$ spin-conserving local operators with $\{\ket{\Phi_n}\}$ as degenerate eigenstates. 
Using arguments similar to the ferromagnetic and AKLT towers of QMBS, we can show that strictly local operators in $\tmA^{(XY)}_{\scar}$ with support in a contiguous region can be constructed from generators completely within that region. 
However, also similar to the other QMBS, the construction of extensive local operators can be more complicated, i.e., there are Type II symmetric Hamiltonians that cannot be expressed as a sum of strictly local operators in the algebra; we again attribute this to the ``non-local" nature of the commutant $\tmC^{(XY)}_{\scar}$.  
We can use these considerations to understand the results of a systematic numerical search for $U(1)$ conserving on-site and nearest-neighbor Hamiltonians (or terms) for which the states $\{ \ket{\Phi_n}\}$ are degenerate, carried out in \cite{mark2020eta}.
In particular, they found the terms listed in group A of Tab.~I there and the terms \#1--\#7 and \#9--\#12 in Tab.~II there.
Among them are 13 linearly independent strictly local operators, in the span of entries \#1--\#7 and \#9--\#11 of Tab.~II [a convenient basis spanning these terms is shown in App.~\ref{app:S1XY}, see Eq.~(\ref{eq:tmAS1kpi})],\footnote{Note that the entries \#3, \#4, and \#5 in Tab.~II in \cite{mark2020eta} can be expressed in terms of \#9, \#10, and \#11 respectively; also, \#6 contains two linearly independent operators and \#11 contains four.
The identity as well as all one-site operators of Tab.~I there are included in the span of the operators \#1-\#4 in Tab.~II.} and one extensive local operator, \#12, which we list here for easy reference:
\begin{equation}
\# 12 = \sum_j i( \ketbra{-,+}{+,-} \,-\, \ketbra{+,-}{-,+})_{j,j+1} ~.
\label{eq:S1asasum}
\end{equation}
Since any nearest-neighbor strictly local operator from the bond algebra can be constructed from nearest-neighbor generators, the 13 independent strictly local terms can be understood from Eq.~(\ref{eq:spin1XYpair}); indeed, we find that the dimension of the two-site algebra, i.e., $\lgen S^x_j S^x_{j+1} + S^y_j S^y_{j+1}, (S^z_j)^2, (S^z_{j+1})^2, (S^z_j + S^z_{j+1})(1 - S^z_j S^z_{j+1}) \rgen$, is $13$, and we have verified that the 13 terms obtained in \cite{mark2020eta} span this two-site algebra (see App.~\ref{subsec:SMproj} for some details).
This exercise can also be repeated to construct three- and more-site strictly local $U(1)$-symmetric operators that annihilate the QMBS, without resorting to numerical searches.
The construction of the extensive local operator \#12 is not so straightforward, and its derivation proceeds similar to that of the DMI term in the case of the spin-1/2 ferromagnetic tower (see Sec.~\ref{subsubsec:FMQMBSTypeII}).
For simplicity, we first unitarily transform the $\pi$-bimagnon QMBS Eq.~(\ref{eq:spin1XYQMBS}) to the $0$-momentum-bimagnon tower [see App.~\ref{subsec:unitaryspin1xy} for details of this transformation], and the operator \#12 with consecutive sites on different sublattices exactly maps onto itself up to an overall unimportant sign.
The bond algebra corresponding to the $0$-bimagnon tower exactly maps onto that of the spin-1/2 ferromagnetic tower [we refer the readers to App.~D of \cite{mark2020eta} and our App.~\ref{app:S1XY} for the details], and the operator \#12 maps onto the DMI term of Eq.~(\ref{eq:DMIHamil}).
We can then directly follow the derivation of the DMI term in the algebra $\tmA^{\FM}_{\scar}$.
To begin, we first numerically verify that the three-site term [similar to $D^\alpha_{j,j+1,j+2}$ in Eq.~(\ref{eq:FMscarsimplegens})] can be generated using the nearest-neighbor generators of the bond algebra for the $0$-bimagnon QMBS on sites $\{j,j+1\}$ and $\{j+1,j+2\}$. 
We then apply permutation operators, which are also in the bond algebra, on these terms and eventually derive the cyclic extensive local term with PBC.
Since the $\pi$-bimagnon QMBS of Eq.~(\ref{eq:spin1XYQMBS}) and the $0$-bimagnon QMBS are unitarily related, the same derivation hence works for the extensive local term in the former case, up to some extra sign factors obtained from the sublattice transformation.
Moreover, it is also possible to use the analogy to the spin-1/2 cyclic DMI term and repeat the arguments in App.~\ref{subsec:DMIimpossible} essentially verbatim to show that this extensive local term of Eq.~(\ref{eq:S1asasum}) for the $0$-bimagnon QMBS cannot be expressed as a sum of strictly local terms that annihilate the QMBS; and a similar statement is true for the extensive local term in the $\pi$-bimagnon case due to the sublattice transformation.
This shows that the \#12 of Eq.~(\ref{eq:S1asasum}) is a Type II operator in $\mA^{(\XY)}_{\scar}$.
Finally, due the similar structures of this algebra $\mA^{(\XY)}_{\scar}$ to the ferromagnetic case, we conjecture the statements Conjs.~\ref{conj:exhaustivechars} and \ref{conj:equalspacing} in this case too.
\subsubsection{Generalization to \texorpdfstring{$\eta$}{}-pairing QMBS}
We now briefly discuss
the $\eta$-pairing QMBS in the electronic Hubbard and related models~\cite{moudgalya2020eta, mark2020eta}. 
To recap, we are considering an electronic Hilbert space of spin-1/2 fermions, and the QMBS states on $L$ sites in one dimension are given by
\begin{equation}
    \ket{\Xi_n} \defn (\eta^\dagger_{\pi})^n \ket{\Omega},\;\;\;\eta^\dagger_{\pi} \defn \sumal{j}{}{(-1)^j \cd_{j, \uparrow} \cd_{j,\downarrow}},
\label{eq:etaQMBS}
\end{equation}
where $\{\cd_{j,\sigma}\}$ and $\{c_{j,\sigma}\}$ for $\sigma \in \{\uparrow, \downarrow\}$ are the spin-$\sigma$ fermionic creation and annihilation operators, and $\ket{\Omega}$ is the vacuum. 
Similar to the ferromagnetic and spin-1 XY towers, $\ket{\Xi_n}$ can also be expressed starting from the fully filled state $\ket{\bar{\Omega}}$ by repeated actions of $\eta_\pi \defn (\eta^\dagger_\pi)^\dagger$. 
$\ket{\bar{\Omega}}$ and $\ket{\Omega}$ are the highest and lowest ladder states of the pseudospin $SU(2)$ symmetry generated by the $\eta^\dagger_\pi$ and $\eta_\pi$ operators~\cite{yang1989eta,vafek2017entanglement}, hence this tower of QMBS spans a complete multiplet of this $SU(2)$.
Moreover, as discussed in \cite{pakrouski2020many, moudgalya2022from}, these states are singlets of certain Lie groups or ``pre-bond" algebras, e.g., see \#1c or \#3b of Tab.~III in \cite{moudgalya2022from}.
The representations of these groups or algebras also contain the operators that permute the sites of the lattice, which enforces the fact that these states lack spatial structure, a feature first pointed out in \cite{pakrouski2020many}.
Reference \cite{mark2020eta} illustrated a correspondence between the QMBS $\{\ket{\Xi_n}\}$ of Eq.~(\ref{eq:etaQMBS}) and the spin-1 XY QMBS $\{\ket{\Phi_n}\}$ of Eq.~(\ref{eq:spin1XYQMBS}).
In addition to the correspondence between the states, they also offered a correspondence between classes of parent Hamiltonians containing the QMBS, between electronic spin $SU(2)$ symmetric operators for $\{\ket{\Xi_n}\}$ and $U(1)$ spin conserving operators for $\{\ket{\Phi_n}\}$.  
Here we exploit the correspondence to directly construct the local algebra of spin $SU(2)$-symmetric electronic operators with $\{\ket{\Xi_n}\}$ as QMBS, using the local algebras of $U(1)$ spin conserving operators shown in Eqs.~(\ref{eq:spin1XYpair}) and (\ref{eq:spin1XYnondegpair}). 
The results of \cite{mark2020eta} directly suggest the following substitutions between the operators on the spin-1 (XY model) and spin-1/2 fermion (Hubbard model) Hilbert spaces: $S^x_j S^x_{j+1} + S^y_j S^y_{j+1} \leftrightarrow T^{(r)}_{j,j+1} \defn \sum_{\sigma}{(\cd_{j, \sigma} c_{j+1, \sigma} + h.c.)}$ and $S^z_j \leftrightarrow K_j - 1$ where $K_j \defn n_{j,\uparrow} + n_{j,\downarrow}$, and we obtain the bond and commutant algebra pair
\begin{align}
    &\tmA^{(\hub)}_{\scar} = \lgen \{T^{(r)}_{j,j+1}\}, \{(K_j - 1)^2\},\nn \\
    &\qquad\qquad\{(K_j + K_{j+1} - 2) (K_j K_{j+1} - K_j - K_{j+1})\} \rgen, \nn \\
    &\tmC^{(\hub)}_{\scar} = \lgen \{\ketbra{\Xi_m}{\Xi_n}\}, \{S^\alpha_{\tot}\} \rgen, 
\label{eq:etapair}
\end{align}
where $\{T^{(r)}_{j,j+1}\}$ denotes the real free-fermion hopping terms and $(K_j - 1)^2$ is the on-site Hubbard term.
The bond algebra $\tmA^{(\hub)}_{\scar}$ of Eq.~(\ref{eq:etapair}) can thus be interpreted as an enlargement of the so-called Hubbard algebra $\mA^{(r)}_{\hub}$ with real hoppings, see \cite{moudgalya2022from} for a discussion of the various types of Hubbard models in the algebra language (particularly Tab.~IV there).
A few comments on these algebras are in order. 
First, the degeneracy of $\{\ket{\Xi_n}\}$ can be lifted by adding the total number operator $N_{\tot} \defn \sum_{j, \sigma \in \{\uparrow, \downarrow\}}{\cd_{j,\sigma} c_{j,\sigma}}$ to the algebra $\tmA^{(\hub)}_{\scar}$, which results in an algebra similar to Eq.~(\ref{eq:spin1XYnondegpair}).
Second, locality considerations for the construction of strictly and extensive local operators are identical to those in the spin-1 XY case, discussed in Sec.~\ref{subsubsec:locspin1XY}.
Hence the results of a numerical search for spin $SU(2)$-symmetric on-site and nearest-neighbor terms for which $\{\ket{\Xi_n}\}$ are degenerate eigenstates, performed in \cite{mark2020eta},  can be understood using the algebra of $\tmA^{(\hub)}_{\scar}$.
They found 14 linearly independent such strictly local terms, given by the terms in group A in Tab.~I there and \#1--\#7 and \#9--\#11 in Tab.~III there,\footnote{The terms in Tab.~I are expressible in terms of those in Tab.~III.
Further, \#3--\#5 can be expressed as linear combinations of \#9--\#11; also \#1, \#6, \#9, and \#10 correspond to two linearly independent operators, while \#11 to four of them.} and these span the 14-dimensional two-site bond algebra generated by the generators of Eq.~(\ref{eq:etapair}) with support on sites $\{j,j+1\}$. 
The derivation of the Type II symmetric DMI-like Hamiltonian of \#12 in Tab.~III of \cite{mark2020eta} 
is also similar to that of the analogous term in the spin-1 XY case shown in Eq.~(\ref{eq:S1asasum}).   
Finally, similar ideas can be used to derive local algebras that consist of terms breaking the spin $SU(2)$ symmetry while preserving the same QMBS, e.g., those that only preserve the total spin $S^z_{\tot}$ such as the ones discussed in \cite{moudgalya2020eta, moudgalya2021review}; or to derive local algebras with related QMBS, e.g., the zero-momentum $\eta$-pairing states~\cite{mark2020eta, moudgalya2022from} or the spin ferromagnetic multiplet in this Hilbert space. 
These derivations follow by rather straightforward guesswork or mappings to algebras discussed here, hence we do not write down their explicit expressions.
\section{Implications for Thermalization}\label{sec:eth}
We now discuss a few implications of this local and commutant algebra interpretation of QMBS, and raise a few questions on them in the context of thermalization. 
\subsection{Towards a Definition for QMBS}\label{subsec:QMBSdefn}
Thinking about QMBS in this framework motivates a 
precise definition for QMBS, which has so far been absent from the literature.  
\subsubsection{Necessary Condition and Violation of ETH}
As we demonstrated with the help of several examples, the QMBS are singlets of algebras generated by local operators that do not commute with each other.
This motivates the following necessary condition for QMBS.
\begin{condition}
Any exact QMBS eigenstate is a common eigenstate of multiple non-commuting local operators.
\end{condition}
\noindent 
Note that the local operators involved can either be strictly local or extensive local.
Any state that satisfies this condition can be embedded into the middle of the spectrum of a generic local Hamiltonian constructed from the said local operators, although as we discuss below its atypicality is not guaranteed only by this condition.  
While this requirement might appear sweeping, it is clear from the literature that numerous examples of QMBS satisfy this definition~\cite{papic2021review, moudgalya2021review}, see also e.g., \cite{shibata2020onsager, iadecola2020quantum, banerjee2021quantum, biswas2022scars, schindler2022exact} for a partial list.
The class of states that satisfy this condition includes tensor network states (e.g., MPS and PEPS) with finite bond dimension but is not limited to them, e.g., the QMBS can have entanglement growing logarithmically with system size~\cite{moudgalya2018b, vafek2017entanglement, schecter2019weak, moudgalya2020eta}, or even proportional to the system size (i.e., as a volume law)~\cite{mori2017thermalization,langlett2022rainbow, srivatsa2022mobility}.
This property of QMBS also has 
direct implications for the Eigenstate Thermalization Hypothesis (ETH)~\cite{deutsch1991quantum, srednicki1994chaos}.
According to ETH, since the reduced density matrix of an eigenstate in the bulk of the spectrum should resemble the Gibbs density matrix expressed in terms of the Hamiltonian, a single eigenstate should contain all the information about the Hamiltonian~\cite{garrison2018does}.
Indeed, when a systematic ``correlation matrix" method~\cite{chertkov2018computational, qi2019determininglocal} to search for local operators for which a given state is an eigenstate is applied to a generic eigenstate of some local Hamiltonian without any local conserved quantities, the local Hamiltonian can generically be uniquely reconstructed from that eigenstate~\cite{qi2019determininglocal}.\footnote{For Hamiltonians with conventional (i.e., extensive local) conserved quantities, we would expect this reconstruction to yield linear combinations of the original Hamiltonian and possibly some of the conserved quantities. In the case of Abelian symmetries, this is not in contradiction with the ETH applied to such systems, where one would expect the reduced density matrix to resemble a generalized Gibbs density matrix. The situation is less clear for systems with non-Abelian symmetries, where the implications of ETH are more subtle and a subject of active study~\cite{murthy2022nonabelian}.}
However, requiring the aforementioned 
property on QMBS, such a unique reconstruction is not possible even in principle, since they are by definition simultaneous eigenstates of multiple non-commuting local operators.
This feature distinguishes QMBS from other generic eigenstates even for a particular Hamiltonian, and implies that the QMBS necessarily violate the ETH. 
While this is a necessary property of QMBS, it is clear that it is not sufficient since there are examples of states that are expected to satisfy ETH and satisfy the above property.
For example, any non-integrable $SU(2)$-symmetric Hamiltonian possess exponentially many eigenstates $\{\ket{E_n}\}$ that satisfy $\vec{S}^2_{\tot}\ket{E_n} = 0$ [these are usually referred to as the spin singlets of the $SU(2)$ symmetry], and they are simultaneous eigenstates of all the total spin operators $\{S^\alpha_{\tot}\}$, which do not commute with each other.
\subsubsection{Sufficient Conditions}
For singlets of any local algebra to be referred to as QMBS, we also require that the decomposition of Eq.~(\ref{eq:Hilbertdecomp}) corresponds to that of Eq.~(\ref{eq:scarfactorize}), i.e., we require the existence of a ``large" thermal block that spans most of the Hilbert space. 
This condition can be made more precise by studying the local algebra generated by the multiple local operators reconstructed from that state using procedures such as the correlation matrix methods~\cite{chertkov2018computational, qi2019determininglocal}; the QMBS state is one of the singlets of this algebra by construction.\footnote{Since we are interested in reconstructing a locally generated algebra, we require reconstruction procedures to search only over (strictly or extensive) local operators with a maximum range $r_{\max}$ independent of the system size $L$.}
Since the existence of the ``thermal" block of Eq.~(\ref{eq:scarfactorize}) is equivalent to requiring that the decomposition of Eq.~(\ref{eq:Hilbertdecomp}) for this local algebra has a ``large" irreducible representation, we arrive at the following sufficient condition for QMBS.
\begin{condition}
Any state can be made a QMBS eigenstate of some Hamiltonian if the dimensions $\{D_\lambda\}$ of the irreducible representations of its ``parent algebra" (algebra generated by all the bounded-range strictly local and extensive local operators that have the state as an eigenstate) satisfy
\begin{equation}
\frac{\max_\lambda D_\lambda}{\dim(\mH)} \rightarrow 1 \quad \text{~as~} L \rightarrow \infty.
\label{eq:maxDlambda}
\end{equation}
\end{condition}
\noindent Note that Eq.~(\ref{eq:maxDlambda}) is satisfied by the tower examples of QMBS discussed in this work, since they have $\max_\lambda D_\lambda = \dim(\mH) - \mathcal{O}(L^p)$, and also by some examples of QMBS like embedding of $2^L$ scar states inside a $3^L$-dimensional Hilbert space in \cite{shiraishi2017eth}; however we cannot rule out other possibilities.\footnote{In particular, there might be cases where some QMBS can only exist in the presence of some conventional continuous on-site unitary symmetry.
If such a case exists, the algebra of reconstructed local operators would likely satisfy $\max_\lambda D_\lambda/D \sim 1/\text{poly}(L)$ due to the presence of the conventional symmetry.}
Nevertheless, with 
the requirement of Eq.~(\ref{eq:maxDlambda}), the QMBS eigenstates, which transform under one-dimensional representations of this local algebra generated by finite-range local parent operators, are atypical eigenstates of generic Hamiltonians constructed from that local algebra. 
The aforementioned $SU(2)$ spin singlets of a non-integrable $SU(2)$-symmetric Hamiltonian---hence eigenstates of non-commuting ``parent Hamiltonians" $S_{\tot}^\alpha,~\alpha \in \{x,y,z\},$---do not satisfy this QMBS condition.
Given a generic spin-$1/2$ non-integrable local $SU(2)$-symmetric Hamiltonian $H$ and a generic choice of its singlet eigenstate $\ket{E_n}$, we generically expect the \textit{only} local operators that can be reconstructed are $H$ and $\{S^\alpha_{\tot}\}$.
The parent algebra $\lgen H, \{S^\alpha_{\tot}\}\rgen$ does not have any exponentially large irreducible representations, since for a generic $H$ its commutant and center is $\lgen H, \vec{S}^2_{\tot}\rgen$, which implies that $H = \oplus_\lambda c_\lambda \mathds{1}_{D_\lambda}$, where $D_\lambda$'s are the dimensions of the irreps of the reconstructed algebra ($d_\lambda = 1$ since the expected commutant is Abelian).
This should be equivalent to the full diagonalization of $H$, which, by virtue of being a generic $SU(2)$-symmetric Hamiltonian, has an eigenstate degeneracy of at most $L+1$, which immediately shows that $\max_\lambda(D_\lambda) = L + 1$, violating Eq.~(\ref{eq:maxDlambda}).
Finally, we also note that this definition of QMBS also implies that scarriness of a state is also a Hilbert space dependent feature and sometimes depends on quantities we are interested in, and we discuss some such ``edge cases" in App.~\ref{subsec:QMBSremarks}.
\subsubsection{Sufficiency of the Shiraishi-Mori Structure}
We now show that the ``Shiraishi-Mori" condition of Eq.~(\ref{eq:targetdefn}), i.e., demanding that the candidate state is in the common kernel of some set of strictly local projectors, is sufficient for Eq.~(\ref{eq:maxDlambda}) to be satisfied, although we cannot prove that these are necessary.
In particular, in App.~\ref{app:QMBSsufficient} we prove the following Lemma.
\begin{restatable}{lem}{smrecsuf}\label{lem:SMsufficient}
In a system of size $L$, if among the reconstructed parent operators bounded by some finite range $r_{\max}$ we have a ``dense" set of $\mO(L)$ strictly local operators $\{A_{[j]}\}$ covering the entire lattice such that the separation between neighboring $A_{[j]}$'s is bounded by an $L$-independent number $\ell_{\max}$, then Eq.~(\ref{eq:maxDlambda}) is satisfied for a parent algebra generated by operators of some finite range $r'_{\max} \geq r_{\max}$.
\end{restatable}
To provide some intuition for Lem.~\ref{lem:SMsufficient}, we note that since the reconstruction of any strictly local $A_{[j]}$ also implies the reconstruction of all its powers, we can w.l.o.g.\ assume $A_{[j]}$ to be a projector, say $P_{[j]}$. 
The state can then be expressed as a part of the kernel of strictly local projectors, i.e., in a target space $\mT$.
The result of Lem.~\ref{lem:smsuf} proves that we can find some finite $r'_{\max} \geq r_{\max}$ such that $\max_\lambda D_\lambda = \dim(\mH) - \dim(\mT)$, hence it satisfies Eq.~(\ref{eq:maxDlambda}) as long as $\dim(\mT)$ scales slower than the Hilbert space dimension.
In App.~\ref{subsec:singleQMBS}, we prove this using the fact that they are the common kernel of a set of ``dense" strictly local projectors [see Eq.~(\ref{eq:fracTbound})].
We conclude this discussion with some remarks.
First, 
in the above we only considered a subset of all possible reconstructed parent operators, and that was sufficient to guarantee Eq.~(\ref{eq:maxDlambda});
additional parent operators only increase $\max_\lambda D_\lambda$ for the algebra of reconstructed local operators, hence Eq.~(\ref{eq:maxDlambda}) still holds.
Second, the fact that the Shiraishi-Mori structure is a sufficient condition for QMBS already shows that all the examples we considered in Sec.~\ref{sec:scarexamples} are QMBS in this definition.
Finally, as we mention in Sec.~\ref{subsec:SMformalism}, the results of \cite{yao2022bounding} along with the  Shiraishi-Mori structure can be used to show that the entanglement entropy of the candidate QMBS, or any other state in $\mT$, over any extensive contiguous subregion is smaller than the Page value~\cite{page1993} expected in generic eigenstates in the middle of the spectrum of a non-integrable local Hamiltonian.
\subsection{QMBS Projectors as Generalized Symmetries}\label{subsec:QMBSgensym}
The commutant language makes QMBS projectors on par with regular conserved quantities as well as the exponentially many ones in the context of Hilbert space fragmentation~\cite{moudgalya2021hilbert}, incorporating all these phenomena under the umbrella of ``generalized" symmetries.
These generalized symmetries are beyond the usual on-site symmetries, and also beyond other exotic symmetries such as subsystem or higher-form symmetries that are being explored in several different contexts in the literature~\cite{mcgreevy2022generalized}. 
Naively, if the QMBS projectors are considered to be non-local symmetries, then QMBS are not examples of ergodicity breaking since ergodicity is usually defined for a given symmetry sector~\cite{mondainicomment2018}.
However, QMBS clearly do not fit into the usual framework of quantum statistical mechanics, e.g., it is not clear if analogues of the Gibbs ensembles can be defined for such systems.
Moreover, these conserved quantities are qualitatively very different from the conventional ones, and this has implications for the dynamics starting from generic initial states.
For one, an important difference is in the distribution of $\{D_\lambda\}$'s, i.e., the sizes of the various quantum number sectors, which is highly ``skewed" in the case of QMBS conserved quantities.
For example, while the presence of a $Z_2$ symmetry in a spin-1/2 Hilbert space would lead to two sectors of sizes $2^{L-1}$, the presence of a single QMBS would also lead to two sectors, but now with dimensions $2^L - 1$ and $1$.\footnote{For more examples, spin-1 systems with a $U(1)$ symmetry with conserved $S^z_{\tot}$ would possess $\sim L$ sectors of rather ``uniform" typical sizes $\sim 3^L/L$, while the example of embedding $2^L$ scar states in a spin-1 Hilbert space considered in Apps.~\ref{app:SMexistence} and \ref{app:QMBSsufficient} would possess one ``thermal" sector with dimension $3^L - 2^L$ and a non-thermal sector with dimension $2^L$ for degenerate scars or $2^L$ non-thermal sectors of dimension $1$ for non-degenerate scars.}
This is essentially a formal way to state the intuition that the presence of QMBS only ``affects" a small part of the Hilbert space, and the dynamics of most initial states remain unchanged.
This difference is also sometimes evident from the Mazur bound for the infinite-temperature autocorrelation function~\cite{mazurbound1969, dhar2020revisiting}.
Given a set of operators $\{Q_\alpha\}$ in the commutant that are mutually orthogonal [defined as $\text{Tr}(Q_\alpha^\dagger Q_\beta) \propto \delta_{\alpha,\beta}$ corresponding to infinite-temperature ensembles], the time-averaged autocorrelation function of an operator under the dynamics of any Hamiltonian in the corresponding local algebra $\mA$ is lower-bounded as~\cite{moudgalya2021hilbert}
\begin{equation}
    \lim_{\tau \rightarrow \infty} \frac{1}{\tau} \int_0^\tau{dt\ \langle A(t) A(0)\rangle} \geq \sumal{\alpha}{}{\frac{\opbraket{A}{Q_\alpha} \opbraket{Q_\alpha}{A}}{\opbraket{Q_\alpha}{Q_\alpha}}},
\label{eq:mazurbound}
\end{equation}
where $A(t) = e^{i H t} A(0) e^{-i H t}$, and the overlap is defined as $\opbraket{A}{B} \defn \text{Tr}(A^\dagger B)/\dim(\mH)$. %
We can then quantify the ``importance" of various operators $\{Q_\alpha\}$ in the commutant for the dynamics of the operator $A$ in terms of their contribution to the R.H.S.\ of Eq.~(\ref{eq:mazurbound}). 
For strictly local operators $A$, the contribution of local conserved quantities such as $S^z_{\tot}$ scales with system size as $\sim 1/L$, whereas for QMBS eigenstates and the corresponding projectors or ket-bra operators as conserved quantities, this is $\sim \exp(-c L)$, analogous to the contribution of ``frozen states" in the case of Hilbert space fragmentation~\cite{moudgalya2021hilbert}. 
Finally, while the QMBS eigenstates can exist for a variety of systems, they are considered to break ergodicity only if their existence cannot be explained by conventional symmetries. 
For example, the states of the ferromagnetic multiplet of the Heisenberg Hamiltonian are not referred to as examples of QMBS as long as $SU(2)$ symmetry is present, while \textit{the same} eigenstates become QMBS once the global $SU(2)$ symmetry is broken.
These inconsistencies in the definitions, which also exist in systems with Hilbert space fragmentation~\cite{moudgalya2021hilbert}, call for a more precise definition of ergodicity and its breaking in isolated quantum systems.
\section{Conclusion and Outlook}\label{sec:conclusion}
In this work, we studied Quantum Many-Body Scars (QMBS) in the language of local and commutant algebras.
In particular, we propose that there is a local algebra, i.e., an algebra generated by strictly local and/or extensive local terms, such that its commutant algebra (i.e., the centralizer) is spanned by projectors onto the QMBS eigenstates.
We demonstrated this 
with explicit examples of QMBS from the literature, including general MPS states, spin-1/2 ferromagnetic tower of states, the AKLT tower of states, and the spin-1 XY $\pi$-bimagnon and electronic $\eta$-pairing towers of states.
In previous works, we showed that Hilbert space fragmentation~\cite{moudgalya2021hilbert} and several conventional symmetries~\cite{moudgalya2022from} can be understood in this language, and in each of these cases, there is a local algebra such that the commutant algebra contains all the conserved quantities that explain the origin of dynamically disconnected subspaces, which are the Krylov subspaces in fragmented systems and the usual symmetry sectors in systems with conventional symmetries.
This work hence casts QMBS in the same framework, attributing the origin of the dynamically disconnected QMBS subspace to unconventional conserved quantities in the commutant algebra, hence demonstrating the similarity between the underlying mathematical structures responsible for QMBS phenomena and those in fragmentation phenomena and in conventional symmetry physics.
Understanding QMBS in this language has a number of advantages. 
First, as we discussed in Sec.~\ref{sec:QMBSunified}, this framework unifies several of the previously introduced unified formalisms for QMBS~\cite{moudgalya2021review, papic2021review, chandran2022review}, particularly the symmetry-based ones~\cite{moudgalya2021review}, and provides a common language in which they can all be related to one another. 
This language generalizes the idea that QMBS are singlets of certain Lie groups, introduced in the Group Invariant formalism~\cite{pakrouski2020many}, to the idea that QMBS are singlets of certain local algebras which need not have any simple underlying Lie group structure.
For Lie groups and bond algebras generated by free-fermion terms, which was the focus of \cite{pakrouski2020many}, these two pictures coincide, and we illustrated this equivalence in \cite{moudgalya2022from}.\footnote{Note that even when the generators of the bond algebras are free-fermion terms, the algebras also contain interacting terms that can be constructed by linear combinations of products of free-fermion terms.}
Reference \cite{pakrouski2020many} also highlighted that several examples of the QMBS eigenstates lack spatial structure, and this was attributed to the presence of the permutation group as a subgroup of the parent Lie group.
However, we find that this occurs more generally since the presence of the permutation group within the local algebra does not require parent Lie group structure.
Additionally, this interpretation of QMBS also has explicit connections to decoherence-free subspaces, noiseless subsystems, and dark states studied in different contexts in the literature~\cite{lidar2003decoherencereview, bartlett2007reference, wu2005holonomic,  holbrook2005noiseless}, and it would be interesting to explore these connections further.
It would also be interesting to better understand how so-called Spectrum Generating Algebra mechanisms, discussed in \cite{mark2020unified, moudgalya2020eta}, or the related scarred Hamiltonians derived from spherical tensor operators~\cite{tang2020multi} fit into this framework.
Second, this language allows the application of the Double Commutant Theorem (DCT), which guarantees that the local algebra is the exhaustive algebra of ``symmetric" operators, i.e., the set of all operators that commute with the conserved quantities in the commutant. 
In the case of conventional symmetries, this allowed us to formally construct \textit{all} Hamiltonians with a given set of symmetries~\cite{moudgalya2022from}. 
Analogously, the local algebra for the QMBS case allows us to build \textit{all} Hamiltonians with the desired set of QMBS in their spectrum; this also shows that there are usually multiple local perturbations that exactly preserve the QMBS. 
For standard examples of QMBS, we showed how ideas from the Shiraishi-Mori formalism~\cite{shiraishi2017eth} can be generalized to construct generators for the exhaustive local algebras with the commutants spanned by the QMBS projectors (or QMBS ket-bra operators for degenerate scars), and all Hamiltonians with the same QMBS should be expressible in terms of these generators.
This leads to an exhaustive characterization of Hamiltonians with a given set of QMBS, which motivated conjectures on the general structure of such Hamiltonians.
For example, for many examples of towers of QMBS, we conjecture that any extensive local Hamiltonians with those QMBS necessarily have the QMBS as \textit{equally-spaced} towers of eigenstates.

The application of DCT also allowed us to precisely understand the  distinction between the two broad classes of Hamiltonians with QMBS: (i) ``Shiraishi-Mori"-like Hamiltonians for which the QMBS are eigenstates of each strictly local term defined in a precise sense, and (ii) intrinsically ``as-a-sum annihilator"~\cite{mark2020eta, odea2020from} or ``beyond Shiraishi-Mori"~\cite{mark2020unified, odea2020from, tang2020multi} Hamiltonians where all the individual terms collaborate such that the QMBS are annihilated by the full extensive local Hamiltonian and not by the individual terms.
In the algebra language, this distinction originates from two types of symmetric Hamiltonians that can be constructed within any bond algebra generated by strictly local terms: Type I and Type II symmetric Hamiltonians, where the former can be expressed as a sum of strictly local symmetric terms whereas the latter cannot.
This also resolves a long-standing open question on the connection between the AKLT Hamiltonian and the Shiraishi-Mori formalism~\cite{mark2020unified, odea2020from, tang2020multi}.
In this work, we found a set of three-site projectors such that the common kernel is completely spanned by the QMBS tower of states of the AKLT model, which led to the exhaustive algebra of all Hamiltonians that contain the AKLT QMBS. 
The generators of this algebra can then be used to construct ``Shiraishi-Mori-like" Hamiltonians with the AKLT QMBS as well as Hamiltonians such as the AKLT Hamiltonian itself that lie beyond.
This then reveals the bigger picture in the landscape of QMBS Hamiltonians.
For the AKLT tower of QMBS, it happened historically that the Hamiltonian beyond the Shiraishi-Mori construction---the celebrated AKLT chain---was known first~\cite{moudgalya2018a, moudgalya2018b} and the projectors required for constructing Shiraishi-Mori-like Hamiltonians have only been illustrated in this paper.
In contrast, for many other QMBS towers such as the spin-1/2 ferromagnetic, spin-1 $\pi$-bimagnon, and electronic $\eta$-pairing, the Shiraishi-Mori-like Hamiltonians were constructed first~\cite{iadecola2018exact, moudgalya2020eta, mark2020eta} and as-a-sum Hamiltonians that lie beyond, such as the DMI-like terms, were discovered in later works using more systematic searches~\cite{mark2020eta, odea2020from, ren2021deformed}.
Hence from the algebra perspective, there is no fundamental difference between the AKLT and the other towers of QMBS---in both cases there are Hamiltonians of both kinds.
However, this framework is not without its caveats.
A major caveat that remains is that while the DCT guarantees that a given operator with QMBS belongs to the local algebra, it does not provide an expression for it. 
Such expressions can be complicated, e.g., in the case of the ferromagnetic tower of QMBS, we showed how the extensive local DMI term can be generated from the strictly local generators of the algebra by means of a highly non-local expression; we were also able to prove that there is no rewriting as a \textit{sum} of symmetric strictly local terms.
This is an example of a Type II operator, and while we were able to rule out their existence for bond algebras with on-site unitary symmetries~ \cite{moudgalya2022from} this is no longer the case for QMBS commutants, as evidenced by several examples.
Other examples of extensive local operators that necessarily involve Type II operators include the $S^z_{\tot}$ for the AKLT ground state QMBS, and the AKLT Hamiltonian itself for the AKLT tower QMBS; in both cases we have not been able to find a useful expression or procedure to construct them from the strictly local generators.
Given a set of strictly local generators, it is hence extremely desirable to develop a systematic understanding of the full space of, or equivalence classes of, Type II operators that can be constructed, and even in simple examples of QMBS we numerically observe several equivalence classes that we have not been able to picture/understand in simple terms.
Results in \cite{tang2020multi, rozon2023broken, sanada2023quantum, omiya2023fractional} provide novel perspectives on potentially Type II Hamiltonians with QMBS, which would be interesting to explore further.
Developing this understanding is crucial for a truly ``exhaustive" construction of Hamiltonians with a given set of QMBS and for proving conjectures on the spectra of Hamiltonians.
Perhaps a first step is to determine a simpler set of generators for all of these QMBS local algebras, analogous to the ferromagnetic tower case, which can in principle be done by numerical checks on small system sizes, e.g., for $L = 4$.
The lack of a deeper understanding of these issues also presents a significant obstacle for using this framework to find new examples of QMBS, for which less systematic methods proposed in the earlier literature~\cite{mark2020unified, moudgalya2020large, moudgalya2020eta, mark2020eta, pakrouski2020many, odea2020from, ren2020quasisymmetry,pakrouski2021group, ren2021deformed} have been highly successful in practice.
Another matter of concern might be the lack of a formal representation theory of the algebras involved, since these no longer have simple generating group structures in them.  
Nevertheless, going beyond groups is essential for understanding some types of QMBS, such as the spin-1 AKLT models, which do not fall into previously proposed mechanisms based on the representation theory of groups~\cite{pakrouski2020many, pakrouski2021group}.
Moreover, algebras beyond-group structures also naturally appear in other physically interesting systems, such as fragmentation~\cite{moudgalya2021hilbert} and categorical symmetries~\cite{mcgreevy2022generalized}, and bond algebras of the type we are interested in have already been understood in terms of more abstract objects such as fusion categories and have representation theories that are well understood in some cases, e.g., in Temperley-Lieb spin chains~\cite{readsaleur2007, aufgebauer2010quantum}.
We hope that this work would eventually motivate the development of such a representation theory of even more general bond algebras, including those that capture quantum scars.
In the meantime, while the more standard approach to studying symmetries is to consider abstract objects such as groups or categories, construct their formal representation theory in complete generality, and then apply it to the Hilbert space of interest, we wish to advocate a different approach where we can start with a concrete Hilbert space and think of symmetries in terms of simultaneous block-diagonalization of matrices.
Since we are only interested in a physics of a given system with a given Hilbert space rather than a whole class of systems across varying Hilbert spaces, we believe the approach we suggest is more practical for physical and numerical purposes, and can be applied even if the formal representation theory or the abstract objects are not known. 
Finally, the local and commutant algebra language motivates a precise definition of QMBS.
A necessary condition for QMBS is that they should be common eigenstates of multiple non-commuting local operators, hence ruling out the unique recovery of the local parent Hamiltonian from the QMBS eigenstate, which implies the violation of the conventional form of ETH for systems with Abelian symmetries, although its status is not clear for systems with non-Abelian symmetries.
In addition, we showed that if a state can be expressed as one of the ground states of a frustration-free Hamiltonian, or, equivalently, as a part of some target space in the Shirashi-Mori formalism, then it satisfies certain sufficient conditions, phrased precisely in terms of algebras generated by the recovered local operators, to be one of the QMBS of some local Hamiltonian.
This calls for a characterization of states that can be made as QMBS of some Hamiltonian, and the answer is clearly beyond conventional tensor network states due to examples of QMBS that exhibit entanglement growth with system size. 
This definition also opens up several examples in the literature that might be examples of QMBS, e.g., the $U$-independent eigenstates in the one and higher dimensional Hubbard models~\cite{bruus1996spectrum, vafek2017entanglement, ren2018slater, ye2018exact}.
Interestingly, most examples of QMBS that we are aware of satisfy the ``Shiraishi-Mori condition," i.e., they can be expressed as the ground state(s) of a frustration-free Hamiltonian, which we showed is a sufficient condition for QMBS.
It is then natural to wonder if this is also a necessary condition for QMBS or if there exist QMBS that do not satisfy this condition.
On a different note, one might also wonder if there could be cases where any such local parent Hamiltonian necessarily has additional conserved quantities, analogous to several other examples in the literature, e.g., any local Hamiltonian with charge and dipole symmetries and with bounded range for all terms necessarily has fragmentation, i.e., exponentially many other conserved quantities~\cite{sala2020fragmentation, khemani2020localization, moudgalya2021hilbert}.
In the light of this enhanced understanding of QMBS, it would also be interesting to revisit the approximate QMBS of the PXP model. 
For example, do the approximate QMBS in the PXP model (approximately) satisfy the proposed definition of QMBS?
Indeed, compelling evidence for the satisfaction of the necessary conditions for QMBS was presented in a recent work~\cite{yao2022bounding}, which applied the correlation matrix method~\cite{qi2019determininglocal} to the PXP QMBS, and numerically found multiple local terms for which these states are approximate eigenstates.
Moreover, one might wonder if there is a point proximate to the PXP model with a larger commutant that might explain the approximate QMBS subspace in the PXP model~\cite{serbyn2020review, papic2021review, windt2022squeezing, desaules2022hypergrid, omiya2022nature}, or if the proposed deformations of the PXP model to ``integrability"~\cite{khemani2019int} or to perfect revivals~\cite{choi2018emergent, bull2020quantum} can be attributed to some strange commutants. 
Beyond Hamiltonians with exact QMBS, we can ask if exact Floquet QMBS~\cite{mizuta2020exact, sugiura2021manybody, pai2019dynamical, sengupta2021phases} can also be understood within a similar framework. 
Of course, while the local algebra allows us to straightforwardly construct Floquet unitaries with the desired QMBS, this does not exclude the existence of intrinsically Floquet QMBS that are not possible in any Hamiltonian systems. 
Beyond many-body physics itself, it would be interesting to check if some similar properties hold for quantum scars in single-particle systems~\cite{heller1984scars, berry1989quantum}, which would provide a good justification for the name ``Quantum Many-Body Scars." 
\section*{Acknowledgements}
We thank David Huse, Hosho Katsura, Igor Klebanov, Nick O'Dea, Kiryl Pakrouski, and Marcos Rigol for useful discussions.
S.M.\ particularly thanks Nicolas Regnault and B.~Andrei Bernevig for previous collaborations on related topics, while O.I.M.\ thanks particularly Daniel Mark and Cheng-Ju~Lin.
This work was supported by the Walter Burke Institute for Theoretical Physics at Caltech; the Institute for Quantum Information and Matter, an NSF Physics Frontiers Center (NSF Grant PHY-1733907); and the National Science Foundation through grant DMR-2001186.
A part of this work was done at the Aspen Center for Physics, which is supported by National Science Foundation grant PHY-1607611.
This work was partially supported by a grant from the Simons Foundation.
\bibliography{refs}
\appendix 
\onecolumngrid
\section{Proof of Existence of Exhaustive Shiraishi-Mori Bond Algebras for Scars}\label{app:SMexistence}
In this appendix, we prove that given a set of states of the form Eq.~(\ref{eq:targetdefn}) that completely span the common kernel of a set of strictly local projectors, we can always construct a Shiraishi-Mori bond algebra of the form of Eq.~(\ref{eq:SMdegbondalg}) generated by strictly local terms corresponding to these states as degenerate QMBS, satisfying Eq.~(\ref{eq:SMdegbondalg}).
For the sake of illustration, we restrict ourselves to one-dimensional systems of size $L$, but these arguments can also be extended to higher dimensions. 
\smsuf*

\begin{proof}
Before proceeding to the general case, we start with a simple example to illustrate the main idea.
\textit{Simple example:} Consider a spin-1 chain with on-site states labelled $\ket{0},\ket{+},\ket{-}$, and on-site projectors $\{P_{[j]} = \ketbra{0}_j\}$, hence $r_{\max} = 1$.
The target space is then given by $\mT = \text{span}\{ \ket{\sigma_1,\sigma_2,\dots,\sigma_L}, \sigma_j \in \{+,-\}\}$, i.e., all states annihilated by all on-site operators $\ketbra{0}_j$.
We then construct the following operators of range $r'_{\max} = 2$ that annihilate the target space: $\ketbra{0}_j \otimes h_{j+1}$ and $h_{j-1} \otimes \ketbra{0}_j$, where $h_{j+1}$ can be an arbitrary on-site operator acting on site $j+1$ (e.g., $\ketbra{\alpha}{\beta}_{j+1}$ with arbitrary $\alpha, \beta \in \{0,+,-\}$), and similarly for $h_{j-1}$ on site $j-1$; and these are the operators $\{\wh_{[j]}\}$ mentioned in the claim.
We now show that the algebra generated by these two-site operators $\tmA_{\SM} = \lgen \{ \ketbra{0}_j \otimes h_{j+1}, h_{j-1} \otimes \ketbra{0}_j \} \rgen$ (where we assume for simplicity that we have a set of $h_k$'s that can generate all on-site operators), acts irreducibly in $\mT^{\perp}$, the space spanned by basis product states with at least one on-site $\ket{0}$.
We do this by first noting that the multiplication of the generators $\ketbra{0}_j \otimes \ketbra{\alpha}{0}_{j+1}$ and $\ketbra{0}_{j+1} \otimes \ketbra{\alpha'}{0}_{j+2}$ shows that $\ketbra{0}_j \otimes \ketbra{\alpha,\alpha'}{0,0}_{j+1,j+2}$ is in the bond algebra $\tmA_{\SM}$.
Next, combining with the Hermitian conjugate of $\ketbra{0}_j \otimes \ketbra{\beta,\beta'}{0,0}_{j+1,j+2}$ gives that $\ketbra{0}_j \otimes \ketbra{\alpha,\alpha'}{\beta,\beta'}_{j+1,j+2}$ is in $\tmA_{\SM}$ with arbitrary $\alpha,\alpha',\beta,\beta' \in \{0,+,-\}$.
By repeating this procedure including generators acting on $[j-1,j]$, $[j+2,j+3]$, etc., we can generate $\ketbra{0}_j \otimes O_{\text{compl}(j)}$ where $O_{\text{compl}(j)}$ can be any operator acting on the sites other than $j$, i.e., on the complement of $j$ in the whole system.
This is true for all $j$, and combing such operators for distinct $j$ and $j'$, we can see that any two basis states that have at least one on-site $\ket{0}$ can be connected to each other by the operators in the bond algebra $\tmA_{\SM}$, which completes the proof of the claim in this example.
Incidentally, this $\tmA_{\SM}$ is the exhaustive bond algebra for a Shiraishi-Mori-type family embedding the $2^L$ states $\{ \ket{\sigma_1,\sigma_2,\dots,\sigma_L}, \sigma_j \in \{+,-\}\}$ as degenerate scars in the $3^L$-dimensional Hilbert space.
\textit{General case:} We now consider the general case where we are given a set of projectors $\mP \defn \{P_{[j]}\}$ and its target space $\mT$, and as we will see, the proof of the claim in this case is qualitatively similarly to the above spin-1 example.
We first divide the full set of projectors $\mP$ into subsets $\{\mP_\alpha\}$ such that $\mP = \bigcup_\alpha \mP_\alpha$, the supports of the projectors within each subset $\mP_\alpha$ do not overlap, and the projectors within each $\mP_\alpha$ are ``dense" on the lattice (i.e., the distance between neighboring projectors is bounded by an $L$-independent number).
For example, in a one-dimensional system with an even system size $L$ and nearest-neighbor projectors labelled as $P_{[j]} = P_{j,j+1}$, and the non-overlapping subsets are $\mP_e \defn \{P_{2k, 2k+1}\}$ and $\mP_o \defn \{P_{2k-1, 2k}\}$.
Denoting the kernel and its orthogonal complement of $\mP_\alpha$ as $\mT_\alpha$ and $\mT_{\alpha}^\perp$, the full target space $\mT$ can be expressed as $\mT = \bigcap_{\alpha}{\mT_\alpha}$. 
Focusing on a single such subset $\mP_\alpha$, we now show that a bond algebra generated by strictly local terms can be constructed such that it is irreducible in $\mT_\alpha^{\perp}$.
Consider diagonalizing $P_{[j]}$'s in $\mP_\alpha$ over the Hilbert space on its support $R_j$ on the lattice:
\begin{equation}
P_{[j]} = \sum_{\nu=D^0_{R_j} + 1}^{D_{R_j}} \sketbra{u^{(j;\nu)}}{u^{(j;\nu)}}_{R_j} ~,
\label{eq:projdiag}
\end{equation}
where $D_{R_j}$ is the corresponding Hilbert space dimension, $\{\sket{u^{(j;\nu)}}_{R_j},\;\;1 \leq \nu \leq D_{R_j}\}$ the corresponding orthonormal eigenvectors, and $D^0_{R_j}$ is the number of zero eigenvalues of $P_{[j]}$ (note that we have used the fact that all the eigenvalues of $P_{[j]}$ are either $0$ or $1$ and also labelled the zero eignevalues as $1 \leq \nu \leq D^0_{R_j}$). 
We then obtain that the projector onto any eigenvector of $P_{[j]}$ with eigenvalue $1$ annihilates all states in $\mT$, i.e., 
\begin{equation}
\sketbra{u^{(j;\nu)}}{u^{(j;\nu)}}_{R_j} \ket{\psi} = 0 \quad \text{~if~~~} \nu \geq D^0_{R_j} + 1,\;\;\ket{\psi} \in \mT~.
\label{eq:unuannih_sm}
\end{equation}
Since $P_{[j]}$ is not zero, there exists at least one such $\nu$ that satisfies Eq.~(\ref{eq:unuannih_sm}), and we assume all $\nu$ in the discussion below are such.
The projectors in Eq.~(\ref{eq:unuannih_sm}) are analogous to the projectors $\ketbra{0}_j$ in the spin-1 illustration above.
Proceeding as in that example, the operators $\sketbra{u^{(j;\nu)}}{u^{(j;\nu)}}_{R_j} \otimes O_{\text{nb}(j)}$ also annihilate the target space, where $O_{\text{nb}(j)}$ can be an arbitrary operator acting in a region neighboring but not overlapping with $R_j$ [$\text{nb}(j)$ w.r.t.\ $R_j$ is an analog of $j \pm 1$ w.r.t. $j$ in the spin-1 illustration].
We consider appropriately chosen operators of this form to be the $\{\wh_{[j]}\}$ of range $r'_{\max} \geq r_{\max}$.
For large enough $r'_{\max}$, since the distance between neighboring projectors within the subset $\mP_\alpha$ is bounded by an $L$-independent number,  the support of $O_{\text{nb}(j)}$ will cover the support of the neighboring $P_{[j']}$ in the subset (analogous to the support of $\ketbra{0}_j \otimes \ketbra{\alpha}{\beta}_{j+1}$ covering the support of $\ketbra{0}_{j+1}$ in the spin-1 example).
For example, in the one-dimensional case and focusing on $P_{[j]} \in \mP_e$, if $P_{[j=2k]} = P_{2k, 2k+1}$, $O_{\text{nb}(j)}$ can be any operator with support on sites $\{2k+2, 2k+3\}$ such that it covers the support of $P_{[j'=2k+2]} = P_{2k+2, 2k+3}$.
Similarly to the spin-1 example, it is easy to see that the bond algebra $\tmA_{\SM}$ generated using operators such as $\{ \sketbra{u^{(j;\nu)}}{u^{(j;\nu)}}_{R_j} \otimes O_{\text{nb}(j)} \}$ contains $\sketbra{u^{(j;\nu)}}{u^{(j;\nu)}}_{R_j} \otimes O_{\text{compl}(R_j)}$, where $O_{\text{compl}(R_j)}$ is an arbitrary operator in the entire lattice complement of $R_j$ which we denote as $\text{compl}(R_j)$.
Consider now combining such operators for distinct $j$ and $j'$, i.e., for disjoint regions $R_j$ and $R_{j'}$.
It is easy to see that the bond algebra $\tmA_{\SM}$ then contains the operator $\sket{u^{(j;\nu)}}_{R_j} \otimes \ket{v}_{R_{j'}} \otimes \ket{x}_{\text{compl}(R_j \cup R_{j'})} 
\bra{w}_{R_j} \otimes \sbra{u^{(j';\nu')}}_{R_{j'}} \otimes \bra{y}_{\text{compl}(R_j \cup R_{j'})}$, where $\ket{v}_{R_{j'}}$ is an arbitrary state on $R_{j'}$, $\ket{w}_{R_j}$ is an arbitrary state on $R_j$, and $\ket{x}_{\text{compl}(R_j \cup R_{j'})}$ and $\ket{y}_{\text{compl}(R_j \cup R_{j'})}$ are arbitrary states on the complement of the union of $R_j$ and $R_{j'}$.
Hence, all states of the form $\sket{u^{(j;\nu)}}_{R_j} \otimes \ket{z}_{\text{compl}(R_j)}$, with an arbitrary state $\ket{z}_{\text{compl}(R_j)}$ on the complement of $R_j$, are ``connected," i.e., the ket-bra operator formed from any pair of such states is in the bond algebra $\tmA_{\SM}$.
This shows that all states in the $\mT_\alpha^\perp$ are connected under the action of operators from the algebra $\tmA_{\SM}$. 
We also note that $\mT_\alpha$ forms an exponentially small fraction of all states in the entire Hilbert space.
Indeed, it is easy to see that for large $L$ this fraction is bounded as
\begin{equation}
\frac{\dim(\mT_\alpha)}{\dim(\mH)} = \prod_{j, P_{[j]} \in \mP_\alpha} \left(\frac{D^0_{R_j}}{D_{R_j}}\right) \leq p_\alpha^L\;\;\text{for some}\;\; p_\alpha < 1, 
\label{eq:dimTalpbound}
\end{equation}
where we have used that $D^0_{R_j}/D_{R_j} < 1$ while the number of such $j$ within the subset $\mP_\alpha$ is a finite fraction of $L$.
Finally, in order to prove that all states in  $\mT^{\perp}$ are connected, we are left to show that states from any pair of different $\mT^\perp_\alpha$ and $\mT^\perp_{\alpha'}$ are connected.
For this, it is sufficient to show that $\mT^\perp_\alpha \cap \mT^\perp_\alpha \neq \vec{0}$, since if the two subspaces share at least one common non-zero vector, then any pair of states in $\mT^\perp_\alpha$ and $\mT^\perp_{\alpha'}$ can be connected by ket-bra's via such a common vector. %
This condition translates to $\mT_\alpha + \mT_{\alpha'} \neq \mH$, which is necessarily true for large system sizes as evident from the dimension counting of $\mT_\alpha$ and $\mT_{\alpha'}$, each being an exponentially small fraction of $\mH$.
This concludes the proof that all states in $\mT^\perp$ are connected, i.e., the algebra $\tmA_{\SM}$ is irreducible in this space. 
\end{proof}
\section{Bond Algebra for the AKLT Ground State(s) as Scars}\label{app:SMalgebra}
In this appendix, we provide many results on the bond algebras $\tmA^{\AKLT}_{\scar}$ corresponding to the case with AKLT ground states as QMBS, discussed in Sec.~\ref{subsec:MPSscar}.
%
In particular, we show that these algebras act irreducibly in the space orthogonal to the four OBC AKLT ground states or the unique PBC AKLT ground state, hence any operator $\ketbra{\psi_\alpha}{\psi_\beta}$ for states $\ket{\psi_\alpha}$ and $\ket{\psi_\beta}$ orthogonal to the AKLT ground states is in the algebra.
In addition, we also show that $S_z^\tot$ is a Type II operator in the PBC $\tmA^{(\AKLT)}_{\scar}$.
In the following, we use the fact that the ground states of the AKLT are the unique states in the common kernel of the AKLT projectors $\{P^{\AKLT}_{j,j+1}\}$, and hence are the only singlets  annihilated by the $\tmA^{\AKLT}_{\scar}$. 
For PBC this is a unique ground state $\ket{G}$,  discussed in Sec.~\ref{subsec:MPSscar}, whereas for OBC there are four ground states $\ket{G_{\sigma\sigma'}}$ with $\sigma,\sigma' \in \{\uparrow, \downarrow\}$ denoting the configurations of the emergent edge spin-1/2's~\cite{aklt1987rigorous, moudgalya2018a}.
Note that everywhere below, when we say that an algebra $\mA$ annihilates some states, we always exclude the identity, and it is implicit that this exclusion is a subalgebra of $\mA$.
Throughout this section, we work with the total angular momentum states formed by two spin-1's on sites $j$ and $j+1$, labelled by the total spin $J$ and its $z$-component $m$, and we denote them by$\ket{T_{J, m}}_{j,j+1}$.
Explicit expressions for these states in the spin-1 language are given by
\begin{gather}
    \ket{T_{2,\pm 2}} \defn \ket{\pm \pm},\;\;\;\ket{T_{2,\pm 1}} \defn \frac{1}{\sqrt{2}}(\ket{\pm 0} + \ket{0 \pm}), \;\;\;\ket{T_{2,0}} \defn \frac{1}{\sqrt{6}}(\ket{+ -} + 2 \ket{0 0} + \ket{- +})\nn \\
    \ket{T_{1, \pm 1}} \defn \frac{1}{\sqrt{2}}(\ket{\pm 0} - \ket{0 \pm}),\;\;\;\ket{T_{1,0}} \defn \frac{1}{\sqrt{2}}(\ket{+ -} - \ket{- +}),\;\;\ket{T_{0,0}} \defn \frac{1}{\sqrt{3}}(\ket{+ -} - \ket{0 0} + \ket{- +}).
\label{eq:Tstates}
\end{gather}
\subsection{OBC AKLT Ground States as Scars}\label{subsec:OBCAKLTGS}
We start with the bond algebra generated by two-site terms of the form $\{ \ketbra{T_{2,m}}{T_{2,m'}}_{j,j+1}, m,m' \in \{-2,-1,0,1,2\} \}$.
We denote the algebra as 
\begin{equation}
    \mA_{1,2,\dots,L}^\text{SMobcAKLT} \defn \lgen \{\ketbra{T_{2, m}}{T_{2,m'}}_{j, j+1},\;1 \leq j \leq L - 1\}\rgen = \lgen \{P^{\AKLT}_{j,j+1} h_{j,j+1} P^{\AKLT}_{j,j+1},\;1 \leq j \leq L - 1\} \rgen,
\label{eq:2sitealgebraOBC}
\end{equation} 
where we observe the second equality numerically for $L \geq 3$ and a sufficiently generic choice of $h_{j,j+1}$.
That is, a single generator per bond is in principle sufficient for $L \geq 3$, while the analytical arguments below for simplicity assume a larger set of generators per bond listed in the first expression.\footnote{Note that to establish the second expression, it is enough to check the equality for $L=3$, since this can then be applied on any three consecutive sites to show that the algebras are equal for all system sizes.}
There are precisely four states annihilated by this algebra $\mA_{1,2,\dots,L}^\text{SMobcAKLT}$, namely the four OBC AKLT ground states $\{\ket{G_{\sigma\sigma'}}_{1, 2, \dots, L},\; \sigma,\sigma' \in \{\uparrow, \downarrow\}\}$, and this follows from the fact that these states span the common kernel of the projectors $\{P^{\AKLT}_{j,j+1}\}$, which is a known fact for any $L$.
(Here and below, when we say ``annihilated by the algebra $\mA_{1,2,\dots,L}^\text{SMobcAKLT}$'' we mean annihilated by the above generators, i.e., excluding the identity.)
Denoting the on-site spin-1 states as $\ket{+}$, $\ket{0}$, and $\ket{-}$, using $\ket{T_{2,2}}_{j,j+1} = \ket{+,+}_{j,j+1}$ it is easy to see that the projector onto the ferromagnetic state $\ketbra{+,+, \dots,+}_{1,2,\dots,L}$ belongs to $\mA_{1,2,\dots,L}^\text{SMobcAKLT}$.
We then prove the following Lemma.
\begin{lemma}\label{lem:OBCAKLT}
For any $L$, the algebra $\mA_{1,2,\dots,L}^\text{SMobcAKLT}$ acts irreducibly in the orthogonal complement to the four OBC AKLT ground states $\{\ket{G_{\sigma\sigma'}}_{1, 2, \dots, L}\}$.
Denoting an orthonormal basis in this space as $\{ \ket{\psi_\alpha}, \alpha=1,\dots, 3^L-4 \}$, this is equivalent to the statement that 
$\ketbra{\psi_\alpha}{+,+,\dots,+}_{1,2,\dots,L} \in \mA_{1,2,\dots,L}^\text{SMobcAKLT}$ for all $\alpha$, since all operators of the form $\ketbra{\psi_\alpha}{\psi_\beta}$ can be generated from these and their Hermitean conjugates.
%
\end{lemma}
\begin{proof}
We proceed by induction, assuming this holds for some $L = k$, we show that it is true for $L = k + 1$. 
We begin the induction from $L=2$, where the span of the four AKLT ground states $\{\ket{G_{\sigma\sigma'}}_{1, 2}\}$ is simply the span of $\{ \ket{T_{1,-1/0/1}}_{1,2}, \ket{T_{0,0}}_{1,2} \}$.
Hence by definition in Eq.~(\ref{eq:2sitealgebraOBC}), $\mA_{1,2}^\text{SMobcAKLT}$ acts irreducibly in the space spanned by $\{ \ket{T_{2,m}}_{1,2}, m \in -2,-1,0,1,2 \}$ orthogonal to $\{\ket{G_{\sigma\sigma'}}_{1, 2}\}$.
For induction, we assume the claim holds for $L = k$, or $k$ consecutive sites in general.
This then implies that the irreducibility holds for the algebras $\mA_{1,2,\dots,k}^\text{SMobcAKLT}$ and $\mA_{2,3,\dots,k+1}^\text{SMobcAKLT}$.
Given $\left(\ketbra{\psi_\alpha}{+,+,\dots,+} \right)_{1,2,\dots,k} \in \mA_{1,2,\dots,k}^\text{SMobcAKLT}$ and similar ketbras in $\mA_{2,3,\dots,k+1}^\text{SMobcAKLT}$,
we can then combine with appropriate $\ketbra{T_{2,m}}{T_{2,2}}, m=2,1,0$, near the ends to obtain the following ketbras from $\mA_{1,2,\dots,k,k+1}^\text{SMobcAKLT}$:
\begin{align*}
& \left(\ketbra{\psi_\alpha}{+,+,\dots,+} \right)_{1,2,\dots,k} \left(\ketbra{+,+}{+,+} \right)_{k,k+1} = (\ket{\psi_\alpha}_{1,2,\dots,k} \otimes \ket{+}_{k+1}) \bra{+,+,\dots,+,+}_{1,2,\dots,k,k+1} ~; \\
& \left(\ketbra{\psi_\alpha}{+,+,\dots,+} \right)_{1,2,\dots,k} \left(\big[\ket{+,0} + \ket{0,+} \big] \bra{+,+} \right)_{k,k+1} = (\ket{\psi_\alpha}_{1,2,\dots,k} \otimes \ket{0}_{k+1}) \bra{+,+,\dots,+,+}_{1,2,\dots,k,k+1} ~; \\
& \left(\ketbra{\psi_\alpha}{+,+,\dots,+} \right)_{1,2,\dots,k} \left(\big[\ket{+,-} + 2\ket{0,0} + \ket{-,+} \big] \bra{+,+} \right)_{k,k+1} = (\ket{\psi_\alpha}_{1,2,\dots,k} \otimes \ket{-}_{k+1}) \bra{+,+,\dots,+,+}_{1,2,\dots,k,k+1} ~; \\
& \left(\ketbra{\psi_\alpha}{+,+,\dots,+} \right)_{2,3,\dots,k+1} \left(\ketbra{+,+} \right)_{1,2} = (\ket{+}_1 \otimes \ket{\psi_\alpha}_{2,3\dots,k+1}) \bra{+,+,\dots,+,+}_{1,2,\dots,k,k+1} ~; \\
& \left(\ketbra{\psi_\alpha}{+,+,\dots,+} \right)_{2,3,\dots,k+1} \left(\big[\ket{+,0} + \ket{0,+} \big] \bra{+,+} \right)_{1,2} = (\ket{0}_1 \otimes \ket{\psi_\alpha}_{2,3,\dots,k+1} ) \bra{+,+,\dots,+,+}_{1,2,\dots,k,k+1} ~; \\
& \left(\ketbra{\psi_\alpha}{+,+,\dots,+} \right)_{2,3,\dots,k+1} \left(\big[\ket{+,-} + 2\ket{0,0} + \ket{-,+} \big] \bra{+,+} \right)_{1,2} = (\ket{-}_1 \otimes \ket{\psi_\alpha}_{2,3,\dots,k+1} ) \bra{+,+,\dots,+,+}_{1,2,\dots,k,k+1} ~, 
\end{align*}
where we have used the fact that $\mA^{\text{SMobcAKLT}}_{1, \dots, k+1}$ is generated by the algebras $\mA^{\text{SMobcAKLT}}_{1, \dots, k}$ and $\mA^{\text{SMobcAKLT}}_{k, k+1}$ or $\mA^{\text{SMobcAKLT}}_{2, \dots, k+1}$ and $\mA^{\text{SMobcAKLT}}_{1, 2}$. 
We can now show that kets appearing on the R.H.S., copied below,
\begin{align}
\ket{\psi_\alpha}_{1,2,\dots,k} \otimes \ket{+}_{k+1} ~, \quad
\ket{\psi_\alpha}_{1,2,\dots,k} \otimes \ket{0}_{k+1} ~, \quad
\ket{\psi_\alpha}_{1,2,\dots,k} \otimes \ket{-}_{k+1} ~, \label{eq:append_right} \\
\ket{+}_1 \otimes \ket{\psi_\alpha}_{2,3\dots,k+1} ~, \quad
\ket{0}_1 \otimes \ket{\psi_\alpha}_{2,3\dots,k+1} ~, \quad
\ket{-}_1 \otimes \ket{\psi_\alpha}_{2,3\dots,k+1} ~, \label{eq:append_left}
\end{align}
not all linearly independent, span orthogonal complement to the four AKLT ground states on sites $1,2,\dots,k,k+1$.
To do so, it is sufficient to show that any state orthogonal to the span of the above states is a part of the AKLT ground state manifold, i.e., is annihilated by the algebra $\mA_{1,2,\dots,k,k+1}^{\text{SMobcAKLT}}$.
Consider any $\ket{\phi}_{1,2,\dots,k,k+1}$ orthogonal to the above states, and decompose it has
\begin{align}
\ket{\phi}_{1,2,\dots,k,k+1} = \ket{u_+}_{1,2,\dots,k} \otimes \ket{+}_{k+1} + \ket{u_0}_{1,2,\dots,k} \otimes \ket{0}_{k+1} + \ket{u_{-}}_{1,2,\dots,k} \otimes \ket{-}_{k+1} ~. 
\end{align}
Requiring orthogonality to the states in Eq.~(\ref{eq:append_right}), we conclude that $\ket{u_{+/0/-}}_{1,2,\dots,k}$ are orthogonal to all $\ket{\psi_{\alpha}}_{1,2,\dots,k}$ and hence $\ket{\phi}_{1,2,\dots,k,k+1}$ is annihilated by $\mathcal{A}_{1,2,\dots,k}^\text{SMobcAKLT}$.
By an identical argument using orthogonality to the states in Eq.~(\ref{eq:append_left}), we conclude that $\ket{\phi}_{1,2,\dots,k,k+1}$ is annihilated by $\mA_{2,3,\dots,k+1}^\text{SMobcAKLT}$.
Since $\mA_{1,2,\dots,k,k+1}^\text{SMobcAKLT}$ is completely generated by $\mA_{1,2,\dots,k}^\text{SMobcAKLT}$ and $\mA_{2,3,\dots,k+1}^\text{SMobcAKLT}$, we conclude that $\ket{\phi}_{1,2,\dots,k,k+1}$ is annihilated by $\mA_{1,2,\dots,k,k+1}^\text{SMobcAKLT}$ and hence must be in the span of the four OBC AKLT ground states $\{\ket{G_{\sigma\sigma'}}_{1, 2, \dots, k+1}\}$.
This proves that the states in  Eqs.~(\ref{eq:append_right})-(\ref{eq:append_left}) indeed span the orthogonal complement to the four OBC AKLT ground states. 
Combining the arguments, this proves the claim for $L = k + 1$, completing the induction and hence proving the claim for all $L$.
\end{proof}
\subsection{PBC AKLT Ground State as a Scar}
We now extend the proof to show irreducibility of $\tmA^{\AKLT}_{\scar}$ for the PBC AKLT ground state.
We consider the algebra
\begin{equation}
    \mA_{1,2,\dots,L}^\text{SMpbcAKLT} \defn \lgen \{\ketbra{T_{2, m}}{T_{2,m'}}_{j, j+1},\;\;1 \leq j \leq L\}\rgen = \lgen \{P^{\AKLT}_{j,j+1} h_{j,j+1} P^{\AKLT}_{j,j+1},\;\;1\leq j \leq L\} \rgen,
\label{eq:2sitealgebra}
\end{equation}
where the subscripts are modulo $L$ and we observe the second equality numerically for $L \geq 3$ and a sufficiently generic choice of $h_{j,j+1}$.
There is precisely one state annihilated by (the above generators of) this algebra $\mA_{1,2,\dots,k}^\text{SMpbcAKLT}$, namely the unique PBC AKLT ground state $\ket{G}_{1,2,\dots,L}$, and this follows from the fact that it is the unique state in the common kernel of the projectors $\{P^{\AKLT}_{j,j+1},\;\; 1 \leq j \leq L\}$, which is a known fact for any $L$.
In the following, we use the fact that the algebra $\mA_{1,2,\dots,L}^\text{SMpbcAKLT}$ is generated by the OBC algebras $\mA_{1,2,\dots,L-1,L}^\text{SMobcAKLT}$ and $\mA_{2,3,\dots,L, 1}^\text{SMobcAKLT}$.
\begin{lemma}
For any $L$, the algebra $\mA_{1,2,\dots,L}^\text{SMpbcAKLT}$ acts irreducibly in the orthogonal complement to the PBC AKLT ground state $\ket{G}_{1, 2, \dots, L}$.
Denoting an orthonormal basis in this space as $\{ \ket{\psi_\alpha}, \alpha=1,\dots, 3^L-1\}$, this is equivalent to the statement that $\ketbra{\psi_\alpha}{+,+,\dots,+}_{1,2,\dots,L} \in \mA_{1,2,\dots,L}^\text{SMpbcAKLT}$ for all $\alpha$.
\end{lemma}
\begin{proof}
Consider the sites $1,2,\dots,L-1,L$ as an OBC system, the corresponding bond algebra is $\mA_{1,2,\dots,L-1,L}^\text{SMobcAKLT}$ and $\{\ket{\psi_\alpha}_{1,2,\dots,L-1,L}, \alpha=1,\dots,3^{L-4}\}$ the orthormal basis that spans the orthogonal complement of the four OBC AKLT states $\ket{G_{\sigma\sigma'}}_{1, \dots, L}$ on these $L$ sites. 
Next, we consider the sites $2,3,\dots,L,1$ as an OBC system and do the same; schematically we denote this algebra as
$\mA_{2,3,\dots,L,1}^\text{SMobcAKLT}$ and the basis as $\{\ket{\psi_\alpha}_{2,3,\dots,L,1}\},$ which are just some new states $\{ \ket{\psi_\alpha^\prime}_{1,2,\dots,L-1,L}\}$.
Following the proof of Lem.~\ref{lem:OBCAKLT}, we can show that $\ketbra{\psi_\alpha}{+, +, \dots, +,+}_{1, 2, \dots, L-1, L}$ and $\ketbra{\psi_\alpha}{+, +, \dots, +,+}_{2, 3, \dots, L, 1}$ are in the algebras $\mA^{\text{SMobcAKLT}}_{1, 2, \dots, L-1, L}$ and $\mA^{\text{SMobcAKLT}}_{2, 3, \dots, L, 1}$, and hence are in the algebra $\mA^{\text{SMpbcAKLT}}_{1, 2, \dots, L}$.  
Further note that the ``bra" states that appear in these cases are actually the same, i.e.,  $\bra{+,+,\dots,+,+}_{1,2,\dots,L-1,L} = \bra{+,+,\dots,+,+}_{2,3,\dots,L,1}$, hence the full bond algebra contains the operators $\ketbra{\psi_\alpha}{+,+,\dots,+,+}_{1,2,\dots,L-1,L}$ and $\ketbra{\psi_\alpha^\prime}{+,+,\dots,+,+}_{1,2,\dots,L-1,L}$.
We now show that the states \{$\ket{\psi_\alpha}_{1,2,\dots,L-1,L}$, $\ket{\psi_\alpha^\prime}_{1,2,\dots,L-1,L}\}$ span the orthogonal complement to the AKLT ground state $\ket{G}$.
We show this in a procedure similar to the OBC case, i.e., that any state orthogonal to these states must be a part of the ground state manifold, which in the PBC case is simply $\ket{G}$.
Indeed, consider any $\ket{\phi}_{1,2,\dots,L-1,L}$ orthogonal to both $\ket{\psi_\alpha}_{1,2,\dots,L-1,L}$ and $\ket{\psi_\alpha^\prime}_{1,2,\dots,L-1,L}$.
By the applications of the results in the OBC case, $\ket{\phi}_{1, 2, \dots, L-1, L}$ must be annihilated by both $\mA_{1,2,\dots,L-1,L}^\text{SMobcAKLT}$ and 
$\mA_{2,3,\dots,L,1}^\text{SMobcAKLT}$, hence it must be annihilated by the PBC bond algebra $\mA^{\text{SMpbcAKLT}}_{1, 2, \dots, L}$. 
This completes the proof of  irreducibility for the PBC case.
\end{proof}
\subsection{Impossibility of writing \texorpdfstring{$\boldsymbol{S^z_{\tot}}$}{} as a sum of strictly local symmetric terms}\label{subsec:Sztotimpossible}
We now use some known properties of the AKLT ground states to prove that in the bond algebra $\tmA^{\AKLT}_{\scar}$,  $S^z_{\tot}$ is a Type II symmetric operator as defined in Sec.~\ref{subsec:symhamtypes}.
This follows from the following Lemma, and the discussion in Sec.~\ref{subsubsec:DCTAKLT} that due to the DCT, $S^z_{\tot}$ is in the algebra $\tmA^{\AKLT}_{\scar}$.
\begin{lemma}
In the spin-1 chain, the operator $S^z_{\tot} = \sum_j{S^z_j}$ cannot be written as a sum of strictly local terms of range bounded by a finite $r_{\max}$ from the PBC bond algebra $\tmA^{\AKLT}_{\scar}$ of Eq.~(\ref{eq:AKLTSMpair}). 
\end{lemma}
\begin{proof}
We start with an explicit proof by contradiction for $r_{\max}=2$.
Indeed, let us assume that we can write
\begin{equation}
S^z_{\tot} = c \mathds{1} + \sum_{j=1}^L a_{j;m,m'} \ketbra{T_{2,m}}{T_{2,m'}}_{j,j+1} ~,
\label{eq:Sztot_presumed_rmax2}
\end{equation}
where $\ket{T_{2,m}}$'s are total angular momentum states on two spin-1's as defined in Eq.~(\ref{eq:Tstates}), and we have used that ket-bra operators formed from them are the most general two-site operators that annihilate $\ket{G}$.
We can immediately set $c = 0$ since we know that $S^z_{\tot} \ket{G} = 0$.
We also know that $S^z_{\tot} \ket{G_{\uparrow\uparrow}} = \ket{G_{\uparrow\uparrow}}$, where $\ket{G_{\uparrow\uparrow}}$ is one of the OBC AKLT ground states.
We use the action of $S^z_{\tot}$ on these two states to arrive at a contradiction.
We first recall the well-known Matrix Product State (MPS) structures of the OBC and PBC AKLT ground states $\ket{G_{\sigma\sigma'}}$ and $\ket{G}$, which read~\cite{schollwock2011density, moudgalya2018b}
\begin{align}
    &\ket{G_{\sigma\sigma'}} = \sumal{\{m_j\}_{j = 1}^L}{}{[A^{[m_1]} A^{[m_2]} \cdots A^{[m_L]}]_{\sigma\sigma'} \ket{m_1 m_2 \cdots m_L}},\;\;\;\sigma, \sigma' \in \{\uparrow, \downarrow\}, \nn\\
    &\ket{G} = \sumal{\{m_j\}_{j = 1}^L}{}{\text{Tr}[A^{[m_1]} A^{[m_2]} \cdots A^{[m_L]}] \ket{m_1 m_2 \cdots m_L}}, \label{eq:GMPS}
\end{align}
where $m_j \in \{+, 0, -\}$ labels the physical indices on site $j$, $\sigma$ and $\sigma'$ are auxiliary indices that label the emergent boundary spins of the OBC AKLT state, and $\{A^{[m_j]}\}$ are matrices over the two-dimensional auxiliary space, and they read
\begin{equation}
    A^{[+]} = \sqrt{\frac{2}{3}}\begin{pmatrix} 0 & 1 \\
    0 & 0\end{pmatrix},\;\;A^{[0]} = -\sqrt{\frac{1}{3}}\begin{pmatrix} 1& 0\\
    0 & -1 \end{pmatrix},\;\;A^{[-]} = - \sqrt{\frac{2}{3}}\begin{pmatrix}
    0 & 0 \\
    1 & 0\end{pmatrix}.
\label{eq:AKLTMPSdetails}
\end{equation}
Due to the structure of Eq.~(\ref{eq:GMPS}), it is easy to see that the condition that the two-site term $\ketbra{T_{2,m}}{T_{2,m'}}_{j,j+1}$ vanishes on $\ket{G}$ for $1 \leq j \leq L$ implies that it also vanishes on $\ket{G_{\uparrow\uparrow}}$ for $1 \leq j \leq L - 1$.
Hence, all terms in Eq.~(\ref{eq:Sztot_presumed_rmax2}) with $1 \leq j \leq L -1$ --- i.e., that do not go across the ``PBC connection" $[L,1]$ --- annihilate the OBC state $\ket{G_{\uparrow\uparrow}}$, and the assumed Eq.~(\ref{eq:Sztot_presumed_rmax2}) implies that
\begin{equation}
Z_R \defn \sumal{m,m'}{}{a_{L;m,m'}\ketbra{T_{2,m}}{T_{2,m'}}_{L,1}}\;\;\implies\;\;Z_R \ket{G_{\uparrow\uparrow}} = \ket{G_{\uparrow\uparrow}},
\label{eq:Sztot_presumed_red}
\end{equation}
where we define a ``straddling region" $R \defn \{L, 1\}$. 
In addition, the definition of $Z_R$ shows that it vanishes on $\ket{G}$, i.e.,  $Z_R \ket{G} = 0$. 
We can then use these conditions and the MPS structures of  $\ket{G_{\uparrow\uparrow}}$ and $\ket{G}$ in Eq.~(\ref{eq:GMPS}) to express conditions on the action of $Z_R$ on states over the region $R$ as follows:
\begin{align}
    &\ket{G_{\uparrow\uparrow}} = \sumal{\sigma,\sigma' \in \{\uparrow,  \downarrow\}}{}{\sket{G^{R}_{\uparrow\uparrow, \sigma\sigma'}} \otimes \sket{G^{\bar{R}}_{\sigma'\sigma}}}\;\;\implies\;\;Z_{R} \sket{G^{R}_{\uparrow\uparrow,\sigma\sigma'}} = \sket{G^{R}_{\uparrow\uparrow,\sigma\sigma'}}\;\;\forall\;\sigma,\sigma' \in \{\uparrow,\downarrow\},
\nn\\
    &\ket{G} = \sumal{\sigma, \sigma' \in \{\uparrow, \downarrow\}}{}{\sket{G^{R}_{\sigma\sigma'}} \otimes \sket{G^{\bar{R}}_{\sigma'\sigma}}}\;\;\implies\;\;Z_R\sket{G^{R}_{\sigma\sigma'}} = 0\;\;\;\forall\;\sigma,\sigma' \in \{\uparrow,\downarrow\}, \label{eq:Gschmidt}
\end{align}
where $\{\sket{G^{R}_{\uparrow\uparrow, \sigma\sigma'}}\}$ and $\{\sket{G^R_{\sigma\sigma'}}\}$ are the vectors of the MPS $\ket{G_{\uparrow\uparrow}}$ and $\ket{G}$ over region $R$ with auxiliary indices $\sigma$ and $\sigma'$, and their forms can be directly deduced from Eq.~(\ref{eq:GMPS}):
\begin{equation}
    \sket{G^R_{\uparrow\uparrow, \sigma\sigma'}} = \sumal{\{m_L, m_1\}}{}{[A^{[m_L]}]_{\sigma\uparrow} [A^{[m_1]}]_{\uparrow\sigma'} \ket{m_L m_1}},\;\;\;\ket{G^{R}_{\sigma\sigma'}} = \sumal{\{m_L, m_1\}}{}{[A^{[m_L]} A^{[m_1]}]_{\sigma\sigma'} \ket{m_L m_1}}.
\label{eq:Schmidtvec}
\end{equation}
Here $\sket{G^R_{\sigma\sigma'}}$ are the standard OBC AKLT ground states over $R=\{L,1\}$; on the other hand, $\sket{G^R_{\uparrow\uparrow, \sigma\sigma'}}$ are the specific extractions from the $\sket{G_{\uparrow\uparrow}}$.
Also, in the preceding Eq.~(\ref{eq:Gschmidt}) $\sket{G^{\bar{R}}_{\sigma'\sigma}}$ are the standard OBC AKLT ground states over $\bar{R}=\{2,3,\dots,L-2,L-1\}$, specializing Eq.~(\ref{eq:GMPS}) to such region.
Note that these ``parts" over $R$ and $\bar{R}$ are \textit{not} the same as the Schmidt vectors of the respective wavefunctions over the region $R$, since they are not guaranteed to be orthogonal, but nevertheless Eq.~(\ref{eq:Gschmidt}) holds due to the linear independence of $\{\sket{G^{\bar{R}}_{\sigma'\sigma}}\}$.
Using Eq.~(\ref{eq:Gschmidt}), we can then arrive at
\begin{equation}
    \braket{G^R_{\sigma\sigma'}}{G^R_{\uparrow\uparrow,\tau\tau'}} = \bra{G^R_{\sigma\sigma'}} Z_R \ket{G^R_{\uparrow\uparrow,\tau\tau'}} = 0\;\;\;\forall \sigma,\sigma',\tau,\tau' \in \{\uparrow, \downarrow\}.
\label{eq:GRorth}
\end{equation}
Eq.~(\ref{eq:GRorth}) is then a contradiction if (at least one element of) the overlap ``matrix" on the L.H.S. is non-zero.
This matrix can straightforwardly be computed for $r_{\max}=2$, where the $\sket{G^R_{\sigma\sigma'}}$ are in the span of $\sket{T_{1,m \in \{+,0,-\}}}$ and $\sket{T_{0,0}}$ on the two sites $R=\{L,1\}$; it is also straightforward to extract $\sket{G^R_{\uparrow\uparrow,\sigma\sigma'}}$, e.g., $\sket{G^R_{\uparrow\uparrow,\uparrow\uparrow}} = (1/3) \ket{00}_{L,1}$, and this clearly has non-zero overlap with $\sket{T_{0,0}}_{L,1}$.
This is contradiction, and hence proves the invalidity of the assumption Eq.~(\ref{eq:Sztot_presumed_rmax2}).
We can generalize this proof to arbitrary fixed $r_{\max}$ as follows.
Suppose we can write
\begin{equation}
S^z_{\tot} = c \mathds{1} + \sum_{j=1}^L \sum_\alpha a_{j;\alpha} T^{(r_{\max})}_{j;\alpha}
\label{eq:Sztot_presumed_rmaxgen}
\end{equation}
where $T^{(r_{\max})}_{j;\alpha} \in \tmA^{\AKLT}_{\scar}$ is a strictly local annihilator of $\ket{G}$ with support contained in $[j,j+r_{\max}-1]$ (modulo $L$).
Again, by acting with this on $\ket{G}$ we conclude $c=0$, and we apply this presumed expression to the OBC ground state $\ket{G_{\uparrow\uparrow}}$ with $S^z_{\tot} = 1$.
As a consequence of the MPS structure of Eq.~(\ref{eq:GMPS}), the terms $\{T^{(r_{\max})}_{j,\alpha},\;\;1 \leq j \leq L - r_{\max} + 1\}$ that do not straddle the PBC link $[L,1]$ annihilate $\ket{G_{\uparrow\uparrow}}$.
Hence analogous to Eq.~(\ref{eq:Sztot_presumed_red}), we obtain (assuming $L \gg r_{\max}$)
\begin{equation}
Z_R \defn \sum_{j=L-r_{\max}+2}^L \sum_\alpha a_{j;\alpha} T^{(r_{\max})}_{j;\alpha}\;\;\implies\;\;Z_R \ket{G_{\uparrow\uparrow}} = \ket{G_{\uparrow\uparrow}} ~,
\label{eq:Sztot_presumed_rmax}
\end{equation}
where we define a ``straddling region" $R \defn \{L - r_{\max} + 2, \dots, L, 1, \cdots, r_{\max} - 1\}$.
Equation~(\ref{eq:Gschmidt}) then follows directly for this new region $R$, and analogous to Eq.~(\ref{eq:Schmidtvec}) the corresponding vectors on the region $R$ are defined as
\begin{align}
    &\sket{G^R_{\uparrow\uparrow, \sigma\sigma'}} = \sumal{\{m_j,\;j \in R \}}{}{[A^{[m_{L - r_{\max} + 2}]} \cdots A^{[m_L]}]_{\sigma\uparrow} [A^{[m_1]} \cdots A^{r_{\max} - 1}]_{\uparrow\sigma'} \ket{m_{L - r_{\max} + 2}\cdots m_L m_1 \cdots m_{r_{\max} - 1}}}, \nn\\
    &\ket{G_{\sigma\sigma'}} = \sumal{\{m_j,\;j \in R\}}{}{[A^{[m_{L - r_{\max} + 2}]}\cdots A^{[m_L]} A^{[m_1]} \cdots A^{r_{\max} - 1}]_{\sigma\sigma'} \ket{m_{L - r_{\max} + 2}\cdots m_L m_1 \cdots m_{r_{\max} - 1}}}.
\label{eq:SchmidtvecR}
\end{align}
Equation ~(\ref{eq:GRorth}) then follows and can be shown to be a contradiction for any finite $r_{\max}$ for $L \gg r_{\max}$ using straightforward AKLT transfer matrix calculations.
This hence shows the invalidity of the assumption of Eq.~(\ref{eq:Sztot_presumed_rmaxgen}), hence completing the proof.
\end{proof}
\section{Bond Algebra for the Spin-1/2 Ferromagnetic Tower as Scars}
In this appendix, we provide details on the bond algebra corresponding to the spin-1/2 ferromagnetic tower of states $\{\ket{\Psi_n}\}$ of Eq.~(\ref{eq:FMtower}) as scars, discussed in Sec.~\ref{subsec:FMtowerscar}. 
As discussed in Sec.~\ref{subsec:FMtowerscar}, ideas from the Shiraishi-Mori formalism can be used to construct the algebra of Hamiltonians that leave the ferromagnetic states degenerate, and they are of the form shown in Eq.~(\ref{eq:FMSMalgebra}).
Further, we found numerical evidence that generators with support on three or more sites, e.g., those of the form of Eq.~(\ref{eq:FMdegalgebra}), are necessary to obtain the desired commutant $\tmC^{\FM}_{\scar}$.
In order to show the irreducibility of this bond algebra $\tmA^{\FM}_{\scar}$ in the orthogonal complement of the scars $\{\ket{\Psi_n}\}$, it is convenient to use a simpler expression involving multiple generators per each three-segment, using two-site projectors $\{P_{j,j+1} = \ketbra{\mS}_{j,j+1}\}$ that annihilate the ferromagnetic states:
\begin{equation}
    \tmA^{\FM}_{\scar} \defn \lgen \{\ketbra{\mS}{\mS}_{j,j+1} \otimes \ketbra{\sigma}{\sigma'}_{j+2}\} ~, \quad
    \{\ketbra{\sigma}{\sigma'}_j \otimes \ketbra{\mS}{\mS}_{j+1,j+2}\}\rgen,
\label{eq:FMtowernaive}
\end{equation}
where $j=1,2,\dots,L-2$ for $L$ sites, $\ket{\mS}_{k,\ell} \defn \frac{1}{\sqrt{2}} (\ket{\uparrow,\downarrow} - \ket{\downarrow,\uparrow})_{k,\ell}$ is a spin-singlet on sites $k$ and $\ell$, and $\sigma,\sigma' \in \{\uparrow, \downarrow\}$.
To verify the equivalence of the expressions of Eq.~(\ref{eq:FMtowernaive}) and those of Eq.~(\ref{eq:FMdegalgebra})
, it is sufficient to check that for $L = 4$ these generators give the same algebra, which is indeed the case from numerical experiments for sufficiently generic $h_{j-1}$ and $h_{j+2}$ in Eq.~(\ref{eq:FMdegalgebra}).
\subsection{Proof of irreducibility}\label{subsec:FMirreducible}
We now show that $\tmC^{\FM}_{\scar}$ is the full commutant of the bond algebra $\tmA^{\FM}_{\scar}$ of Eq.~(\ref{eq:FMtowernaive}) by proving the following Lemma.
\begin{lemma}
Denoting the bond algebra $\tmA^{\FM}_{\scar}$ of Eq.~(\ref{eq:FMtowernaive}) on $L$ site system $1,2,\dots,L$ as $\mA_{1,2,\dots,L}$, for any $L$ there is a set of states $W_L = \{\ket{\psi_\alpha}_{1,2,\dots,L} \}$ (not required to be orthonormalized) such that $\mA_{1,2,\dots,L} = \lgen\{ \ketbra{\psi_\alpha}{\psi_{\alpha'}}_{1,2,\dots,L}, ~ \ket{\psi_{\alpha}}, \ket{\psi_{\alpha'}} \in W_L \}\rgen$ and any state orthogonal to $W_L$ is annihilated by $\mA_{1,2,\dots,L}$.
\end{lemma}

\begin{proof}
The set $W_L$ is constructed such that it also contains the specific states $\{\ket{\uparrow,\dots,\uparrow}_{1,\dots,j-1} \otimes \ket{\mS}_{j,j+1} \otimes \ket{\uparrow,\dots,\uparrow}_{j+2,\dots,L}$, $j=1,\dots,L-1\}$.
This is enough to conclude that $\mA_{1,2,\dots, L}$ acts irreducibly in the orthogonal complement to the ferromagnetic multiplet $\{(S^-_{\tot})^n\ket{F}\}$, $\ket{F} = \ket{\uparrow \cdots \uparrow}_{1, 2, \dots, L}$, which is precisely the set of states annihilated by this algebra (which is a well-known fact for the important projectors involved, namely $\{ \ketbra{\mS}_{j,j+1}, j=1,\dots,L-1 \}$).
We proceed inductively similar to the AKLT case discussed in App.~\ref{app:SMalgebra}.
For $L = 3$, we take $W_3 = \{ \ket{\mS}_{1,2} \otimes \ket{\uparrow}_3, 
\ket{\uparrow}_1 \otimes \ket{\mS}_{2,3},
\ket{\mS}_{1,2} \otimes \ket{\downarrow}_3, 
\ket{\downarrow}_1 \otimes \ket{\mS}_{2,3} \}$, which are linearly independent and span the space orthogonal to the four ferromagnetic states on the three sites.
It is easy to verify that the corresponding $\ketbra{\psi_\alpha}{\psi_{\alpha'}}$ indeed span $\mA_{1,2,3}$.
Proceeding by induction, we suppose the claim is true for $L = k$ sites.
We can take any $\ket{\psi_\alpha}_{1,2,\dots,k} \in W_k$ and a convenient fixed $\ket{\uparrow,\dots,\uparrow}_{1,\dots,k-2} \otimes \ket{\mS}_{k-1,k} \in W_k$ (here using the requirement that the $W_k$ contains the specific states), so that $\ket{\psi_\alpha}_{1,2,\dots,k} \bra{\uparrow,\dots,\uparrow}_{1,\dots,k-2} \otimes \bra{\mS}_{k-1,k} \in \mA_{1,2,\dots,k}$, and then combine with the generators of $\tmA^{\FM}_{\scar}$ acting on sites $k-1,k,k+1$ to construct operators that belong to $\mA_{1,2,\dots,k,k+1}$:
\begin{equation}
\ket{\psi_\alpha}_{1,2,\dots,k} \bra{\uparrow,\dots,\uparrow}_{1,\dots,k-2} \otimes \bra{\mS}_{k-1,k} \Big(\ketbra{\mS}{\mS}_{k-1,k} \otimes \ketbra{\sigma}{\up}_{k+1} \Big) = \ket{\psi_\alpha}_{1,2,\dots,k} \otimes \ket{\sigma}_{k+1}  \bra{\uparrow,\dots,\uparrow}_{1,\dots,k-2} \otimes \bra{\mS}_{k-1,k} \otimes \bra{\up}_{k+1} ~. 
\end{equation}
Hence, defining $(k+1)$-site states $\sket{\tilde{\psi}_{\alpha;\sigma}}_{1,2,\dots,k,k+1} \defn \ket{\psi_\alpha}_{1,2,\dots,k} \otimes \ket{\sigma}_{k+1}$ we have that all $(\sketbra{\tilde{\psi}_{\alpha;\sigma}}{\tilde{\psi}_{\alpha';\sigma'}})_{1,2,\dots,k,k+1} \in \mA_{1,2,\dots,k,k+1}$.
We can repeat the same exercise with the $k$-site bond algebra $\mA_{2,\dots,k,k+1}$ containing $\ket{\psi_\alpha}_{2,\dots,k,k+1} \bra{\mS}_{2,3} \otimes \bra{\uparrow,\dots,\uparrow}_{4,\dots,k+1}$, multiplying on the right by $\ketbra{\sigma}{\up}_1 \otimes \ketbra{\mS}{\mS}_{2,3}$ to obtain elements of $\mA_{1,2,\dots,k,k+1}$, leading us to define $(k+1)$-site states $\sket{\doubletilde{\psi}_{\sigma;\alpha}}_{1,2,\dots,k,k+1} \defn \ket{\sigma}_1 \otimes \ket{\psi_\alpha}_{2,\dots,k,k+1}$ to conclude that all $(\sketbra{\doubletilde{\psi}_{\sigma;\alpha}}{\doubletilde{\psi}_{\sigma';\alpha'}})_{1,2,\dots,k,k+1} \in \mA_{1,2,\dots,k,k+1}$.
We now define
\begin{equation}
    W_{k+1} \defn \{ \sket{\tilde{\psi}_{\alpha;\sigma}}_{1,2,\dots,k,k+1} \} \cup
    \{ \sket{\doubletilde{\psi}_{\sigma;\alpha}}_{1,2,\dots,k,k+1} \} ~.
\end{equation}
Note that the two sets being joined here happen to have common states, e.g., $\ket{\uparrow,\dots,\uparrow}_{1,\dots,j-1} \otimes \ket{\mS}_{j,j+1} \otimes \ket{\uparrow,\dots,\uparrow}_{j+2,\dots,k+1}$, $j=2,\dots,k-1$, i.e., with the singlet flanked by some non-zero number of $\uparrow$'s on both sides.
This is where the requirement that the $W_k$ contains the specific states is used again, and it is also easy to see that the corresponding requirement will be satisfied for $W_{k+1}$.
The fact that the two sets being joined to form $W_{k+1}$ have at least one common element (true for $k \geq 3$) implies that ket-bra operators constructed from any pair of states in $W_{k+1}$ belong to $\mA_{1,2,\dots,k,k+1}$.
It now remains to show that any state orthogonal to $W_{k+1}$ is annihilated by $\mA_{1,2,\dots,k,k+1}$.
The argument here is identical to that used in the AKLT case in App.~\ref{app:SMalgebra}.
Any $\ket{\phi}_{1,2,\dots,k,k+1} = \ket{u_\uparrow}_{1,2,\dots,k} \otimes \ket{\uparrow}_{k+1} + \ket{u_\downarrow}_{1,2,\dots,k} \otimes \ket{\downarrow}_{k+1}$ orthogonal to $W_{k+1}$ has its $k$-site parts  $\ket{u_\uparrow}_{1,2,\dots,k}$ and $\ket{u_\downarrow}_{1,2,\dots,k}$ orthogonal to $W_k$ and hence annihilated by $\mA_{1,2,\dots,k}$, which implies that $\ket{\phi}_{1,2,\dots,k,k+1}$ is annihilated by $\mA_{1,2,\dots,k}$.
Similarly, $\ket{\phi}_{1,2,\dots,k,k+1}$ can be shown to be annihilated by $\mA_{2,\dots,k,k+1}$, which shows that $\ket{\phi}_{1,2,\dots,k,k+1}$ is annihilated by $\mA_{1,2,\dots,k,k+1} = \lgen \mA_{1,2,\dots,k}, ~ \mA_{2,\dots,k,k+1} \rgen$.
This also implies that the ket-bra operators constructed from any pair of states in $W_{k+1}$ in fact span $\mA_{1,2,\dots,k,k+1}$, completing the inductive step.
\end{proof}
\subsection{Generation of the DMI term}\label{subsec:DMIgeneration}
As discussed in Sec.~\ref{subsec:FMtowerscar}, 
the full PBC DMI term of Eq.~(\ref{eq:DMIHamil}) to be generateable from the generators of $\tmA^{\FM}_{\scar}$ due to the DCT.
We now explicitly show that it
can be expressed in terms of the generators of $\tmA^{\FM}_{\scar}$ in Eq.~(\ref{eq:FMtowernaive}).
We show this in multiple steps discussed below.
\subsubsection{The three-site cyclic DMI term from three-site generators}\label{subsec:threeDMInaive}
We start with the three-site DMI term, defined as
\begin{equation}
    D^\alpha_{j,k,\ell} \defn (\vec{S}_j \times \vec{S}_k)\cdot\widehat{\alpha} + (\vec{S}_k \times \vec{S}_\ell)\cdot\widehat{\alpha} + (\vec{S}_\ell \times \vec{S}_j)\cdot\widehat{\alpha},
\label{eq:threesiteDMIdefn}
\end{equation}
and express this in terms of the generators of $\tmA^{\FM}_{\scar}$ in Eq.~(\ref{eq:FMtowernaive}).
We first focus on the $\widehat{\alpha} = \widehat{z}$ component and expand $D^z_{1,2,3}$ as
\begin{align}
& D_{1,2,3}^{z} = S_1^x S_2^y - S_1^y S_2^x + S_2^x S_3^y - S_2^y S_3^x + S_3^x S_1^y - S_3^y S_1^x = \nn \\
& = \frac{-i}{3} \Big[ \ketbra{\mS_{12},\uparrow_3}{\uparrow_1,\mS_{23}} - \ketbra{\mS_{12},\downarrow_3}{\downarrow_1,\mS_{23}} + \text{cycl.perm.} \Big] + \text{h.c.}.
\label{eq:DMIzexpr}
\end{align}
The equality between the two lines can be checked, e.g., by writing out each of them in the ket-bra notation.
The second line can be expressed in terms of spin operators using
\begin{equation}
\ketbra{\mS_{12},\uparrow_3}{\uparrow_1,\mS_{23}} - \ketbra{\mS_{12},\downarrow_3}{\downarrow_1,\mS_{23}} = -2 \ketbra{\mS}_{12} \Big(S_3^z + S_1^z \Big) \ketbra{\mS}_{23} = -2 \Big(\frac{1}{4} - \vec{S}_1 \cdot \vec{S}_2 \Big) \Big(S_3^z + S_1^z \Big) \Big(\frac{1}{4} - \vec{S}_2 \cdot \vec{S}_3 \Big),
\label{eq:DMIspinexp}
\end{equation}
where the first equality can be checked by writing out the R.H.S. in the ket-bra notation, and we have also used $\ketbra{\mS}_{k,\ell} = \frac{1}{4} - \vec{S}_k \cdot \vec{S}_\ell$.
Further, the terms obtained by cyclic permutations of the spins can be generated from the above term by applying two-spin exchanges between sites 1 and 2 and between sites 2 and 3.
\footnote{
Explicitly, we have the two-site exchange operator $P_{k,\ell}^\text{exch} \defn 2\vec{S}_k \cdot \vec{S}_\ell + \frac{1}{2}$, which can be expressed in terms of the generators of $\tmA^{\FM}_{\scar}$ in Eq.~(\ref{eq:FMtowernaive}) and is in that algebra.  
We then have a three-site permutation operator $P_{1,2,3}^\text{exch} \defn P_{1,2}^\text{exch} P_{2,3}^\text{exch} : \ket{\alpha,\beta,\gamma}_{1,2,3} \to \ket{\gamma,\alpha,\beta}_{1,2,3}$; this is a unitary operator, and equivalently its conjugate action can be used to ``translate" on-site operators in a cyclic manner, $P_{1,2,3}^\text{exch} O_k (P_{1,2,3}^{\text{exch}})^\dagger = O_{k+1}$ on the three-site loop $1,2,3$; it is also in $\tmA^{\FM}_{\scar}$.}
Using Eq.~(\ref{eq:DMIspinexp}) and applying cyclic permutations, Eq.~(\ref{eq:DMIzexpr}) can be expressed in terms of the generators of $\tmA^{\FM}_{\scar}$ in Eq.~(\ref{eq:FMtowernaive}).
Note that Eqs.~(\ref{eq:DMIzexpr})-(\ref{eq:DMIspinexp}) are for the $\widehat{\alpha} = \widehat{z}$ component.
For the $\widehat{\alpha} = \widehat{x}$ or $\widehat{\alpha} = \widehat{y}$ components, we obtain similar expressions with the $S^z_j$'s replaced by $S^x_j$'s or $S^y_j$'s respectively.
\subsubsection{A simple set of generators for the bond algebra}\label{subsec:FMsimple}
We next obtain an alternate set of generators for the bond algebra $\tmA^{\FM}_{\scar}$ by enlarging the algebra $\mA_{SU(2)} = \lgen \{\vec{S}_j \cdot \vec{S}_{j+1}\} \rgen$ with three-site DMI terms $\{D^\alpha_{j,j+1,j+2}\}$ within all contiguous clusters of three sites on a one-dimensional chain, i.e., we show that
\begin{equation}
    \tmA^{\FM}_{\scar} = \lgen \{\vec{S}_j\cdot\vec{S}_{j+1}\},\;\;\{D^\alpha_{j-1, j, j+1},~\alpha \in \{x, y, z\}\}\rgen.  
\label{eq:FMtowerbond}
\end{equation}
As discussed earlier, all the shown generators in Eq.~(\ref{eq:FMtowerbond}) can be expressed in terms of generators of $\tmA^{\FM}_{\scar}$ in Eq.~(\ref{eq:FMtowernaive}).
To show the converse, we use the identity:
\begin{equation}
    \left(\frac{1}{4} - \vec{S}_1\cdot\vec{S}_2\right) D_{1,2,3}^{z} \left(\frac{1}{4} - \vec{S}_2\cdot\vec{S}_3\right)\left(\frac{1}{4} - \vec{S}_1\cdot\vec{S}_2\right) =  (\ketbra{\mS})_{12} D_{1,2,3}^{z} (\ketbra{\mS})_{23} (\ketbra{\mS})_{12} = \frac{3i}{2} (\ketbra{\mS})_{12} \otimes \sigma_3^z,
    \label{eq:FMtower_DMI_to_naive}
\end{equation}
where we have used
\footnote{The schematic ket-bra notation here can be made precise by appending arbitrary ket's and bra's in  appropriate places on the missing sites, e.g., the first equation would read
$\bra{\mS}_{12} \otimes \bra{u}_3 D_{1,2,3}^{z} \ket{v}_1 \otimes \ket{\mS}_{23} = -3i\big(\bra{u}_3 \ket{\uparrow}_3 \bra{\uparrow}_1 \ket{v}_1 - \bra{u}_3 \ket{\downarrow}_3 \bra{\downarrow}_1 \ket{v}_1 \big)$.}
\begin{equation}
\bra{\mS}_{12} D_{1,2,3}^{z} \ket{\mS}_{23} = -3i\big(\ket{\uparrow}_3 \bra{\uparrow}_1 - \ket{\downarrow}_3 \bra{\downarrow}_1 \big),\;\;\;\;\;
\bra{\mS}_{23} \ket{\mS}_{12} = -\frac{1}{2} \big(\ket{\uparrow}_1 \bra{\uparrow}_3 + \ket{\downarrow}_1 \bra{\downarrow}_3 \big) ~.
\end{equation}
By considering equation similar to Eq.~(\ref{eq:FMtower_DMI_to_naive}) replacing, e.g., $D^z \to D^x$, we can clearly use the generators in Eq.~(\ref{eq:FMtowerbond}) to produce the ones in Eq.~(\ref{eq:FMtowernaive}), showing equivalence of these two sets of generators in Eqs.~(\ref{eq:FMtowernaive}) and (\ref{eq:FMtower_DMI_to_naive}).
Of course, not all the generators in Eq.~(\ref{eq:FMtowerbond}) are independent.
For example, since the Heisenberg terms can be used to generate the two-site exchange operators
\begin{equation}
    P^{\exch}_{j, k} \defn 2 \vec{S}_j\cdot\vec{S}_k + \frac{1}{2},
\label{eq:Pexchdefn}
\end{equation}
a single three-site DMI term $D^\alpha_{j,j+1,j+2}$ can be used to generate any three-site DMI term $D^\alpha_{j_1, j_2, j_3}$ using exchanges.
In fact, we also numerically observe that the addition of the Heisenberg terms is not necessary in the generators of Eq.~(\ref{eq:FMtowerbond}), i.e., we observe  $\tmA^{\FM}_{\scar} = \lgen \{D^\alpha_{j-1, j, j+1},~\alpha \in \{x, y, z\}\} \rgen$
for systems of size $L \geq 4$. 
However, we have not been able to find simple expressions for the Heisenberg terms in terms of the DMI terms.
\subsubsection{Full DMI term from the three-site generators}\label{subsec:fullDMI}
Finally, we inductively show that the DMI term on any contiguous $k$ sites can be generated from the generators of $\tmA^{\FM}_{\scar}$ in Eq.~(\ref{eq:FMtowerbond}), where the $k$-site DMI terms $D^\alpha_{j_1, \cdots, j_k}$ with support on an arbitrary set of $k$ sites $\{j_1,\cdots, j_k\}$ are defined as
\begin{equation}
    D^\alpha_{j_1, \cdots, j_k} \defn \sumal{l = 1}{k-1}{\hat{\alpha}\cdot (\vec{S}_{j_l} \times \vec{S}_{j_{l+1}})} + \hat{\alpha}\cdot(\vec{S}_{j_k} \times \vec{S}_{j_1}),\;\;\;\alpha \in \{x, y, z\}. 
\label{eq:ksiteDMI}
\end{equation}
We first show the generation of the four-site DMI term.
Starting from the three-site terms $D^\alpha_{j, j+1, j + 2}$ and $D^\alpha_{j+1, j+2, j + 3}$, we can generate $D^\alpha_{j, j+1, j + 2, j + 3}$ using the two-site exchange operator $P^{\exch}_{j+2, j+3}$ 
[which can be expressed in terms of $\vec{S}_{j+2}\cdot\vec{S}_{j+3}$ according to Eq.~(\ref{eq:Pexchdefn})] 
as
\begin{equation}
    D^\alpha_{j, j+1, j + 3} = P^{\exch}_{j+2, j+3} D^\alpha_{j, j+1, j+2} P^{\exch}_{j+2, j+3},\;\;\;D^\alpha_{j, j+1, j+2, j+3} = D^\alpha_{j,j+1,j+3} + D^\alpha_{j+1,j+2,j+3}.
\label{eq:DMI4sitegen}
\end{equation}
Similarly, we can express the $(k+1)$-site DMI term $D^\alpha_{j, j+1, \cdots, j + k}$ using the $D^\alpha_{j, j + 1, \cdots, j+ k-1}$ (which is assumed to be generated by induction) and the 3-site DMI term $D^\alpha_{j + k -2, j + k-1, j+k}$ as follows
\begin{equation}
    D^\alpha_{j,j+1, \cdots, j+ k-2, j+k} = P^{\exch}_{j+k-1, j+k} D^\alpha_{j, j+1, \cdots, j+k-1} P^{\exch}_{j+k-1,j+k},\;\;\;D^\alpha_{j,j+1, \cdots, j+k} = D^\alpha_{j,j+1, \cdots, j+k-2, j+k} + D^\alpha_{j + k-2, j+k-1,j+k}.
\label{eq:DMIksitegen}
\end{equation}
Using this procedure, it is easy to see that ``closed-loop" DMI terms of all ranges can be generated.
This also includes the full PBC DMI term $D^\alpha_{1, \cdots, L}$, which is the Hamiltonian $H_{\DMI}$ of Eq.~(\ref{eq:DMIHamil}), whose final expression in terms of the generators of $\tmA^{\FM}_{\scar}$ of Eq.~(\ref{eq:FMtowerbond}) reads
\begin{equation}
    H_{\DMI} = \sumal{j = 1}{L-2}{\left[\prodal{k = 1}{L-j-2}{P^{\exch}_{L-k, L-k+1}}\right] D^\alpha_{j,j+1,j+2}\left[\prodal{k = 1}{L-j-2}{P^{\exch}_{L-k, L-k+1}}\right]^\dagger}, 
\label{eq:DMIfinalexp}
\end{equation}
where $P^{\exch}_{j,k}$ is defined in Eq.~(\ref{eq:Pexchdefn}).
\subsection{Impossibility of writing the full DMI Hamiltonian as a sum of strictly local symmetric terms}\label{subsec:DMIimpossible}
Here we prove the following Lemma.
\begin{lemma}
The PBC DMI Hamiltonian $H_{\alpha-\DMI}$ of Eq.~(\ref{eq:DMIHamil}) cannot be expressed as a sum of strictly local 
terms from the bond algebra $\tmA_{\scar}^{\FM}$, or w.l.o.g.\ strictly local terms that annihilate all states in the ferromagnetic multiplet.
\end{lemma}
\begin{proof}
Consider a product state with a spin texture that makes a complete rotation independent of $L$, e.g., by an angle of $4\pi$ in the $S_x$-$S_y$ plane, as we go around the chain:
\begin{equation}
\ket{\Psi^{\tex}} = e^{-i q \sum_{j=1}^L j S_j^z} \ket{\Psi^{\FM,+\hat{x}}} ~, \quad \quad 
q = \frac{4\pi}{L} ~, \quad\quad 
\ket{\Psi^{\FM,+\hat{x}}} \defn \bigotimes_j \ket{\rightarrow}_j ~,\;\;\;\ket{\rightarrow} \defn \frac{\ket{\uparrow} + \ket{\downarrow}}{\sqrt{2}}.
\end{equation}
The textured state is chosen such that the local contributions to the $H_{z-\DMI}$ have a uniform non-zero value proportional to $q$ at small $q$:
\begin{equation}
\bra{\Psi^\tex} S_j^x S_{j+1}^y - S_j^y S_{j+1}^x \ket{\Psi^\tex} = \frac{\sin(q)}{4}, \implies \bra{\Psi^\tex} H_{z-\DMI} \ket{\Psi^\tex} = \frac{L \sin(q)}{4} \to \pi ~\text{for}~ L \to \infty ~,
\label{eq:HzDMI_in_Psitextured}
\end{equation}
where we have used the identities
\begin{gather}
    e^{i \varphi S^z_j} S^x_j e^{-i \varphi S^z_j} = \cos(\varphi) S^x_j - \sin(\varphi) S^y_j,\;\;\;e^{i\varphi S^z_j} S^y_j e^{-i\varphi S^z_j} = \cos(\varphi) S^y_j + \sin(\varphi) S^x_j,\nn\\
    \bra{\rightarrow}S^x_j\ket{\rightarrow} = \frac{1}{2},\;\;\;\bra{\rightarrow}S^y_j\ket{\rightarrow} = 0.
\end{gather}
Suppose now there exists a decomposition of the DMI term as $H_{z-\DMI} =  \sum_{R}{}{\hh_R}$, where the sum runs over an extensive number of strictly local terms $\hh_R$ labelled by $R$, such that each $\hh_R$ annihilates the ferromagnetic multiplet and has support bounded by a fixed number independent of $L$; we also assume that the norm of each $\hh_{R}$ is bounded by a fixed number independent of $L$.
Note that below we will also abuse notation and refer to $R$ also as finite region where $h_R$ acts centered around a site labelled by $j_R$.
Under the above assumptions, we will show that the expectation value of the R.H.S.\ evaluated in $\ket{\Psi^\tex}$ scales as $\sim C L q^2$ for small $q \sim 1/L$ with some fixed $C$ independent of $L$.
This expectation value vanishes when $L \to \infty$, in contradiction to Eq.~(\ref{eq:HzDMI_in_Psitextured}).
The proof uses the fact that each $\hh_R$ annihilates any perfect ferromagnetic state, which implies that its expectation value in the textured state is bounded by an $\mO(q^2)$ value.
Specifically, denoting the restriction of product states to that region by a subscript $R$, we have
\begin{equation}
\bra{\Psi^{\tex}} \hh_{R} \ket {\Psi^{\tex}} =
\bra{\Psi^{\tex}}_{R} \hh_{R} \ket {\Psi^{\tex}}_{R} =
\sbra{\tilde{\Psi}^{\FM}}_{R}
e^{i q \sum_{j' \in R} (j'-j_R) S_{j'}^z}\hh_{R} e^{-i q \sum_{j' \in R} (j'-j_R) S_{j'}^z}
\sket{\tilde{\Psi}^{\FM}}_{R} ~,
\label{eq:hRexpect}
\end{equation}
where $j_R$ is some fixed site in the region $R$, e.g., the site in the middle of the region, and we have defined
\begin{equation}
\sket{\tilde{\Psi}^{\FM}}_{R} \defn e^{-i q j_R \sum_{j' \in R} S_{j'}^z} \sket{\Psi^{\FM, +\hat{x}}}_{R} = e^{-i q j_R S^z_{\tot,R}} \sket{\Psi^{\FM, +\hat{x}}}_{R},
\end{equation}
where $S^z_{\tot, R}$ is the total $z$-spin operator restricted to the region $R$.
Hence $\sket{\tilde{\Psi}^{\FM}}$ is a uniform ferromagnetic state with spins pointing in the same direction as spin at site $j_R$ in the state $\ket{\Psi^\tex}$; in particular it is annihilated by $\hh_R$.
Furthermore, note that $|j'-j_R|$ for any $j' \in R$ is bounded by the size of $R$, which is bounded by an $L$-independent number.
Consider now the series expansions of the operator in Eq.~(\ref{eq:hRexpect}) in powers of $q$
\begin{equation}
    e^{i q \sum_{j' \in R} (j'-j_R) S_{j'}^z}\hh_{R} e^{-i q \sum_{j' \in R} (j'-j_R) S_{j'}^z} = \hh_{R} + iq \Big[\sum_{j' \in R} (j'-j_R) S_{j'}^z, \hh_{R} \Big] + \mO(q^2) ~.
\end{equation}
Taking the expectation value of this operator in the state $\sket{\tilde{\Psi}^{\FM}}$, note that the $\mO(q^0)$ and $\mO(q^1)$ terms vanish since the state is annihilated by $\hh_R$!
Further, since $R$ is a region containing a fixed number of sites we expect the series to be convergent, with the norm of the $\mO(q^2)$ ``remainder" bounded by $M q^2$, where $M$ is an $L$-independent number (since the norm of $\hh_R$ is also assumed bounded by an $L$-independent number).
Hence $|\bra{\Psi^{\tex}} \hh_{R} \ket {\Psi^{\tex}}| \leq M q^2$ for some $L$-independent number $M$, and then $|\bra{\Psi^{\tex}} \sum_{R} \hh_{R} \ket {\Psi^{\tex}}| \leq C L q^2$ with some $L$-independent number $C$, as claimed earlier, thus completing the proof.
\end{proof}
\section{Bond Algebra for the AKLT Tower as Scars}\label{app:AKLTtowerann}
In this appendix, we provide details of the bond algebra for the tower of QMBS eigenstates discovered in the AKLT spin chain, discussed in Sec.~\ref{subsec:AKLTscar}. 
\subsection{Review of AKLT Tower of States}\label{subsec:AKLTtowerreview}
We first provide a review of 
known results on models that share these QMBS as eigenstates.
The one-dimensional AKLT Hamiltonian is defined as
\begin{equation}
    H^{(\Upsilon)}_{\AKLT} \defn \sumal{j = 1}{L_\Upsilon}{P^{\AKLT}_{j,j+1}},\;\;\Upsilon \in \{p, o\},\;\;\;P^{\AKLT}_{j,j+1} \defn \sumal{m = -2}{2}{(\ketbra{T_{2,m}})_{j,j+1}},
\label{eq:AKLTHamil}
\end{equation} 
where $\Upsilon = p$ and $\Upsilon = o$ stand for PBC and OBC respectively, $L_o = L -1 $ and $L_p = L$, the subscripts are modulo $L$ for PBC, and $\ket{T_{J, m}}_{j,j+1}$ are the total angular momentum states of two spin-1's, defined in Eq.~(\ref{eq:Tstates}).
$P^{\AKLT}_{j,j+1}$ is hence the projector onto the states of total angular momentum $2$, whose expression in terms of spin variables is shown in Eq.~(\ref{eq:AKLTprojdefn}).
Reference \cite{moudgalya2018a} solved for a tower of states in the spectrum of $H^{(\Upsilon)}_{\AKLT}$, which are given by $\{\ket{\psi_n} \defn (Q^\dagger)^n\ket{G}\}$ for PBC and $\{\ket{\psi_{n, \uparrow\uparrow}} \defn (Q^\dagger)^n\ket{G_{\uparrow\uparrow}}\}$ for OBC, where $\ket{G}$ and $\ket{G_{\uparrow\uparrow}}$ are the AKLT ground states for PBC and OBC respectively, discussed in Sec.~\ref{subsec:MPSscar}.
These QMBS eigenstates have the following eigenvalues under $H^{(\Upsilon)}_{\AKLT}$ and $S^z_{\tot}$:
\begin{gather}
    H^{(p)}_{\AKLT}\ket{\psi_n} = 2n \ket{\psi_n},\;\;S^z_{\tot} \ket{\psi_n} = 2n \ket{\psi_n},\nn \\
    H^{(o)}_{\AKLT}\ket{\psi_{n,\uparrow\uparrow}} = 2n\ket{\psi_{n, \uparrow\uparrow}},\;\;S^z_{\tot} \ket{\psi_{n,\uparrow\uparrow}} = (2n +1)\ket{\psi_{n,\uparrow\uparrow}},
\label{eq:towerspins}
\end{gather}
where the spin eigenvalues are a consequence of the operator $Q^\dagger$ being a spin 2 raising operator, as shown in  Eq.~(\ref{eq:psiPBCQdagdefn}).
References~\cite{mark2020unified, moudgalya2020large} further showed that a family of nearest-neighbor Hamiltonians with the same set of QMBS for PBC and OBC is given by
\begin{equation}
    H^{(\Upsilon)}_{\AKLT\text{-fam}} = \sum_j{}{\left[\mE (T^{(2,2)}_{j,j+1} + T^{(2,1)}_{j,j+1}) + \sumal{l,m = -2}{0}{\alpha_{l,m}(j) (\ket{T_{2,l}}\bra{T_{2,m}})_{j,j+1}}\right]},\;\;T^{(J,m)}_{j,j+1} \defn (\ketbra{T_{J, m}}{T_{J,m}})_{j,j+1},
\label{eq:AKLTscarfamily}
\end{equation}
where $\{\alpha_{l,m}(j)\}$ are arbitrary constants.
Similar to $H^{(\Upsilon)}_{\AKLT}$, the QMBS have the eigenvalues $\{2n\mE\}$ under $H^{(\Upsilon)}_{\AKLT\text{-fam}}$ and hence form a tower of equally spaced states with spacings $2 \mE$. 
Note that Eq.~(\ref{eq:AKLTscarfamily}) reduces to the AKLT model $H^{(\Upsilon)}_{\AKLT}$ of Eq.~(\ref{eq:AKLTHamil}) for $\mE = 1$ and $\alpha_{l,m}(j) = \delta_{l,m}$.
Further, we can use Eq.~(\ref{eq:towerspins}) to construct a family of Hamiltonians $\tH^{(\Upsilon)}_{\AKLT\text{-fam}}$ for which the QMBS eigenstates are all degenerate by simply subtracting $S^z_{\tot}$ times a constant from $H^{(\Upsilon)}_{\AKLT\text{-fam}}$, i.e., 
\begin{equation}
    \tH^{(p)}_{\AKLT\text{-fam}} \defn H^{(p)}_{\AKLT\text{-fam}} - \mE S^z_{\tot},\;\;\;\tH^{(o)}_{\AKLT\text{-fam}} \defn H^{(o)}_{\AKLT\text{-fam}} - \mE (S^z_{\tot} - 1).
\label{eq:degHamils}
\end{equation}
When restricted to the AKLT case with PBC, Eq.~(\ref{eq:degHamils}) reduces to Eq.~(\ref{eq:AKLTdeghamil}).
\subsection{Bond Algebra for the PBC AKLT QMBS}\label{subsec:PBCAKLT}
We now provide some details on the bond algebra for the QMBS of the PBC AKLT model discussed in Sec.~\ref{subsec:AKLTscar}.
To do so, we first need to find an appropriate set of projectors such that the AKLT QMBS $\{\ket{\psi_n} = (Q^\dagger)^n \ket{G}\}$ completely span their common kernel. 
\subsubsection{Two-site projectors}
We start by constructing nearest-neighbor projectors that are required to vanish on the states $\{\ket{\psi_n}\}$, which can be computed by performing a Schmidt decomposition of the states over the two sites $\{j,j+1\}$ (i.e., with the subsystems being the two sites and the rest of the system), computing the linear span of these Schmidt states, and constructing projector out of that subspace.
For the AKLT tower of states, this can be worked out analytically in the MPS formalism, starting with the MPS representation of the ground state $\ket{G}$ shown in Eq.~(\ref{eq:GMPS}), e.g., we can deduce that the linear span of Schmidt states over sites $\{j,j+1\}$ is given by
\begin{equation}
    \mS_{j,j+1} = \text{span}\{\sumal{m_j, m_{j+1}}{}{\sumal{\tau}{}{(A^{[m_j]})_{\sigma\tau} (A^{[m_{j+1}]})_{\tau\sigma'}}\ket{m_j m_{j+1}}}\},
\end{equation}
where $1 \leq \sigma, \sigma',\tau \leq 2$, and $(A^{[m_j]})_{\sigma\tau}$ denotes the matrix elements of $A^{[m_j]}$. 
Projectors that vanish on this subspace can then be constructed directly, which gives $P^{\AKLT}$ discussed in Sec.~\ref{subsec:MPSscar} and App.~\ref{app:SMalgebra}; this is equivalent to the parent Hamiltonian construction discussed in \cite{perezgarcia2007matrix, schuch2010peps, moudgalya2020large}.
We can repeat the exercise for the AKLT tower of states with the observation that the entire QMBS subspace $\{\ket{\psi_n}\}$ can be represented as the span of a one-parameter family of MPS, i.e.,
\begin{gather}
    \text{span}\{\ket{\psi_n}\} = \text{span}\{e^{\xi Q^\dagger}\ket{G}\} = \text{span}\{\prodal{j}{}{e^{\xi(-1)^j  (S^+_j)^2}}\ket{G}\} = \text{span}\{\sumal{\{m_j\}}{}{\Tr[B^{[m_1]} B^{[m_2]} \cdots B^{[m_L]}]\ket{m_1 \dots m_L}}\},\nn \\
    B^{[m_j]} \defn \sumal{n_j}{}{[e^{\xi (-1)^j (S^+_j)^2}]_{m_j, n_j} A^{[n_j]}} = A^{[m_j]} + (-1)^j \xi \sumal{n_j}{}{[(S^+_j)^2]_{m_j, n_j} A^{[n_j]}},\;\;m_j,n_j \in \{+, 0, -\},
\label{eq:BMPSdefn}
\end{gather}
where $[\bullet]_{m_j,n_j}$ denotes the matrix elements of $\bullet$. 
The linear span of all the Schmidt states over $\{j,j+1\}$ in the states of Eq.~(\ref{eq:BMPSdefn}) is the span of the two-site MPS, i.e., 
\begin{equation}
    \mS_{j,j+1} = \text{span}_{\xi}\{\sumal{m_j, m_{j+1}}{}{\sumal{\tau}{}{(B^{[m_j]})_{\sigma\tau} (B^{[m_{j+1}]})_{\tau\sigma'}}\ket{m_j m_{j+1}}}\},
\end{equation}
where the subscript $\xi$ indicates that the span is taken over all values of $\xi$, which in this case equals the span of vectors that appear in the above expression with coefficients $\xi^0$ and $\xi^1$ [none of the higher powers appear in the expansion].
%
It is easy to show that this is precisely the span of the OBC towers of QMBS on the two sites, i.e.,  
\begin{equation}
    \mS_{j,j+1} \defn \text{span}\{\ket{\psi_{n,\sigma\sigma'}}_{j,j+1},\;\;\sigma,\sigma' \in \{\uparrow, \downarrow\}\},\;\;\;\ket{\psi_{n,\sigma\sigma'}}_{j,j+1} \defn \frac{1}{\mN_n}\left(\sumal{k = j}{j+1}{(-1)^k (S^+_k)^2}\right)^n\ket{G_{\sigma\sigma'}}_{j,j+1}
\label{eq:twositeSchmidt}
\end{equation}
where $\{\ket{G_{\sigma\sigma'}}_{j,j+1}\}$ are the four OBC AKLT ground states on the two sites $\{j,j+1\}$; this expression is a consequence of the raising operator $Q^\dagger$ being a sum of on-site terms.
The projectors out of the subspace $\mS_{j,j+1}$ then turn out to have the compact expression~\cite{mark2020unified,moudgalya2020large}
\begin{equation}
    \Pi_{j,j+1} = \sumal{m = -2}{0}{T^{(2,m)}_{j,j+1}},
\label{eq:2siteproj}
\end{equation}
where $\{T^{(J,m)}_{j,j+1}\}$ are projectors onto total angular momentum states, defined in Eq.~(\ref{eq:AKLTscarfamily}).
The common kernel of these projectors can be computed using efficient methods discussed in \cite{yao2022bounding, moudgalya2022numerical}, and for small system sizes we numerically observe that its dimension grows as
\begin{equation}
    \dim\ker(\{\Pi_{j,j+1}\}) = 2^L + 2\;\;\text{for even $L$}.
\label{eq:2sitekernelPBC}
\end{equation}
The exponential growth with system size implies that the projectors vanish on many more states than $\{\ket{\psi_n}\}$, hence these are not the desired projectors.
This fact can be seen easily by noting that any product state containing only the ``$-$'' or ``$0$'' spin-1 on-site states and such that no two ``$0$"'s are next to each other is annihilated by the above projectors; however, we do not have an analytical argument for the precise count of all states in the common kernel including non-product ones.
\subsubsection{Three-site projectors}
We hence move on to look for three-site projectors $\{\Pi_{[j,j+2]}\}$ that vanish on the states $\{\ket{\psi_n}\}$.
These projectors can be computed in a way similar to the two-site projectors, by studying the Schmidt decomposition of $\{\ket{\psi_n}\}$ over the three sites $\{j,j+1,j+2\}$ and the linear span of the Schmidt states $\mS_{[j,j+2]}$. 
This can be worked out in the MPS language using the same method as described for the two-site projectors, and similar to Eq.~(\ref{eq:twositeSchmidt}) this can be shown to be the subspace spanned by the OBC AKLT towers of QMBS on the three sites, i.e., 
\begin{equation}
    \mS_{[j,j+2]} \defn \text{span}\{\ket{\psi_{n,\sigma\sigma'}}_{[j,j+2]},\;\;\sigma,\sigma' \in \{\uparrow, \downarrow\}\},\;\;\;\ket{\psi_{n,\sigma\sigma'}}_{[j,j+2]} \defn \frac{1}{\mN_n}\left(\sumal{k = j}{j+2}{(-1)^k (S^+_k)^2}\right)^n\ket{G_{\sigma\sigma'}}_{[j,j+2]},
\label{eq:threesiteSchmidt}
\end{equation}
which is the span of the four towers of states originating from the four three-site ground states.
In the spin language, we have $\mS_{[j,j+2]} = \text{span}\{\ket{s_a}_{[j,j+2]}, a \in \{1,\dots,8\}\}$, where the (unnormalized) states are defined as
\begin{align}
    &\ket{s_1} \defn \ket{J = 1, J_z = 1} = - \ket{+ 0 0} + \ket{0 + 0} - \ket{0 0 +} + 2 \ket{+ - +}, \nn \\
    &\ket{s_2} \defn \ket{J = 1, J_z = 0} = - \ket{+ 0 -} + \ket{- + 0} + \ket{0 + -} - \ket{- 0 +} -\ket{0 0 0} + \ket{0 - +} + \ket{+ - 0}, \nn \\
    &\ket{s_3} \defn \ket{J = 1, J_z = -1} = - \ket{- 0 0} + \ket{0 - 0} - \ket{0 0 -} + 2 \ket{- + -}, \nn \\
    &\ket{s_4} \defn \ket{J = 0, J_z = 0} = \ket{+ 0 -} - \ket{+ - 0} - \ket{0 + -} + \ket{0 - +} + \ket{- + 0} - \ket{- 0 +}, \nn \\
    &\ket{s_5} \defn \ket{J = 3, J_z = 3} = \ket{+ + +}, \nn \\
    &\ket{s_6} \defn \ket{J = 2, J_z = 2} = \ket{+ + 0} - \ket{0 + +}, \nn \\
    &\ket{s_7} \defn \ket{J = \times, J_z = 2} = \ket{+ 0 +}, \nn \\
    &\ket{s_8} \defn \ket{J = \times, J_z = 1} = -\ket{+ 0 0} - \ket{0 + 0} -\ket{0 0 +} + 2 \ket{+ + -} + 2\ket{- + +},
\label{eq:3sitesubspace}
\end{align}
where $J$ and $J_z$ denote the total angular momentum quantum numbers of the spin-1's on three consecutive sites. 
The first four states $\ket{s_1}$-$\ket{s_4}$ are the OBC AKLT ground states on the three sites, and the remaining states $\ket{s_5}$-$\ket{s_8}$ are obtained by acting the raising operator $Q^\dagger$ on the four ground states.
Note that unlike the other states, $\ket{s_7}$ and $\ket{s_8}$ are not eigenstates of the total spin angular momentum on the three sites; further, $\ket{s_1}$ and $\ket{s_8}$ are not mutually orthogonal.
$\Pi_{[j,j+2]}$ is then the projector onto the subspace orthogonal to $\{ \ket{s_a}, a \in \{1,\dots,8\}\}$ on the three sites $\{j,j+1,j+2\}$.
Similar to the two-site case, the kernel of these projectors $\{\Pi_{[j,j+2]}\}$ can be computed using efficient methods discussed in \cite{yao2022bounding, moudgalya2022numerical}.
Here we find that up to fairly large system sizes, the dimension of the common kernel is
\begin{equation}
    \dim\ker(\{\Pi_{[j,j+2]}\}) = \twopartdef{\frac{L}{2} + 3}{L = 2 \times \text{even}}{\frac{L}{2} + 1}{L = 2 \times \text{odd}}, 
\label{eq:3sitekernelPBC}
\end{equation}
and in App.~\ref{subsec:AKLTanalyticPBC} we analytically prove an upper bound on the dimension of this kernel.
For system sizes $L = 2 \times \text{odd}$, the common kernel is precisely spanned by the states of the tower $\{\ket{\psi_n}\}$ and the spin-1 ferromagnetic state given by
\begin{equation}
    \ket{F} \defn \ket{+ + \dots + +}.
\label{eq:spin1ferromagnetic}
\end{equation}
Note that we count the ferromagnetic state separately since it is not a part of the tower defined by Eq.~(\ref{eq:psiPBCQdagdefn}) for these system sizes, i.e., $(Q^\dagger)^{L/2} \ket{G} = 0$ for $L/2$ odd, so the tower does not reach the ferromagnetic state. 
For system sizes $L = 2 \times \text{even}$, the kernel is larger than the span of the tower of states $\{\ket{\psi_n}\}$ (which now includes the ferromagnetic state $\ket{F}$) by two states. 
For system sizes $L \leq 8$, we numerically determine that these are states with spin $S^z = L - 1$ and momenta $k = \pm \pi/2$, which exists only for these system sizes.
We can immediately solve for these states using this information, and they are given by
\begin{equation}
    \ket{1_{k}} = \sumal{j = 1}{L}{e^{i k j} S^-_j \ket{F}},\;\;\;k \in \{-\frac{\pi}{2}, \frac{\pi}{2}\},  
\label{eq:1kstates}
\end{equation}
and it is easy to verify that they are annihilated by the projectors $\Pi_{[j,j+2]}$, or, equivalently, by any three-site state orthogonal to the $\{\ket{s_j}\}$ in Eq.~(\ref{eq:3sitesubspace}).
Note that these states are the same as the  $\ket{1_{k}}$ eigenstates of the AKLT model, solved for in \cite{moudgalya2018a}.
Further, we numerically observe that these extra states are not eliminated from the common kernel even if we consider four-site projectors $\{\Pi_{[j,j+3]}\}$ that are required to annihilate $\{\ket{\psi_n}\}$.
In general, we can conclude that the common kernel of $\{\Pi_{[j,j+2]}\}$ is spanned by 
$\{\ket{\psi_n}\}$ and $\{\ket{\phi_m}\}$, all of which are eigenstates of the PBC AKLT model. 
This form of the kernel motivates the construction fo the PBC bond algebra of Eq.~(\ref{eq:PBCAKLTtowerpair}), discussed in Sec.~\ref{subsec:AKLTscar}.
\subsection{Bond Algebra for the OBC AKLT QMBS}\label{subsec:OBCAKLT}
Moving on to the OBC AKLT model, we again wish to construct a set of local projectors such that the common kernel is completely spanned by the tower of states $\{\ket{\psi_{n, \uparrow\uparrow}}\}$. 
\subsubsection{Two-site projectors}
Similar to the PBC case, we can start with nearest-neighbor projectors and look for two-site projectors that annihilate the tower. 
Using the same procedure, i.e., constructing the subspace spanned by the Schmidt states of $\{\ket{\psi_{n, \uparrow\uparrow}}\}$, we obtain the same set of projectors $\{\Pi_{j,j+1}\}$ as Eq.~(\ref{eq:2siteproj}) in the bulk of the system. 
Similar to the PBC case, the common kernel of these projectors for OBC also grows exponentially with system size:
\begin{equation}
    \dim\ker(\{\Pi_{j,j+1}\}) = 3 \times 2^{L-1},
\label{eq:2sitekerneldim}
\end{equation}
and the exponential growth can be understood using similar arguments as in the PBC case. 
Note that strictly speaking, the projectors on the boundary should be chosen differently if we require the kernel to only contain the tower $\{\ket{\psi_{n, \uparrow\uparrow}}\}$ and not the towers $\ket{\psi_{n, \sigma\sigma'}}$.
However, we observe and it is easy to show that the choice of the boundary projectors does not affect the exponential nature of the scaling of the kernel dimension with system size. 
For example, if we include the two-site boundary projectors $\Pi^{(l)}_{1,2}$ and $\Pi^{(r)}_{L-1,L}$ we discuss below in the next section, we find that $\dim\ker(\Pi^{(l)}_{1,2}, \{\Pi_{j,j+1}\}, \Pi^{(r)}_{L-1, L}) = 6 \times 2^{L-4}$.
\subsubsection{Three-site bulk projectors + Two-site boundary projectors}
Hence we move on to three-site projectors, we again get the same set of projectors $\{\Pi_{[j,j+2]}\}$ in the bulk of the system as the PBC case. 
The common kernel of these projectors can be computed numerically, and we find that
\begin{equation}
    \dim\ker(\{\Pi_{[j,j+2]}\}) = 2L + 2,
\label{eq:3sitekerneldim}
\end{equation} 
which was also observed in \cite{yao2022bounding}.
In App.~\ref{subsec:PsiOBC} we analytically prove an upper bound on the dimension of this kernel, and it appears that the kernel is spanned completely by $\{\ket{\psi_{n,\sigma\sigma'}}\}$, i.e., the four OBC towers of states, and possibly a few extra states such as $\ket{F}$ which might not be part of the towers for certain system sizes.
Following the ideas discussed in Sec.~\ref{subsec:AKLTscar}, this actually implies the existence of a bond and commutant algebra pair
\begin{equation}
    \tmA_{\scar} = \lgen \{\Pi_{[j,j+2]} h_{[j,j+2]} \Pi_{[j,j+2]}\} \rgen,\;\;\;\tmC_{\scar} = \lgen \{\ketbra{\psi_{n,\sigma\sigma'}}{\psi_{m,\tau\tau'}}\}\rgen,\;\;\;\sigma,\sigma',\tau,\tau' \in \{\uparrow, \downarrow\},
\label{eq:alltowerpair}
\end{equation}
for a generic choice of $h_{[j,j+2]}$.
In order to limit the kernel to only one of the towers $\{\ket{\psi_{n, \uparrow\uparrow}}\}$, we need additional projectors acting on the boundaries. 
We can start by adding on-site projectors $(\ket{-}\bra{-})_1$ and $(\ket{-}\bra{-})_L$ to the list of three-site projectors $\{\Pi_{[j,j+2]}\}$ in order to ``enforce" only the tower $\{\ket{\psi_{n,\uparrow\uparrow}}\}$ to constitute the common kernel. 
However, we numerically find that some states from the other towers survive, i.e., they are annihilated by the edge projectors, and these states are not eigenstates of the OBC AKLT model (for which, we remind, only states in the tower $\{\ket{\psi_{n,\uparrow\uparrow}} \}$ are eigenstates while the other three towers are generally not eigenstates).
We then construct left and right two-site boundary projectors and require them to vanish on the towers of states $\ket{\psi_{n, \uparrow\uparrow}}$. 
These are respectively the projectors out of the linear span of the Schmidt states over the sites $\{1, 2\}$ and over the sites $\{L-1, L\}$, and the corresponding subspaces are spanned by the (unnormalized) three states $\{\ket{l_a}_{1,2}, a=1,2,3\}$ and  $\{\ket{r_a}_{L-1,L}, a=1,2,3\}$ defined as
\begin{align}
    &\ket{l_1} \defn \ket{+ 0} - \ket{0 +},\;\;\ket{l_2} = \ket{+ - } - \ket{0 0},\;\;\ket{l_3} \defn \ket{ +  +}; \nn\\
    &\ket{r_1} \defn \ket{+ 0} - \ket{0 +},\;\;\ket{r_2} = \ket{- +} - \ket{0 0},\;\;\ket{r_3} \defn \ket{ +  +}.
\label{eq:lr2sitesubspace}
\end{align}
Referring to the left and right projectors out of the above subspaces as $\Pi^{(l)}_{1,2}$ and $\Pi^{(r)}_{L-1,L}$ respectively, we numerically observe that the dimension of the common kernel of the projectors $\{\Pi^{(l)}_{1,2}, \{\Pi_{[j,j+2]}\}, \Pi^{(r)}_{L-1,L}\}$ is given by
\begin{equation}
    \dim\ker(\Pi^{(l)}_{1,2}, \{\Pi_{[j,j+2]}\}, \Pi^{(r)}_{L-1,L}) = \twopartdef{\frac{L}{2} + 1}{L\text{ is even}}{\frac{L+1}{2}}{L\text{ is odd}}, 
\label{eq:AKLTOBCkerneldim}
\end{equation}
and in App.~\ref{subsec:AKLTanalyticOBC} we analytically prove an upper bound on the dimension of this kernel.
We also numerically verify that this kernel is completely spanned by $\{\ket{\psi_{n, \uparrow\uparrow}}\}$ for odd system sizes, and by $\{\ket{\psi_{n, \uparrow\uparrow}}\}$ and $\ket{F}$, the ferromagnetic state, for even system sizes, all of which are eigenstates of the OBC AKLT model.
This form of the kernel motivates the construction of the OBC bond algebra given by
\begin{equation}
\tmA^{(o)}_{\scar} = \lgen \Pi^{(l)}_{1, 2} h^{(l)}_{[1]}\Pi^{(l)}_{1, 2},\; \{\Pi_{[j,j+2]} h_{[j]} \Pi_{[j,j+2]}\},\;  \Pi^{(r)}_{L-1, L} h^{(r)}_{[L]}\Pi^{(r)}_{L-1, L}\rgen, \;\;\; \tmC^{(o)}_{\scar} = \lgen \{\ket{\psi_{n,\uparrow\uparrow}}\bra{\psi_{m, \uparrow\uparrow}}\}\rgen,
\label{eq:ACpairobc}
\end{equation}
where $h^{(l)}_{[1]}$, $h_{[j]}$, $h^{(r)}_{[L]}$  are sufficiently generic operators with support in the vicinity of sites indicated in the subscripts.
\section{Some Analytical Results on the Singlets of the AKLT Tower Bond Algebras}\label{app:AKLTanalytic}
Here we collect some analytical results on the singlets of the bond algebras described in Sec.~\ref{subsec:AKLTscar} and App.~\ref{app:AKLTtowerann}, more specifically, upper bounds on the number of states annihilated by the Shiraishi-Mori projectors involved in the constructions of the bond algebras.
\subsection{Three-site bulk projectors with OBC}\label{subsec:PsiOBC}
Consider states annihilated by the projectors $\{\Pi_{[j,j+2]}$, $j=1,\dots,L-2\}$, in OBC.
As discussed in App.~\ref{subsec:PBCAKLT}, each $\Pi_{[j,j+2]}$ projects onto the subspace orthogonal to $\{\ket{s_a}_{[j,j+2]}, a \in \{1, \dots, 8\} \}$ of Eq.~(\ref{eq:3sitesubspace}); it is easy to check that this subspace is spanned by the following $19$ states $\{\ket{u_a}_{[j,j+2]}, a \in \{1, \dots, 19\} \}$:
\begin{align}
& \ket{u_1} \defn \ket{-\!-\!-}, \quad 
\ket{u_2} \defn \ket{0\!-\!-}, \quad
\ket{u_3} \defn \ket{-0-}, \quad
\ket{u_4} \defn \ket{-\!-\!0}, \quad
\ket{u_5} \defn \ket{+\!-\!-}, \quad
\ket{u_6} \defn \ket{-\!-\!+},
\nn \\
& \ket{u_7} \defn \ket{-00} + \ket{0\!-\!0}, \quad
\ket{u_8} \defn \ket{00-} + \ket{0\!-\!0}, \quad
\ket{u_{9}} \defn \ket{-0+} + \ket{0\!-\!+}, \quad
\ket{u_{10}} \defn \ket{+0-} + \ket{+\!-\!0}, \nn\\
& \ket{u_{11}} \defn \ket{00+} - \ket{+00}, \quad
\ket{u_{12}} \defn \ket{+\!+\!-} - \ket{-\!+\!+}, \quad
\ket{u_{13}} \defn \ket{+\!+\!0} + \ket{0\!+\!+},\nn\\
& \ket{u_{14}} \defn \ket{0\!-\!+} - \ket{-\!+\!0}, \quad
\ket{u_{15}} \defn \ket{0\!+\!-} - \ket{+\!-\!0}, \quad \ket{u_{16}} \defn \ket{-\!+\!-} - 2\ket{0\!-\!0}, \nn\\
& \ket{u_{17}} \defn \ket{+\!-\!0} + \ket{-\!+\!0} + 2\ket{000}, \quad
\ket{u_{18}} \defn \ket{+\!-\!+} + \ket{-\!+\!+} + 2\ket{00+}, \quad
\ket{u_{19}} \defn 2 \ket{0\!+\!0} + \ket{+\!+\!-} - \ket{+\!-\!+}.
\label{eq:19dimspan}
\end{align}
Note that not all of the states are normalized nor are all of them are mutually orthogonal; however, they are linearly independent and span the desired subspace.
Below we require the annihilation by projectors onto each of these states to derive conditions on the states in the common kernel, we find that 
the two-site projectors $\ketbra{T_{2,m}}_{j,j+1}, m \in \{-2,-1,0\}$, defined in Eq.~(\ref{eq:Tstates}) can be expressed in terms of ket-bra operators of the form $\{\ketbra{u_a}{u_b}\}$, which simplifies some of the analysis.
Requiring the annihilation of $\ket{\Psi}$ by $\Pi_{[j,j+2]}$ is equivalent to requiring annihilation by projectors $\{\ketbra{u_a}_{[j,j+2]}, a \in \{1, \dots, 19\}\}$.
Then consider expansion in the computational basis, $\ket{\Psi} = \sum_{\{m_j\}} \Psi(m_1,m_2,\dots,m_L) \ket{m_1,m_2,\dots,m_L}$.
Requiring annihilation by $\{\ketbra{u_a}, a \in \{1,2,4,5,6\}\}$ is equivalent to
\begin{align}
\Psi(\dots, -, -, \dots) = 0 \label{eq:--}
\end{align}
for any location along the chain of the two consecutive sites hosting a ``$--$'' pattern somewhere in the chain with any configuration on the rest of the chain marked with dots.
This is equivalent to the condition of annihilation by two-site $\{\ketbra{--}_{j,j+1}\}$, and it is easy to show that these two-site projectors $\ketbra{T_{2,-2}}_{j,j+1}$ can be expressed in terms of the three-site ket-bra operators $\{ \ketbra{u_a}{u_b}_{[j,j+2]}, a,b \in \{ 1,2,4,5,6\}\}$.
Further requiring annihilation by $\{\ketbra{u_a}, a \in \{3,7,8,9,10\}\}$, combined with the previous conditions from $\ketbra{u_2}$ and $\ketbra{u_4}$, gives the condition
\begin{align}
\Psi(\dots, -,0, \dots) = -\Psi(\dots, 0,-, \dots) \label{eq:-0}
\end{align}
for any location of the two consecutive sites, which again reflects the fact that we can express the two-site projectors $\{\ketbra{T_{2,-1}}_{j,j+1}\}$ in terms of the corresponding three-site ket-bra operators.
Further requiring annihilation by $\{\ketbra{u_a}, a \in \{11,12,13,14,15,16\}\}$ produces three-site conditions
\begin{align}
& \Psi(\dots, 0,0,+, \dots) = \Psi(\dots, +,0,0, \dots)~, \label{eq:00+} \\
& \Psi(\dots, +,+,-, \dots) = \Psi(\dots, -,+,+, \dots) ~, \label{eq:++-} \\
& \Psi(\dots, +,+,0, \dots) = -\Psi(\dots, 0,+,+, \dots) ~, \label{eq:++0} \\
& \Psi(\dots, 0,-,+, \dots) = \Psi(\dots, -,+,0, \dots) ~, \label{eq:0-+} \\
& \Psi(\dots, 0,+,-, \dots) = \Psi(\dots, +,-,0, \dots) ~, \label{eq:0+-} \\
& \Psi(\dots, -,+,-, \dots) = 2 \Psi(\dots, 0,-,0, \dots) \label{eq:-+-} ~.
\end{align}
Further requiring annihilation by $\ketbra{u_{17}}$ and $\ketbra{u_{18}}$, combined with the previous conditions from $\ketbra{u_{16}}$ and $\ketbra{u_8}$, gives condition
\begin{align}
\Psi(\dots, +,-, \dots) + \Psi(\dots, -,+, \dots) + 2\Psi(\dots, 0,0, \dots) = 0 ~. \label{eq:+-}
\end{align}
This two-site condition again reflects that we can express the two-site projectors $\{\ketbra{T_{2,0}}_{j,j+1}\}$ in terms of the corresponding three-site ket-bra operators.
Finally, requiring annihilation by $\ketbra{u_{19}}$ gives condition
\begin{align}
2\Psi(\dots, 0,+,0, \dots) + \Psi(\dots, +,+,-, \dots) - \Psi(\dots, +,-,+ \dots) = 0 ~. \label{eq:0+0}
\end{align}
For the most part, we will be using conditions Eqs.~(\ref{eq:-0})-(\ref{eq:-+-}) that specifically relate the amplitudes on two configurations related by a simple ``local move" that is easy to read off for each condition.
In particular, we have moves that exchange nearest-neighbor ``$-$" and $0$; hop $00$ past a $+$; hop $++$ past a ``$-$" or a $0$; hop $0$ past a $-+$ or $+-$; and relate $-\!+\!-$ and $0-0$.
Our strategy below is to use these moves to relate the amplitude on any given configuration (a.k.a. product state in the computational basis, often referred to simply as ``state" below) to the amplitudes on some specific ``reference" configurations and count the number of independent reference configurations.
For the OBC chain, we will prove the following Lemma.

\begin{lemma}\label{lem:OBCAKLTkercount}
For $\ket{\Psi}$ annihilated by $\{\Pi_{[j,j+2]}, j=1,\dots,L-2\}$,
$\Psi(m_1,m_2,\dots,m_L)$ for any $\ket{m_1,m_2,\dots,m_L}$ can be related to amplitudes on states in the following three families: \\
i) $\ket{-,0,0,\dots,0,0,0}$, 
$\ket{+,-,0,\dots,0,0,0}$, 
$\ket{+,+,-,\dots,0,0,0}$, $\dots$,
$\ket{+,+,+,\dots,+,-,0}$, 
$\ket{+,+,+,\dots,+,+,-}$; \\
ii) $\ket{-,+,0,\dots,0,0,0}$, 
$\ket{+,-,+,\dots,0,0,0}$, $\dots$,
$\ket{+,+,+,\dots,-,+,0}$, 
$\ket{+,+,+,\dots,+,-,+}$; \\
iii) $\ket{+,+,+,\dots,+,+,+}$, 
$\ket{+,+,+,\dots,+,+,0}$, 
$\ket{+,+,+,\dots,+,0,+}$. \\
There are $L$ states in the first family, $L-1$ states in the second family, and $3$ states in the third family, for the total of $2L+2$ states; hence $\dim \ker(\{\Pi_{[j,j+2]} \}) \leq 2L + 2$ in OBC.
\end{lemma}
\begin{proof}
Consider first the case where at least one $m_j = -1$.
In the following analysis, we recall that having a configuration $--$ anywhere along the chain immediately implies that the amplitude is zero due to Eq.~(\ref{eq:--}); hence the amplitude of any configuration that can be ``connected" to a configuration with $--$ is also zero, and we implicitly consider only cases where this combination never occurs.
Using the moves $m\!+\!+ \to +\!+\!m$ for $m \in \{-,0\}$, and $00m \to m00$ for $m \in \{-, +\}$, we can push all instances of $+\!+$ to the left and all instances of $00$ to the right and relate the original configuration to a configuration of the form $+\!\dots\!+\![\text{singletons}]0\dots0$, where for each site in the middle ``singletons'' region the state on the site differs from those of its neighbors, and we have used the fact that $--$ cannot appear.
In the singletons region, we then try to push all $0$'s as far to the right as possible via moves $0- \to -0$ [Eq.~(\ref{eq:-0})], $0-+ \to -+0$ [Eq.~(\ref{eq:0-+})], and $0\!+\!- \to +\!-\!0$ [Eq.~(\ref{eq:0+-})], ``cleaning up" any instances of $++$ and $00$ that may occur in the process by pushing these to the left or to the right boundary of the singletons region thus shrinking it.
We perform all these moves until no such moves are possible, obtaining a new (possibly smaller) singleton region.
We can then argue that the configuration in the singleton region must reach the form
$[- + - +\!\dots\!(-\text{~or~}+) 0\!+\!0\!+\!\dots\!0\!+]$, recalling that we started with the assumption that there must be at least one $-$ state in the region while the rest of the pattern can in principle be empty.\footnote{The argument goes as follows.
The leftmost and rightmost states cannot be $+$ or $0$ respectively, by definition of the singleton region.
Suppose the leftmost site in the singleton region has state $0$.
The next site cannot be the state $-$ since then we are not done with the $0- \to -0$ moves, so the start of the singleton region must be $0+$.
The next site cannot be a $-$ since then we are not done with the $0\!+\!- \to +\!-\!0$ moves, so the start of the singleton region must be $0\!+\!0$.
The alternating pattern of $0$ and $+$ then must continue, but at some point we must encounter a $-$ since at least one is present, thus reaching contradiction that the discussed moves are no longer possible.
Hence, the singleton region must start with the state $-$.
By the same argument,
we then must continue with alternating $+$ and $-$ until there are no $-$'s left in the singleton region, at which point the rest of the singleton region must have alternating $0$ and $+$ (either of these states is possible starting after the last $-$).}
We now show that we can further ``simplify'' the singleton region by applications of some of the previous moves and moves $-0 \to 0-$ or $-\!+\!0 \to 0\!-\!+$ moving $0$ temporarily to the left.
Specifically, we can perform the following moves at the left boundary of the $0\!+\!0\!+\!\dots$ segment inside the singleton region: 
$-0\!+\!0+ \to 0\!-\!+0+ \to -\!+\!00+ \to -\!+\!+00 \to +\!+\!-00$ or 
$-\!+\!0\!+\!0+ \to 0\!-\!+\!+0+ \to 0\!+\!+\!-\!0+ \to +\!+0\!-\!0+ \to +\!+\!00\!-\!+ \to +\!+\!-\!+\!00$.
We can follow this by again pushing the leftmost $++$ to the left and the rightmost $00$ to the right, effectively reducing the size of the $0\!+\!0\!+\dots$ segment (and hence the singleton region) by $4$.
After repeated applications of these steps, and few extra similar steps if the original size of the segment $0\!+\!0\!+\dots$ was $4n+2$, we can ``simplify" the singleton region to the form $[-\!+\!-\!+\!\dots\!(-\text{~or~}+)]$.
Note that it was crucial to have at least one $-$ to facilitate the required moves to bring the singleton region to this form.
Finally, we can further reduce the size of the singletons region using moves $- + - \to 0\!-\!0 \to -00$ [Eq.~(\ref{eq:-+-}) followed by Eq.~(\ref{eq:-0})] until there is a single $-$ remaining in the singleton region.
At this point we have connected the amplitude of the original configuration to the amplitude on one of the states in the family i) or ii), and this completes the analysis of configurations with at least one $m_j = -1$.
We are left to consider special cases where there are no $-$ states among the $m_j$'s.
If there are at least two $0$'s, we can move any $++$ intervening between them and bring the two $0$'s to be either nearest neighbors $00$ or next-nearest neighbors $0\!+\!0$.
In the first case we can use Eq.~(\ref{eq:+-}) and in the second case Eq.~(\ref{eq:0+0}) to relate the amplitude to those on states with at least one $-$, which are presumed to be already fixed by the preceding construction by amplitudes on the states in the families i) or ii).
Finally, if are no $0$'s or there is precisely one $0$, we can move $++$'s and relate such states to the states in the family iii), which completes the proof of the claim.
For this family, we can in fact write down the corresponding three linearly-independent states in the common kernel:
\begin{equation}
\ket{\text{iii}_0} \defn \ket{F} = \ket{+,+,\dots,+}, \quad 
\ket{\text{iii}_1} \defn (S_1^{-} - S_3^{-} + S_5^{-} - S_7^{-} + \dots) \ket{F}, \quad
\ket{\text{iii}_2} \defn (S_2^{-} - S_4^{-} + S_6^{-} - S_8^{-} + \dots) \ket{F},
\label{eq:famIIIeigstates}
\end{equation}
where the exhibited sums run as long as they stay within the OBC chain.
To establish that these states are indeed in the common kernel, the only non-trivial case to check is the annihilation by $\ket{\dots}\bra{u_{13}}$, which is simple to see.
The consideration of these cases completes the proof of the Lemma.
\end{proof}

Some remarks are in order.
The above claim upper-bounds the dimension of the common kernel of $\{\Pi_{[j,j+2]} \}$ in OBC by $2L+2$, and the numerical study summarized in App.~\ref{subsec:OBCAKLT} [see Eq.~(\ref{eq:3sitekerneldim})] shows that this is in fact exact answer.
However, we do not have a direct analytical argument other than noting that the count basically matches naive count of the states in the four AKLT towers in OBC, up to few states for certain system sizes.
One challenge 
is that a given configuration may be connected to reference configurations by multiple paths (i.e., sequences of moves), and we have to assure the amplitude evaluated along each path is independent of the path.
Specifically, Eqs.~(\ref{eq:-0}) and~(\ref{eq:++0}) contain a minus sign when relating the amplitudes, while Eq.~(\ref{eq:-+-}) contains a factor of $2$.
While in the latter case we can see path independence by relating the factor of $2$ to the change in $N_0$ by $2$, tying this to the configuration property independent of the path taken, it is not clear how to do this in the former cases. 
We can nevertheless make the following simple observations.
The states in the family i) have distinct $S^z_{\tot}$ taking values $-1,0,1,\dots,L-3,L-2$, while the states in the family ii) have distinct $S^z_{\tot}$ taking values $0,1,\dots,L-3,L-2$.
All the moves considered preserve $S^z_{\tot}$, hence the amplitudes of states in these families with distinct $S^z_{\tot}$ should be independent.
Furthermore, states in the families i) and ii) with the same $S^z_{\tot}$ cannot be connected by the moves in Eqs.~(\ref{eq:-0})-(\ref{eq:-+-}):
Indeed, it is easy to see that these moves preserve the parity of the number of $+$'s to the right of the rightmost $-$.  
Hence, we do not need to worry about connections under such moves between any states in the families i) and ii), only about possible ``cycles'' involving only one such state at a time; we also still need to worry whether Eqs.~(\ref{eq:+-}) and (\ref{eq:0+0}) together with cycles may further constrain the amplitudes for the two families; again, we do not attempt full analytic proofs but are content with the obtained upper bound that agrees with the numerical results.
In any case, the numerical study provides the definite answer for the size of the common kernel and its relation to the AKLT towers and any few extra states.
\subsection{Three-site bulk projectors and two-site boundary projectors with OBC }\label{subsec:AKLTanalyticOBC}
We can also see how the two-site boundary projectors added in App.~\ref{subsec:OBCAKLT} reduce the size of the kernel, see Eq.~(\ref{eq:AKLTOBCkerneldim}).
Six states orthogonal to $\{\ket{l_a}_{1,2}, a \in \{1,2,3\} \}$ of Eq.~(\ref{eq:lr2sitesubspace}) are $\{\ket{v_b}_{1,2}, b \in \{1,\dots,6\} \}$ given by
\begin{align}
\ket{v_1} \defn \ket{--}, \quad 
\ket{v_2} \defn \ket{-0}, \quad 
\ket{v_3} \defn \ket{0-}, \quad 
\ket{v_4} \defn \ket{-+}, \quad 
\ket{v_5} \defn \ket{+0} + \ket{0+}, \quad 
\ket{v_6} \defn \ket{+-} + \ket{00}, 
\end{align}
and annihilation by $\Pi_{1,2}^{(l)}$ is equivalent to annihilation by all $\ketbra{v_b}_{1,2}$.
Similarly, six states orthogonal to $\{\ket{r_a}_{L-1,L}, a \in \{1,2,3\} \}$ of Eq.~(\ref{eq:lr2sitesubspace}) are $\{\ket{w_b}_{L-1,L}, b \in \{1,\dots,6\} \}$ given by
\begin{align}
\ket{w_1} \defn \ket{--}, \quad 
\ket{w_2} \defn \ket{-0}, \quad 
\ket{w_3} \defn \ket{0-}, \quad 
\ket{w_4} \defn \ket{+-}, \quad 
\ket{w_5} \defn \ket{+0} + \ket{0+}, \quad 
\ket{w_6} \defn \ket{-+} + \ket{00},
\end{align}
and annihilation by $\Pi_{L-1,L}^{(r)}$ is equivalent to annihilation by all $\ketbra{w_b}_{L-1,L}$.
We consider additionally requiring annihilation by these projectors for each of the three families in Lem.~\ref{lem:OBCAKLTkercount}.
Starting first with the family i), requiring annihilation by $\ketbra{w_2}_{L-1,L}$ and $\ketbra{w_4}_{L-1,L}$, and implementing the ``moves" discussed in App.~\ref{subsec:PsiOBC} gives that the amplitude on any state in this family with at least one $+$ must be zero for any $L$; since any such state can be related to a state with either the $\ket{w_2}$ or $\ket{w_4}$ on sites $\{L-1,L\}$ by pushing $00$ units to the left, and hence must have zero amplitude.
For $L$ even, this includes also the state with no $+$'s, i.e., the very first of the listed states in the family i), while for $L$ odd we can add requiring annihilation by  $\ketbra{w_3}_{L-1,L}$ to conclude that the amplitude on this state is also zero.
Thus, in either case, the family i) does not survive the addition of these boundary projectors.
We now consider the family ii).
Requiring annihilation by $\ketbra{v_4}_{1,2}$ gives that the amplitude must be zero for each state in this family where the $-$ is located on an odd site (labeling the sites $1,2,\dots,L$), where we recall that the amplitudes on configurations connected by moving $++$ units are related.
This eliminates $L/2$ and $(L+1)/2$ states from the family ii) for $L$ even and odd respectively, leaving at most $L/2-1$ and $(L-3)/2$ independent non-zero amplitudes for $L$ even and odd respectively.
Finally, we consider the family iii).
Adding requiring annihilation by $\ketbra{w_5}_{L-1,L}$ relates the amplitudes of the last two states in this family, thus leaving at most 2 independent amplitudes.
Putting everything together, we have at most $(L/2+1)$ independent amplitudes for $L$ even and $(L+1)/2$ independent amplitudes for $L$ odd.
This matches the result of the numerical study in App.~\ref{subsec:OBCAKLT}, see Eq.~(\ref{eq:AKLTOBCkerneldim}).
Again, we do not have a full direct proof for the same reason as before, where we cannot exclude the possibility that different amplitudes acquired along different paths leading to the same reference states would enforce some of the amplitudes in the family ii) to be zero.
However, an indirect argument is that this count matches the count of the states in the AKLT tower $\{ \ket{\psi_{n,\uparrow\uparrow}}\}$, which we know are indeed annihilated by these projectors.
\subsection{Three-site bulk projectors with PBC}\label{subsec:AKLTanalyticPBC}
Here we start with the bulk OBC analysis in App.~\ref{subsec:PsiOBC}
leading to the three families in 
Lem.~\ref{lem:OBCAKLTkercount} there, and add requirement of annihilation by $\Pi_{[j,j+2]}$ with $j=L-1$ and $j=L$, with identifications $L+1 \equiv 1, L+2 \equiv 2$, i.e., going across the PBC connection between the sites $L$ and $1$.
This gives the same requirements, $\Psi$ in Eqs.~(\ref{eq:--})-(\ref{eq:0+0}), but now the exhibited two-site or three-site locations can happen anywhere on the circle formed by the PBC chain, and the amplitudes on configurations can now be related also by local moves that go across the PBC connection; hence we can reduce the number of independent amplitudes using the extra moves, as we will now describe.
Throughout, we assume $L$ is even, which is required for the PBC tower of states to exist.
Consider first the family iii).
The PBC makes no difference for ferromagnetic configuration $\ket{F}$, which is trivially annihilated by all $\ketbra{u_a}_{[j,j+2]}$ on any three sites.
For the other two reference configurations, the corresponding OBC constructions of eigenstates in Eq.~(\ref{eq:famIIIeigstates}), states $\ket{\text{iii}_1}$ and $\ket{\text{iii}_2}$, can nicely fit in PBC if $L = 4n$ and form the $\ket{1_{k=\pm \pi/2}}$ states of Eq.~(\ref{eq:1kstates}), so we have two more states in the common kernel for such $L$.
On the other hand, if $L=4n+2$, the states $\ket{\text{iii}_1}$ and $\ket{\text{iii}_2}$ in Eq.~(\ref{eq:famIIIeigstates}) do not ``fit" in PBC, and we can show that for any $\ket{\Psi}$ in the common kernel its amplitudes on these configurations must be zero.
As an example, we show this for the configuration $\ket{+,+,+\dots,+,+,0}$.
Using Eq.~(\ref{eq:++0}), we hop the $0$ over the $++$'s all the way to the left and then across the PBC, returning to the original configuration but with sign $(-1)^{L/2} = -1$ for $L/2$ odd; hence the amplitude must be zero.
Thus, for the family iii), we have $3$ states in the common kernel if $L=4n$ but only $1$ state if $L=4n+2$.
Turning to the families i) and ii), we first note that all such configurations where the number of $+$ states $N_+$ is even must have zero amplitude, which is consistent with the absence of states with $S^z_{\tot}$ odd from the PBC AKLT tower of QMBS.
We will show this for the family i), while the family ii) will follow easily.
Consider first configurations in the family i) with $N_+$ even and at least two $0$'s.
Working near the single $-$, we perform moves $-00 \to 0\!-\!0 \to -\!+\!-$ [Eqs.~(\ref{eq:-0}) and (\ref{eq:-+-})], we can use Eq.~(\ref{eq:++-}) to hop the left $-$ across $++$'s to the left if $N_+ \neq 0$ (this is where the evenness of the original $N_+$ is important), then we can use Eq.~(\ref{eq:-0}) to move the same $-$ across the PBC across all $0$'s, eventually bringing this $-$ next to the other $-$, and this configuration must have zero amplitude by Eq.~(\ref{eq:--}).
Consider now the family i) case with precisely one $0$,
$\ket{+,+,+,\dots,+,-,0}$, which is the only case left with $N_+$ even.
We can relate it to configuration $\ket{+,+,+,\dots,0,-,+}$ using the following two different sequences of moves.
In the first sequence, we do right hops of $0$ over $++$'s using Eq.~(\ref{eq:++0}), performing $L/2-1$ such moves and acquiring minus sign from each move, for the total factor of $(-1)^{L/2-1}$.
In the second sequence, we first do $-0+ \to 0-+ \to -\!+\!0$ using Eqs.~(\ref{eq:-0}) and (\ref{eq:0-+}) and acquire a minus sign [this move goes across the PBC]; we then perform right hops of $0$ over $++$'s, a total of $L/2-2$ moves that results in a sign factor of $(-1)^{L/2-2}$; finally, we do $0\!+\!- \to + - 0 \to +0-$ using Eqs.~(\ref{eq:0+-}) and (\ref{eq:-0}) and acquire one more minus sign, for the total factor of $(-1)^{L/2}$ for this sequence.
We can then see that the accumulated signs differ for the two sequences, and hence the amplitude must be zero.
The family ii) cases with $N_+$ even can be treated identically to the family i) cases, by first moving $-+$ to the right across all $0$'s using Eq.~(\ref{eq:0-+}), reaching configuration of the form $\ket{+,+,\dots,+,0,\dots,0,-, +}$, which on the PBC circle has the same structure as in the family i) considered above, and the same arguments apply showing that such configurations with $N_+$ even must have zero amplitude.
Thus, from the family i) we are left to consider $L/2$ configurations $\ket{+,-,0,0,0,0,\dots,0,0}$, $\ket{+,+,+,-,0,0,\dots,0,0}$, $\dots$, $\ket{+,+,+,+,+,+,\dots,+,-}$, while from the family ii) we have $L/2$ configurations $\ket{-,+,0,0,0,0,\dots,0,0}$, $\ket{+,+,-,+,0,0,\dots,0,0}$, $\dots$, $\ket{+,+,+,+,+,+,\dots,-,+}$, both with $N_+$ odd.
It is easy to see that the first $L/2-1$ configurations in the family ii) can be related to the corresponding $L/2-1$ configurations in the family i): we can move the right-most $+$ across all $00$'s all the way to the right [using Eq.~(\ref{eq:00+})], then the leftmost $0$ across the $+-$ [using Eq.~(\ref{eq:0+-})] and then across all $++$'s to the left, including across the last $++$ crossing the PBC, to arrive at a configuration in the family i).
Thus, we have related the family i) and family ii) configurations with at least two $0$'s.
Finally, we can show that the amplitude on the last configuration in the family i) and the last configuration in the family ii)---the configurations with no $0$'s---are also related, which we do separately for $L = 4n$ and $L = 4n+2$.
For $L=4n$, we use Eq.~(\ref{eq:+-}) to write
\begin{equation}
\Psi(+,+,+,\dots,+,+,-) + \Psi(+,+,+,\dots,+,-,+) + 2 \Psi(+,+,+,\dots,+,0,0) = 0
\end{equation}
and then show that $\Psi(+,+,+,\dots,+,0,0) = 0$ for such $L$.
By starting with $\ket{+,+,+,\dots,+,0,0}$, we can move the left $0$ across $L/2-1$ instances of $++$ to the left, acquiring minus sign for each move; we now have $00$ across the PBC, which we move to the left across one $+$ [using Eq.~(\ref{eq:00+})], returning to the very initial configuration but with sign $(-1)^{L/2-1} = -1$ for $L/2$ even, hence we must have $\Psi(+,+,+,\dots,+,0,0) = 0$, which implies $\Psi(+,+,+,\dots,+,+,-) = -\Psi(+,+,+,\dots,+,-,+)$.
For $L=4n+2$, we instead use Eq.~(\ref{eq:0+0})
\begin{equation}
2\Psi(+,+,+,\dots,0,+,0) + \Psi(+,+,+,\dots,+,+,-) - \Psi(+,+,+,\dots,+,-,+) = 0
\end{equation}
and show that $\Psi(+,+,+,\dots,0,+,0) = 0$ for such $L$.
Starting with configuration $\ket{+,+,+,\dots,0,+,0}$, we first hop the left $0$ across $L/2-2$ of $++$'s to the left (landing at site $j=2$); then hop the right $0$ across one $++$ to the left (this $0$ then lands at the site $j=L-2$); finally, we hop the other $0$ (currently at site $j=2$) across $++$ to the left, going across the PBC and landing at $j=L$.
We have thus returned to the original configuration after performing $L/2$ moves of the type Eq.~(\ref{eq:++0}), which accumulates sign $(-1)^{L/2} = -1$ for $L/2$ odd.
This proves that the corresponding amplitude must be zero, which implies $\Psi(+,+,+,\dots,+,+,-) = \Psi(+,+,+,\dots,+,-,+)$.
To summarize, from the families i) and ii), we end up with at most $L/2$ independent amplitudes.
Combining with the analysis of the family iii), we conclude that there are at most $L/2+3$ independent amplitudes for $L=4n$ and $L/2+1$ for $L=4n+2$, which appears to match the results of the numerical study in Eq.~(\ref{eq:3sitekernelPBC}).
This is also the total count of the AKLT tower of states $\{\ket{\psi_n}\}$ and the extra states $\{\ket{\phi_m}\}$ discussed in App.~\ref{subsec:PBCAKLT}, which we know are part of the common kernel.
\section{Some Details on the Spin-1 Models with Bimagnon Towers as QMBS}
\label{app:S1XY}
In this appendix, we provide some details behind the discussion in Sec.~\ref{subsec:spin1XYHubbardscar} of the bond algebras for the spin-1 $k=\pi$ and $k=0$ bimagnon scar towers, connecting with previous works~\cite{iadecola2018exact, mark2020eta}.
\subsection{Shiraishi-Mori Projectors and Bond Algebras}\label{subsec:SMproj}
The Shiraishi-Mori projectors, i.e., the set of projectors such that the QMBS span their common kernel, are conveniently defined using the following two-site states $\ket{X_a}_{j,j+1}$ (reusing notation from Apps.~C and D in \cite{mark2020eta}):
\begin{gather}
 \ket{X_1} = \frac{1}{\sqrt{2}}(\ket{1,-1} + \ket{-1,1})~, \quad 
\ket{X_2} = \ket{0,0}~, \quad
\ket{X_3} = \ket{1,0}~, \quad
\ket{X_4} = \ket{0,1}~, \quad
\ket{X_5} = \ket{-1,0}~, \quad
\ket{X_6} = \ket{0,-1}~, \label{eq:ketXs}\\
 \ket{X_7} = \frac{1}{\sqrt{2}}(\ket{1,-1} - \ket{-1,1})~, \quad
\ket{X_8} = \ket{1,1}~, \quad
\ket{X_9} = \ket{-1,-1}~. \nonumber
\end{gather}
For the $k=\pi$ bimagnon tower defined in Eq.~(\ref{eq:spin1XYQMBS}), the Shiraishi-Mori projector is
\begin{equation}
P^{(\XY, \pi)}_{j,j+1} \defn \sum_{a=1}^6 \ketbra{X_a}_{j,j+1} ~.
\label{eq:XYSMproj}
\end{equation}
For two sites $j$ and $j+1$, it is easy to verify that the bond algebra of $U(1)$ spin conserving
operators such that the $k = \pi$ QMBS of Eq.~(\ref{eq:spin1XYQMBS}) are degenerate eigenstates, is spanned by the 13 linearly independent operators $\{\mathds{1}_{j,j+1};~  \ketbra{X_a}{X_b}_{j,j+1},~ a,b \in \{1,2\}, \text{~or~} a,b \in \{3,4\}, \text{~or~} a,b \in \{5,6\} \}$.
To give some examples,
$\ketbra{X_1}{X_2}_{j,j+1} = (S_j^z)^2 (S_{j+1}^z)^2 (S_j^x S_{j+1}^x + S_j^y S_{j+1}^y)/\sqrt{2}$;
$\ketbra{X_3}_{j,j+1} = (S_j^z)^2 \left[ 1 - (S_{j+1}^z)^2 + (S_j^z + S_{j+1}^z)(1 - S_j^z S_{j+1}^z) \right]/2$;
etc.
Thus, we can denote the bond algebra $\tmA^{(\XY)}_{\scar}$ of Eq.~(\ref{eq:spin1XYpair}) also as
\begin{equation}
\tmA^{(\XY,\pi)}_{\scar} \defn \tmA^{(\XY)}_{\scar} = 
\lgen \{ \ketbra{X_a}{X_b}_{j,j+1},~ a,b \in \{1,2\}, \text{~or~} a,b \in \{3,4\}, \text{~or~} a,b \in \{5,6\} \} \rgen ~,
\label{eq:tmAS1kpi}
\end{equation}
where the sites $j$ and $j+1$ belong to the different sublattices of the bipartite chain.
For the $k=0$ bimagnon tower, the Shiraishi-Mori projector is 
\begin{equation}
P^{(\text{XY,0})}_{j, j+1} \defn \sum_{a=2}^7 \ketbra{X_a}_{j,j+1} ~.
\end{equation}
It is very similar to the $k=\pi$ bimagnon tower projector, except that $\ketbra{X_1}_{j,j+1}$ is replaced by $\ketbra{X_7}_{j,j+1}$ (so that requiring annihilation by $\ketbra{X_7}_{j,j+1}$ locally enforces ``$k=0$" $\ket{X_1}_{j,j+1}$ in the wavefunction, as opposed to locally ``$k=\pi$" $\ket{X_7}_{j,j+1}$ enforced by requiring annihilation by $\ketbra{X_1}_{j,j+1}$).
The corresponding $U(1)$ spin 
conserving bond algebra can be defined as
\begin{equation}
\tmA^{(\text{XY,0})}_{\scar} \defn 
\lgen \{ \ketbra{X_a}{X_b}_{j,j+1},~ a,b \in \{7,2\}, \text{~or~} a,b \in \{3,4\}, \text{~or~} a,b \in \{5,6\} \} \rgen ~,
\label{eq:tmAS1k0}
\end{equation}
where we can assume that $j$ and $j+1$ belong to different sublattices just like in Eq.~(\ref{eq:tmAS1kpi}), although this is not necessary for this $k = 0$ bimagnon tower.
\subsection{Unitary transformation between the \texorpdfstring{$k = 0$}{} and \texorpdfstring{$k = \pi$}{} towers}\label{subsec:unitaryspin1xy}
As discussed in Sec.~\ref{subsec:spin1XYHubbardscar}, the $k = \pi$ and $k = 0$ bimagnon towers can be mapped to one another on a bipartite lattice using a unitary transformation $\hat{U} = \prod_j \hat{u}_j$, with $\hat{u}_j = \exp(i \frac{\pi}{2} S^z_j)$ on one sublattice [conjugation by this $u_j$ maps $S_j^{\pm} \to \pm i S_j^{\pm}$, $(S_j^{\pm})^2 \to -(S_j^{\pm})^2$] and $\hat{u}_j = \mathds{1}$ on the other.
Indeed, it is easy to see that this transformation maps $\ket{X_a}_{j,j+1},~2 \leq a \leq 6$, to themselves up to unimportant phases, while it exchanges $\ket{X_1}_{j,j+1}$ and $\ket{X_7}_{j,j+1}$,
hence conjugating using this unitary realizes an isomorphism of the two algebras $\tmA^{(\XY,\pi)}_{\scar}$ and $\tmA^{(\XY,0)}_{\scar}$.
Also, 
term \#12 of Eq.~(\ref{eq:S1asasum}), which annihilates the $k=\pi$ bimagnon tower, maps to itself up to a sign.
Hence, for all our purposes it suffices to focus on the $k=0$ case, and all results can be translated to the $k=\pi$ case by this unitary transformation.
\subsection{Permutation group in the bond algebra}\label{subsec:permspin1xy}
Using Eq.~(\ref{eq:tmAS1k0}) it is easy to see that $\tmA^{(\XY, 0)}_{\scar}$ contains two-site exchange operators $P_{j,j+1}^{\exch} \defn \sum_{\alpha,\beta} \ketbra{\alpha,\beta}{\beta,\alpha}_{j,j+1}$, namely:
\begin{equation}
P_{j,j+1}^{\exch} = \Big(
\ketbra{X_3}{X_4} + \ketbra{X_4}{X_3} + \ketbra{X_5}{X_6} + \ketbra{X_6}{X_5} + \mathds{1} - \ketbra{X_3} - \ketbra{X_4} - \ketbra{X_5} - \ketbra{X_6} - 2 \ketbra{X_7} \Big)_{j,j+1} ~.
\label{eq:Pexchspin1XY}
\end{equation}
This bond algebra hence also contains arbitrary exchange operators $P_{\ell,m}^{\exch}$, and consequently also contains a representation of the permutation group on $L$ sites.
Just like in the spin-1/2 ferromagnetic tower, the presence of the exchange operators in the bond algebra is ultimately responsible for the spatial structure of the singlets of this algebra, i.e., the states of the $k = 0$ bimagnon tower.
In particular, this implies that these states lack of any spatial structure: When any such singlet state is expanded in the computational basis, all permutation-related basis states must have the same amplitude.
Indeed, the properties $(P_{\ell,m}^{\exch})^2 = 1$ and $P_{\ell,m}^{\exch} P_{m,n}^{\exch} P_{\ell,m}^{\exch}  = P_{\ell,n}^{\exch}$ guarantee that any common eigenstate of all such exchange operators must have the same eigenvalue under all the exchange operators; furthermore, this eigenvalue cannot be $(-1)$ for large system sizes.
\subsection{Mapping 
term \#12 to the spin-1/2 DMI term} 
Finally, App.~D in~\cite{mark2020eta} showed that the $k=0$ bimagnon tower and the corresponding extensive local annihilator, Eq.~(\ref{eq:S1asasum}), can be related precisely to the spin-1/2 ferromagnetic tower and the DMI annihilator discussed in Sec.~\ref{subsec:FMtowerscar}.
This is achieved by relating $\ket{+} \leftrightarrow \ket{\uparrow}$, $\ket{-} \leftrightarrow \ket{\downarrow}$, which also relates the spin-1 operators $\frac{1}{2} S_j^z, \frac{1}{2} (S_j^{\pm})^2$ to spin-1/2 operators $S_j^z, S_j^{\pm}$.
Hence the proof in App.~\ref{subsec:DMIimpossible} that the DMI annihilator cannot be represented as a sum of strictly local annihilators carries over to the spin-1 Hamiltonian Eq.~(\ref{eq:S1asasum}), showing that it is an example of a Type II symmetric Hamiltonian for the $k=0$ bimagnon scar tower (and hence also for the $k=\pi$ bimagnon scar tower by the unitary mapping).
\section{Sufficient Conditions for QMBS}\label{app:QMBSsufficient}
In this appendix, we prove Lem.~\ref{lem:SMsufficient}, which provides
%
a set of sufficient conditions that a state $\ket{\psi_{\candt}}$ on a system of size $L$ can satisfy for it to be a QMBS of some local Hamiltonian.
We restate it here for convenience.
\smrecsuf*
\subsection{Existence of an exponentially large block}\label{subsec:explargeblock}
Before proving the main result, we note that for deriving some simple results it is more natural and convenient to assume that the union of the supports of $\{A_{[j]} \}$ completely covers the lattice and that among these there are non-commuting operators occurring at finite density across the lattice\footnote{Indeed, the recovery of a strictly local $A_{[j]}$ that annihilates $\ket{\psi_{\text{canddt}}}$ also implies the recovery of strictly local larger ranged operators of the form $\{O_{[j]} A_{[j]}\}$}.
In such a case, if $r_{\max} \geq 2$ and the $A_{[j]}$'s with overlapping supports do not commute, we can already show that the parent algebra $\mA = \lgen \{A_{[j]}\} \rgen$ possesses an exponentially large block.
%
The dimension of $\mA$ clearly grows exponentially in $L$, since we can construct strings of operators such as $A_{[j_1]} A_{[j_2]} \cdots A_{[j_k]}$ for any $k$ and any positions $j_1, j_2, \dots, j_k$ that are sufficiently far from each other such that the supports of any two $A_{[j]}$'s do not overlap, and there are exponentially many such linearly independent strings.
Concretely, we can divide the lattice into $L/n$ non-overlapping regions $\{R_\ell\}_{\ell = 1}^{L/n}$ of roughly equal size $n \sim \mO(1)$ such that each $R_\ell$ contains at least one pair of non-commuting $A_{[j]}$'s.
We can then lower bound $\dim(\mA)$ in terms of $D_{A,\loc} \defn \dim(\mA_{R_{\ell}})$ (assumed the same for all $\ell$ for simplicity), where $\mA_{R_\ell}$ is the algebra generated by $A_{[j]}$'s that are \textit{completely} within $R_{\ell}$; and it is easy to see that $\dim(\mA) \geq D_{A,\loc}^{L/n}$ by the above ``strings of operators" argument.
Since $\mA_{R_\ell}$ is non-Abelian by definition, $D_{Z,\loc} \defn \dim(\mZ_{R_{\ell}}) < D_{A,\loc}$, where $\mZ_{R_{\ell}}$ is the center of $\mA_{R_{\ell}}$. 
The full center $\mZ$ of $\mA$ should be contained in $\lgen \{\mZ_{R_\ell}\}_{\ell=1}^{L/n} \rgen$, hence we obtain $\dim(\mZ) \leq D_{Z,\loc}^{L/n}$.
We then note a general bound 
\begin{equation}
    \dim(\mA) = \sum_{\lambda = 1}^{\dim(\mZ)}{D^2_\lambda}\;\;\implies\;\;\max_\lambda(D_\lambda) \geq \sqrt{\frac{\dim(\mA)}{\dim(\mZ)}},
\label{eq:genbound}
\end{equation}
and applying this to the recovered algebra $\mA$ we obtain 
\begin{equation}
    \max_\lambda (D_\lambda) \geq \left(\frac{D_{A,\loc}}{D_{Z,\loc}}\right)^{\frac{L}{2n}} \sim \exp(c L),
\end{equation}
implying the existence of an exponentially large block. 
Another perspective on this bound is that each region $R_\ell$ ``gives'' a qudit of dimension (at least) $\sqrt{D_{A,\text{loc}}/D_{Z,\text{loc}}}$ and the bond algebra contains arbitrary operators in the Hilbert space of $L/n$ such qudits.
\subsection{Existence of a ``thermal" block satisfying Eq.~(\ref{eq:maxDlambda})}\label{subsec:singleQMBS}
We now proceed to the main proof of Lem.~\ref{lem:SMsufficient}.
Note that we can assume $A_{[j]}$'s w.l.o.g.\ to be projectors $P_{[j]}$'s, for some value of $r_{\max} \geq 1$, since the recovery of $A_{[j]}$ implies the recovery of all its powers. 
We then wish to show that there exists a finite $r'_{\max} \geq r_{\max}$ such that Eq.~(\ref{eq:maxDlambda}) is satisfied for the (irreps of the) algebra $\tmA_{\SM} \defn \lgen \{\wh_{[j]}\} \rgen$, where $\wh_{[j]}$ are the strictly local reconstructed operators with range bounded by $r'_{\max}$.
This is closely related to the result of Lem.~\ref{lem:smsuf} proven in App.~\ref{app:SMexistence}, hence we have suggestively denoted the reconstructed algebra with the same notation $\tmA_{\SM}$. 
There the operators $\{\wh_{[j]}\}$ that generate such an algebra are obtained by ``dressing" the projectors $P_{[j]}$ by longer range terms, e.g., $\wh_{[j]} = P_{[j]} \otimes O_{\text{nb}}(j)$, where $\text{nb}(j)$ denotes some finite neighborhood of $j$ not including the support of $P_{[j]}$. 
Since the projectors vanish on $\ket{\psi_{\candt}}$ by definition, so do the operators $\wh_{[j]}$ and they would be obtained by the reconstruction procedure applied with a range $r'_{\max}$.
Hence the algebra of local operators reconstructed with this range $r'_{\max}$ necessarily satisfies
\begin{equation}
    \max_\lambda(D_\lambda) \geq \dim(\mH) - \dim(\mT)\;\;\implies\;\;\frac{\max_\lambda D_\lambda}{\dim(\mH)} \geq 1 - \frac{\dim(\mT)}{\dim(\mH)},
\label{eq:SMalgdim}
\end{equation}
where $\dim(\mT)$ is the dimension of the target space $\mT$, the common kernel of the projectors $\{P_{[j]}\}$.  
For the sake of illustration, we return to the example of the spin-1 chain used in App.~\ref{app:SMexistence} with on-site states labelled $\ket{0},\ket{+},\ket{-}$.
Suppose we have a candidate state $\ket{\psi_{\candt}}$ for which we have recovered, using $r_{\max} = 1$, that it is annihilated by all on-site operators $\ketbra{0}_j$, which form the set $\{P_{[j]}\}$ discussed above.
If we perform our test for QMBS procedure with range $r'_{\max} = 2$, we will recover also the following two-site operators that annihilate this state: $\ketbra{0}_j \otimes h_{j+1}$ and $h_{j-1} \otimes \ketbra{0}_j$, where $h_{j-1}$ and $h_{j+1}$ can be arbitrary on-site operators acting on sites $j-1$ and $j+1$; these operators form the set $\{\wh_{[j]}\}$ discussed above.
The common kernel of the projectors $\{\ketbra{0}_j\}$ in this case is the space spanned by product states with no $\ket{0}$'s, hence $\dim(\mT) = 2^L$. 
As we showed in App.~\ref{app:SMexistence}, this algebra acts irreducibly in $\mT^\perp$, the space spanned by basis product states with at least one on-site $\ket{0}$.
Thus, in this case we have $\max_\lambda D_\lambda \geq 3^L - 2^L$ for the irreps of $\tmA_{\SM}$, and hence Eq.~(\ref{eq:maxDlambda}) is satisfied.
We now prove that as long as the projectors $P_{[j]}$ are ``dense" on the lattice,  $\dim(\mT)/\dim(\mH) \leq p^L$  in general for some $p < 1$, hence Eq.~(\ref{eq:maxDlambda}) is satisfied.
We follow the same procedure as in App.~\ref{app:SMexistence}, first dividing the set of projectors $\mP \defn \{P_{[j]}\}$ into a finite number of non-overlapping ``dense" subsets $\{\mP_\alpha\}$ such that $\mP = \bigcup_{\alpha}{\mP_\alpha}$, and studying each of these subsets separately.
In particular, we note that the common kernel $\mT$ can be expressed in terms of the common kernel of these non-overlapping subset of projectors as $\mT = \bigcap_{\alpha} \mT_\alpha$, hence we obtain the bound for its dimension $\dim(\mT) \leq \min_\alpha \dim(\mT_\alpha)$
Combining this with the bound of Eq.~(\ref{eq:dimTalpbound}), for large $L$ we obtain
\begin{equation}
    \frac{\dim(\mT)}{\dim(\mH)} \leq p^L\;\;\text{for some}\;\;p < 1, 
\label{eq:fracTbound}
\end{equation}
which shows that Eq.~(\ref{eq:maxDlambda}) is satisfied as $L \rightarrow \infty$.
\subsection{Hilbert Space Dependence of Scarriness}\label{subsec:QMBSremarks}
In this section, we provide a few concrete examples that illustrate that the notion of scarriness depends on the Hilbert space being considered and quantities we are interested in.
First, we demonstrate an example from the spin-1 illustration discussed in Apps.~\ref{app:SMexistence} and \ref{subsec:singleQMBS}.
Suppose we start with a candidate state $\ket{\Psi_{\text{canddt}}}$ originating as an eigenstate of a generic non-integrable Hamiltonian $H_{\text{spin-1/2}}$ acting in the space of states spanned by $\{ \ket{\sigma_1,\sigma_2,\dots,\sigma_L}, \sigma_j \in \{+,-\}\}$, which we will loosely call the spin-1/2 subspace.
When subjected to the test for QMBS procedure discussed in Sec.~\ref{subsec:QMBSdefn}, it is reasonable to expect that the reconstructed parent operators at finite $r_{\text{max}}$ would be the previously considered $\{ \ketbra{0}_j \otimes h_{j+1}, h_{j-1} \otimes \ketbra{0}_j \}$ plus only the original Hamiltonian $H_{\text{spin-1/2}}$ (note that multiplying an extensive local operator like $H_{\text{spin-1/2}}$ by a strictly local operator gives a very non-local operator and would not be considered by the procedure).
If this is true, then all eigenstates of $H_{\text{spin-1/2}}$ in the spin-1/2 subspace are singlets of the algebra of reconstructed local operators, while the algebra acts irreducibly in the orthogonal complement to these states, i.e., in the space of states spanned by basis states with at least one on-site $\ket{0}$, as discussed in App.~\ref{app:SMexistence}.
Hence this reconstruction on the state demonstrates that while it is generically thermal as a spin-1/2 state [since Eq.~(\ref{eq:maxDlambda}) is not satisfied when it is viewed in the restricted ``spin-1/2" Hilbert space], it is a QMBS as a spin-1 state in the whole Hilbert space since Eq.~(\ref{eq:maxDlambda}) is then satisfied.
Interestingly, this reconstruction also informs that as a QMBS it must come together with all the other eigenstates of $H_{\text{spin-1/2}}$ as QMBS.
Incidentally, so-called Yang-Zhang Slater determinant states in the Hubbard model considered in \cite{yang1990so}, constructed by populating plane wave states of only one spin species, would be qualitatively similar to this spin-1 model example and could also be called QMBS, if we ignore the free-fermion character of the construction involving only the kinetic energy of the fermions (which is easy to remedy if we allow in addition interactions involving only fermions of the same spin species).
Finally, we consider an example of a candidate state that is of the form  $\ket{\Psi_{\text{thermal}}} \otimes \ket{0}$, where $\ket{\Psi_{\text{thermal}}}$ is some thermal state on some Hilbert space, and $\ket{0}$ is the state on a ``dummy" qubit. 
This state could perhaps be called a QMBS with respect to the added single qubit, e.g., the expectation value of any operator on the dummy qubit would be highly non-generic. 
On the other hand, the algebra of reconstructed local operators with a finite $r_{\max}$ would not satisfy Eq.~(\ref{eq:maxDlambda}), consistent with intuition that this should not be a QMBS from the point of view of the bulk of the system. 

\end{document}